\documentclass[3p,twocolumn,usenatbib]{elsarticle}

\usepackage{lineno}
\usepackage[draft]{hyperref}

\usepackage{amsmath}	% Advanced maths commands
\usepackage{amssymb}	% Extra maths symbols

\modulolinenumbers[5]

\journal{New Astronomy Reviews}

\biboptions{authoryear}

\newcommand{\INTEGRAL}{{\it INTEGRAL}}
\newcommand{\Lhx}{L_{\rm hx}}
\newcommand{\Lx}{L_{\rm x}}

\begin{document}

\begin{frontmatter}

\title{The Galactic LMXB Population and the Galactic Centre Region}%\tnoteref{mytitlenote}}

\author[iki,hse]{S. Sazonov\corref{cor}}
\cortext[cor]{Corresponding author}
\ead{sazonov@iki.rssi.ru}

\author[inaf-iasf]{A. Paizis}
\author[inaf-iaps]{A. Bazzano}
\author[iki]{I. Chelovekov}
\author[mpa,iki]{I. Khabibullin}
\author[msu]{K. Postnov}
\author[iki]{I. Mereminskiy}
\author[inaf-iaps]{M. Fiocchi}

\author[esac]{G. B\'{e}langer}
\author[south]{A.J. Bird}
\author[geneva]{E. Bozzo}
\author[lyngby]{J. Chenevez}
\author[palermo]{M. Del Santo}
\author[issi]{M. Falanga}
\author[bologna]{R. Farinelli}
\author[geneva]{C. Ferrigno}
\author[iki]{S. Grebenev}
\author[iki]{R. Krivonos}
\author[esac,estec]{E. Kuulkers}
\author[lyngby]{N. Lund}
\author[esac]{C. Sanchez-Fernandez}
\author[ostia]{A. Tarana}
\author[inaf-iaps]{P. Ubertini}
\author[bamberg]{J. Wilms}

\address[iki]{Space Research Institute, Profsoyuznaya 84/32, 117997 Moscow, Russia}
\address[hse]{National Research University Higher School of Economics, Myasnitskaya ul. 20, 101000 Moscow, Russia}
\address[inaf-iasf]{INAF-IASF, via Alfonso Corti 12, I-20133 Milano, Italy}
\address[inaf-iaps]{INAF-IAPS, via Fosso del Cavaliere, 00133 Roma, Italy}
\address[mpa]{Max-Planck-Institut f\"{u}r Astrophysik, Karl-Schwarzschild-Strasse 1, 85741 Garching, Germany}
\address[msu]{Sternberg Astronomical Institute, M.V. Lomonosov Moscow State University, Universitetskij pr. 13, 119234 Moscow, Russia}
\address[esac]{ESA/ESAC, 28691 Villanueva de la Ca\~{n}ada, Madrid, Spain}
\address[south]{University of Southampton, United Kingdom}
\address[geneva]{University of Geneva, Department of Astronomy, Chemin d'Ecogia 16, 1290, Versoix, Switzerland}
\address[lyngby]{DTU Space, National Space Institute, Technical University of Denmark, Elektrovej 328, DK-2800 Lyngby, Denmark}
\address[palermo]{INAF-IASF Palermo, Via U. La Malfa 153, I-90146 Palermo, Italy}
\address[issi]{International Space Science Institute, Hallerstrasse 6, CH-3012 Bern, Switzerland}
\address[bologna]{INAF, Osservatorio di Astrofisica e Scienza dello Spazio, via Gobetti 101, I-40129 Bologna, Italy}
\address[estec]{ESA/ESTEC, Keplerlaan 1, 2201 AZ Noordwijk, The Netherlands}
\address[ostia]{Liceo Scientifico F. Enriques, via F. Paolini, 196, I-00122, Ostia (Rome), Italy}
\address[bamberg]{Dr. Karl Remeis-Sternwarte and Erlangen Centre for Astroparticle Physics, Universit\"{a}t Erlangen-N\"{u}rnberg, Sternwartstr. 7, 96049, Bamberg, Germany}

\begin{abstract}
Seventeen years of hard X-ray observations with the instruments of the \INTEGRAL\ observatory, with a focus on the Milky Way and in particular on the Galactic Centre region, have provided a unique database for exploration of the Galactic population of low-mass X-ray binaries (LMXBs). Our understanding of the diverse energetic phenomena associated with accretion of matter onto neutron stars and black holes has greatly improved. We review the large variety of \INTEGRAL\ based results related to LMXBs. In particular, we discuss the spatial distribution of LMXBs over the Galaxy and their X-ray luminosity function as well as various physical phenomena associated with Atoll and Z sources, bursters, symbiotic X-ray binaries, ultracompact X-ray binaries and persistent black hole LMXBs. We also present an up-to-date catalogue of confirmed LMXBs detected by \INTEGRAL, which comprises 166 objects. Last but not least, the long-term monitoring of the Galactic Centre with \INTEGRAL\ has shed light on the activity of Sgr A* in the recent past, confirming previous indications that our supermassive black hole experienced a major accretion episode just $\sim$100~years ago. This exciting topic is covered in this review too.  

\end{abstract}

\begin{keyword}
\texttt{X-ray binaries, INTEGRAL observatory, accretion, black holes, neutron stars, Sgr A*}
\end{keyword}

\end{frontmatter}

%\linenumbers

\tableofcontents

%%%%%%%%
\section{Introduction}
\label{s:intro}
%%%%%%%%

X-ray binaries divide into high-mass and low-mass ones (HMXBs and LMXBs, respectively) according to the mass of the companion star feeding the relativistic compact object (a neutron star, NS, or a black hole, BH). In HMXBs, the donor is an early-type star with a mass exceeding $\sim$5\,$M_\odot$, while in LMXBs the donor is less massive than $\sim$1\,$M_\odot$ and is either a main-sequence or evolved late-type star or a white dwarf. There is also a small subclass of intermediate-mass X-ray binaries (IMXBs), with the donor's mass lying between the two limiting values quoted above, such as Hercules X-1. Although huge energy release in the form of X-rays due to accretion onto a relativistic compact object unites HMXBs and LMXBs, these two classes turn out to be quite different as concerns their origin, evolution, abundance in the disc and bulge of the Milky Way and other galaxies, and the spectral and timing properties of their X-ray emission. The \INTEGRAL\ observatory \citep{2003A&A...411L...1W, 2011SSRv..161..149W} has greatly increased our knowledge of the Galactic HMXB and LXMB populations, and the aim of the present review is to summarize \INTEGRAL\ based results related to LMXBs. \INTEGRAL\ achievements pertaining to HMXBs are overviewed by Kretschmar et al. in the same volume.    

LMXBs are interesting for a number of reasons. First, they are ideal for detailed studies of the physics of accretion onto relativistic compact objects (see e.g. \citealt{2007A&ARv..15....1D} for a review). They also allow us to study phenomena (such as type I X-ray bursts) occurring at the surface of NSs and thereby infer the fundamental properties thereof (see \citealt{2016ARA&A..54..401O} for a review). Finally, as LXMBs constitute just a tiny fraction of the old stellar population of the Milky Way and other galaxies, they represent exotic  evolution tracks of binary stars and are crucial for testing models of stellar evolution and supernova explosions (e.g. \citealt{2008ApJS..174..223B,2017hsn..book.1499C}). It is also worth noting that despite the extreme rarity of LMXBs, they completely dominate the X-ray output of the old stellar population of the Milky Way and other galaxies, with other, much more numerous stellar-type X-ray sources, such as cataclysmic variables (CVs) and coronally active stars, contributing just a few per cent to the cumulating X-ray luminosity \citep{2006A&A...450..117S}. 

The history of LMXBs began in 1962 together with that of X-ray astronomy, when the first X-ray source outside of the Solar System and the brightest persistent X-ray source in the sky Scorpio X-1 was discovered by Riccardo Giacconi and his colleagues with a detector on board a sounding rocket \citep{1962PhRvL...9..439G,2003RvMP...75..995G}. Subsequent efforts led to the identification of Sco~X-1 with an optical star of 13th visual magnitude \citep{1966ApJ...146..316S} and in 1967 (shortly before the discovery of radio pulsars), Shklovsky proposed that accretion of gas from a normal companion onto a NS in a close binary system might be behind \citep{1967ApJ...148L...1S}. This idea was not immediately accepted since it was thought that the supernova leading to the formation of the NS would break the binary system apart. 

Systematic population studies of LMXBs and X-ray binaries in general became possible in the 1970s, when the \emph{Uhuru} satellite carried out its famous all-sky X-ray survey and detected about 400 X-ray sources \citep{1978ApJS...38..357F}, largely Galactic X-ray binaries. Since then, several generations of X-ray observatories have followed in orbit and dramatically broadened our knowledge of X-ray binaries. 

Already in the 1970s, it was revealed that apart from soft X-ray photons, accreting NSs and BHs also produce hard X-ray radiation (e.g. \citealt{1979Natur.279..506S}), and this was one of the main drivers for further development of space-borne hard X-ray instruments. The hard X-ray band (above $\sim$10\,keV) is important also because it provides a virtually unobscured view of the entire Milky Way. Massive studies of LMXBs in X-rays became possible in the second half of the 1980s when imaging instruments based on the coded-aperture principle began to operate in orbit. One of the most successful observatories of this kind was \emph{GRANAT} (1989--1998), a predecessor of \INTEGRAL\ in many respects. Among the most important results of \emph{GRANAT} are (i) detailed maps of the Galactic Centre region in the soft (4--20\,keV) and hard (40--150\,keV) X-ray bands, which enabled the discovery of a number of X-ray binaries, and (ii) high-quality broad-band spectra of accreting NSs and BHs (e.g. \citealt{1994ApJ...425..110P,1994ApJS...92..381C}).

The \INTEGRAL\ observatory, launched in 2002, with its broad energy coverage, large field of view, good angular, spectral and timing resolution, is particularly well suited for systematically studying Galactic populations of X-ray sources. With this understanding, \INTEGRAL's observational programme heavily focused on the Galactic part of the sky in the first years of the mission. This policy has been gradually changing towards a more homogeneous sky coverage over the subsequent years, but the exposure accumulated over the Milky Way has been constantly increasing and now (as of 2019) the {\it INTEGRAL} hard X-ray map provides a unique, multi-year long view of the Galaxy (see Krivonos et al. in this volume for a review). A particularly long exposure ($\sim$40\,Ms per position) has been accumulated in the Galactic Centre region of $\sim$30\,deg radius, including the entire Galactic bulge. This is especially valuable for studies of LMXBs, since they belong to the old stellar population. 

In addition, the \INTEGRAL\ observations of the Galactic Centre region have provided exciting new information on the violent activity of the central supermassive black hole (Sgr A*) in its recent past (several hundred years) via the detection of a hard X-ray \lq echo' from the Sgr~B2 giant molecular cloud, which has been intensively followed up by other X-ray missions. The present review covers this topic as well. 

The LMXB theme is very broad and there are a number of distinct subclasses of LMXBs. In this review, we present the relevant \INTEGRAL\ results with particular attention to the LMXB identification, catalogues and population studies (Sections~\ref{s:obs}, \ref{s:sample}, \ref{s:population}). Results on weakly magnetized NS binaries (Section~\ref{s:weakfield}) and X-ray pulsars (Section~\ref{s:pulsars}) are also presented, together with ultracompact binaries (Section~\ref{s:ucxb}), persistent BH LMXBs (Section~\ref{s:persBH}) and the aforementioned Sgr A* (Section~\ref{s:sgrA}). BH X-ray transients and accreting millesecond X-ray pulsars are presented in separate reviews of relevant \INTEGRAL\ results (Motta et al. and Papitto et al., respectively) in the same volume.

%%%%%%%%
\section{Discovery and monitoring of LMXBs with INTEGRAL}
\label{s:obs}
%%%%%%%%

In the first years of the mission, about a third of the \INTEGRAL\ observing time was allocated for the institutes that had developed the instruments, for the INTEGRAL Science Data Centre \citep{2003A&A...411L..53C} and for the Russian scientists in return for providing the Proton launcher. This so-called Core Programme mostly consisted of a deep survey of the central radian of the Galaxy and frequent scans of the Galactic plane. Early results of these campaigns were reported by \cite{2003A&A...411L.349W,2003A&A...411L.363P,2003A&A...411L.369D,2003A&A...411L.373R}. In particular, \cite{2003A&A...411L.363P}, using a sample of 8 bright NS LMXBs (Atoll and Z sources), demonstrated the ability of the JEM-X \citep{2003A&A...411L.231L}, IBIS/ISGRI \citep{2003A&A...411L.131U,2003A&A...411L.141L} and SPI \citep{2003A&A...411L..63V} instruments aboard \INTEGRAL\ to perform broad-band (from $\sim$5 to several hundred keV) X-ray spectroscopy of Galactic X-ray binaries, confirming pre-launch expectations. 

\cite{2004AstL...30..382R} carried out a systematic analysis of IBIS/ISGRI 15--60\,keV maps of the $\sim 35^\circ\times25^\circ$ central region of the Galaxy using deep observations (2\,Ms total time) performed in August and September 2003 (as part of the Guest Observer Programme, complementing the Core Programme). A total of 60 point-like sources were detected above a threshold of 1--2\,mCrab, of which 9 were new discoveries. Of the previously known 51 sources, 38 were LMXBs. Similar systematic studies based on early \INTEGRAL/IBIS observations were carried out in the Galactic plane, where $\sim$60 previously known or newly identified LMXBs as well as a few dozen unidentified sources were found \citep{2004ApJ...607L..33B,2004AstL...30..534M,2006AstL...32..145R}. 

A lot of of further observations of the Milky Way have been performed over the course of the mission. In particular, long-term campaigns of regular (every {\it INTEGRAL} revolution when permitted by visibility constraints) monitoring of the Galactic bulge\footnote{http://integral.esac.esa.int/BULGE/} and Galactic plane\footnote{http://gpsiasf.iasf-roma.inaf.it/} have been carried out since 2005 and 2011, respectively, with the data becoming publicly available nearly immediately to enable rapid follow-up observations. The main goal of these multi-year monitoring programmes is to study the variability and transient activity of hard X-ray sources, in particular LMXBs, on various time scales from hours to years \citep{2007A&A...466..595K}.

The Optical Monitoring Camera (OMC) on board \INTEGRAL\ \citep{2003A&A...411L.261M} provides  simultaneous photometry in  the Johnson V-band. The first decade of \INTEGRAL\ observations resulted in an OMC database containing light curves for more than 70,000 sources brighter than V$\sim$18 and with more than fifty photometric points each. \cite{2012A&A...548A..79A} characterized the potential variability of these sources. They found about 5,000 confirmed variable sources for which they provide (together with other information) the median of the visual magnitude, charts, light curves in  machine  readable  format and, when found, the period and period-folded light curves. Such results for all the sources contained in the updated OMC-variable catalogue (currently comprising about 150 LMXBs) can be browsed and retrieved from http://sdc.cab.inta-csic.es/omc/.

%%%%%%%%%%%
\subsection{IBIS and JEM-X catalogues}

The absolute majority of sources detected by \INTEGRAL\ are point-like and the main role in their detection and monitoring is played by the ISGRI detector of the IBIS instrument. As reviewed by Krivonos et al. in this volume, two main teams, one led by A. Bird and another led by R. Krivonos, have been independently compiling and updating catalogues of sources detected by IBIS over the sky. Below we briefly describe, with a focus on LMXBs, the key catalogues that have been published in the last ten years. Previous \INTEGRAL\ works comprise the \INTEGRAL/IBIS census of the sky beyond 100\,keV \citep{2006ApJ...649L...9B} and the first IBIS/ISGRI soft gamma-ray Galactic plane survey catalogue \citep{2004ApJ...607L..33B}. 

The IBIS 7-year all-sky survey catalogue \citep{2010A&A...523A..61K}, covering the observations performed from December 2002 up to July 2009, contained 521 sources detected in the 17--60\,keV energy band, including 101 LMXBs and 38 unidentified sources. The IBIS catalogue compiled after 1000 \INTEGRAL\ orbits (up to the end of 2010, \citealt{2016ApJS..223...15B}) comprised 939 sources detected in the 17--100\,keV energy band, of which 129 were classified as LMXBs and 9 as X-ray binaries (i.e. either HMXBs or LMXBs), with 219 objects remaining unidentified at the time of publishing. The IBIS 9-year Galactic hard X-ray survey \citep{2012A&A...545A..27K} was restricted to the Galactic plane region ($|b|<17.5^\circ$) and contained 402 sources detected in the 17--60\,keV energy band, including 108 LMXBs and 34 unidentified objects. Recently, an updated analysis of the IBIS data for the $|b|<17.5^\circ$ region was carried out by \cite{2017MNRAS.470..512K} based on 14 years (up to March 2017) of observations. A total of 522 sources were detected in the 17--60\,keV energy band, including 72 newly found objects (i.e. not present in any previous \INTEGRAL\ catalogues). Most of these sources are weak, $\sim$0.2--0.5\,mCrab (which is the typical depth of the current IBIS map near the Galactic plane), or $\lesssim 10^{-11}$\,erg\,s$^{-1}$\,cm$^{-2}$ in the 17--60\,keV energy band, and were unidentified at the time of publishing. 

It is also necessary to mention a catalogue of sources detected using the 11-year (up to January 2014) map of the sky obtained with IBIS at energies above 100\,keV \citep{2015MNRAS.448.3766K}. A total of 88 sources were detected with a significance of more than 5$\sigma$ in the 100--150\,keV energy band, of which 38 are LMXBs. In the same study, a similar analysis was also done for the even harder energy band of 150--300\,keV, which led to the detection of 25 sources, more than half of which (13) are LMXBs. This clearly demonstrates that (i) LMXBs efficiently generate not only X-rays but also soft gamma-rays and that (ii) despite the extreme rarity of such objects among the Galactic stars, they, together with active galactic nuclei (AGN), dominate the hard X-ray/soft gamma-ray sky at the flux levels (a few mCrab) accessible to the \INTEGRAL\ observatory. 

Observations with JEM-X, another imaging instrument aboard \INTEGRAL\, have nicely complemented the IBIS/ISGRI Galactic hard X-ray surveys described above. JEM-X is sensitive in the softer, 3--35\,keV, but overlapping energy band, has a fairly large field of view ($13.2^\circ$ diameter at zero response) and a factor of $\sim$3 better angular resolution ($\sim$3$'$ FWHM) compared to IBIS. These characteristics make the JEM-X database a valuable source of information on LXMBs and other Galactic populations of X-ray sources. 

\cite{2015AstL...41..765G} analyzed the JEM-X data obtained from February 2003 to April 2013 and constructed maps of the Galactic Centre region with a radius of $20^\circ$ in the 5--10 and 10--25\,keV energy bands, with a typical depth of 1--2\,mCrab. This represents a significant improvement in terms of area ($\sim 1200$ vs $\sim 100$\,sq.\,deg) over the X-ray maps of similar quality available for this region from previous missions. A total of 83 sources were detected on the time-averaged JEM-X maps and an additional 22 \lq transients' were found on images obtained in individual \INTEGRAL\ pointings or revolutions. All of these sources, except one (IGR~J17452--2909), were already known from previous missions or IBIS observations. As could be expected for the $\sim$10\,keV energy band, the vast majority of objects in the JEM-X Galactic Centre catalogue are LMXBs (73 of the 105 sources), and this sample has proved very useful for studying statistical properties of the Galactic LMXB population (see \S\ref{s:population} below). 

%%%%%%%%%%%
\subsection{Follow-up efforts}

Sometimes, when \INTEGRAL\ discovers (or re-discovers) a source, its likely LMXB origin is strongly suggested by its X-ray behaviour inferred from the \INTEGRAL\ data: e.g. the detection of type-I X-ray bursts allows one to unambiguously classify the object as a NS LMXB, while an X-ray nova-like behaviour implies a BH LMXB origin. However, more often the identification and classification of LMXBs among \INTEGRAL\ sources discovered in the crowded and heavily obscured Galactic plane and Galactic Centre regions is not easy. Given the moderately good localization accuracy (a few arcmin for weak, $\sim$mCrab, sources) provided by IBIS, the process of identification of an \INTEGRAL\ source usually includes several stages: (i) search for a likely association in existing (soft) X-ray catalogues (e.g. the {\it ROSAT} all-sky survey catalogue), which may provide preliminary information on the object's nature (in particular, whether it is persistent or transient) and substantially reduce the localization area for further investigation, (ii) follow-up \lq snapshot' observations with a focusing X-ray telescope, primarily aimed at obtaining an accurate position, and (iii) follow-up observations with ground-based optical (and sometimes radio) telescopes using the improved coordinates. 

Already at the early stage of the \INTEGRAL\ mission, such follow-up efforts, complemented by analyses of archival data of previous missions, led to the identification of several likely NS and BH LMXBs among the first \lq IGR' hard X-ray sources: IGR~J17091--3624, IGR~J17303--0601, IGR~J17464--3213 and IGR~J17597--2201 \citep{2004A&A...426L..41M,2005A&A...430..997L}. Particularly exciting were the discoveries of a new accretion-powered millisecond pulsar IGR~J00291$+$5934 \citep{2005A&A...444...15F} and a new symbiotic X-ray binary IGR~J16194--2810 \citep{2007A&A...470..331M}, which became just the 6th and 5th known objects of these exotic classes, respectively.

Also previously known sources in the crowded Galactic Centre have been studied thanks to the high spatial resolution and sensitivity in the hard X-rays provided by \INTEGRAL. An example is the persistent 1E\,1743.1$-$2843 ($l,b=0.26,-0.03$) whose 20--40\,keV emission has been reported for the first time by \cite{2004ApJ...607L..33B} using \INTEGRAL/IBIS data. The broad-band (\INTEGRAL, \textit{XMM-Newton} and \textit{Chandra}) spectrum has then been studied in detail by \cite{2006A&A...456.1105D} providing the first evidence of emission up to 70\,keV and suggesting that 1E\,1743.1$-$2843 could be a NS LMXB located beyond the Galactic Centre.

Since the optical counterparts of LMXBs are usually dim (in particular compared to HMXBs), the selection of the right counterpart in the optical and near-infrared (NIR) maps of the Milky Way is often non-trivial even when an accurate (a few arcsec) position is available from follow-up X-ray observations. Optical/NIR spectroscopy is usually also difficult. Nevertheless, in such cases, a plausible LMXB origin can often be inferred from an analysis of the source's spectral energy distribution (SED). Specifically, a low optical--NIR/X-ray flux ratio strongly suggests an LMXB (rather than e.g. HMXB, AGN, CV or stellar-coronal) origin. This SED-based approach has been successfully used to unveil a number of previously unknown LMXBs among \INTEGRAL\ sources (e.g.  \citealt[][]{2008A&A...484..783C,2011ApJ...738..183P,2011A&A...533A...3C} and the on-line catalogue of \INTEGRAL\ sources identified through optical and NIR spectroscopy\footnote{http://www.iasfbo.inaf.it/$\sim$masetti/IGR/main.html}). 

Furthermore, the same method allows one to constrain the size of an LMXB through the well-known relation \citep{1994A&A...290..133V} between the size of a binary ($a$) and its optical/NIR and X-ray luminosities ($L_V$ and $L_X$, respectively), \textbf{There was an error here!} $L_{V}\propto L_X^{1/2} a$, which reflects the fact that the optical/NIR emission of LMXBs mainly arises via reprocessing of part of the X-ray radiation in the outer parts of the accretion disc. \cite{2011MNRAS.411..620Z,2015MNRAS.446.2418Z} applied this approach to a dozen of suspected LXMBs in the Galactic bulge selected from the \INTEGRAL\ surveys of the region, including several \lq IGR' sources, and obtained interesting upper limits on their orbital periods. For one object (IGR~J17597--2201), they concluded that it is likely yet another representative of the exotic class of symbiotic LMXBs.

One of the most recent follow-up results associated with new LMXBs found by {\it INTEGRAL} is the discovery in 2018 of the transient source IGR~J17591--2342. This was followed by an extensive programme of X-ray ({\it Chandra}, {\it NICER}, {\it NuSTAR} and {\it Neil Gehrels Swift}), optical, IR and radio observations, which revealed that this object is an accreting millisecond X-ray pulsar (see \citealt{2019ApJ...874...69N} and references therein).

%\cite{2006A&A...456L...5F} -- Swift, RXTE, and INTEGRAL observation of Swift J1922.7-1716.
%\cite{2011MNRAS.415.2373S} -- XMM-Newton and INTEGRAL observations of the very faint X-ray transient IGR J17285-2922/XTE J1728-295 during the 2010 outburst. Likely a LMXB based on X-ray behaviour.
%\cite{2016MNRAS.461.2688P} -- Swift J174540.7-290015: a new accreting binary in the Galactic Centre.

%%%%%%%%
\section{The INTEGRAL LMXB sample}
\label{s:sample}
%%%%%%%%

In Table~\ref{table:lmxb} we show the list of LMXBs detected by \INTEGRAL, consisting of 166 objects. The table has been compiled starting from the latest  \INTEGRAL\ General Reference Catalogue\footnote{https://www.isdc.unige.ch/integral/science/catalogue}(version 43, June 2019). LMXBs have been selected according to the widely used \lq \textit{CLASS}' parameter (14XX and 15XX) relative to known LMXBs that have been detected by \INTEGRAL, be it at a single pointing level or mosaic, as found in the literature. LMXBs detected later than June 2019 have also been added, together with sources that are labeled as unidentified in the catalogue (\textit{CLASS}=9000 and 9999) and that turned out to be LMXBs in later literature. A careful search of the literature has also been performed to select LMXBs among sources that are generically labeled as \lq X-ray binary' (\textit{CLASS}=10XX) and \lq Symbiotic star' (\textit{CLASS}=1940) in the reference catalogue.

In the table, NS as well as (candidate) BHs are given. Source names, coordinates, relative type/comments and distances are provided. \lq IGR' sources (a total of 33 objects discovered by \INTEGRAL) are shown in bold, with the discovery reference indicated. Unless otherwise stated, the information listed in the table is taken from the following catalogues: \citet{2007A&A...469..807L, 2016ApJS..223...15B} and the aforementioned \INTEGRAL\ general reference catalogue. 

\begin{figure*}
\centering
\includegraphics[width=\textwidth]{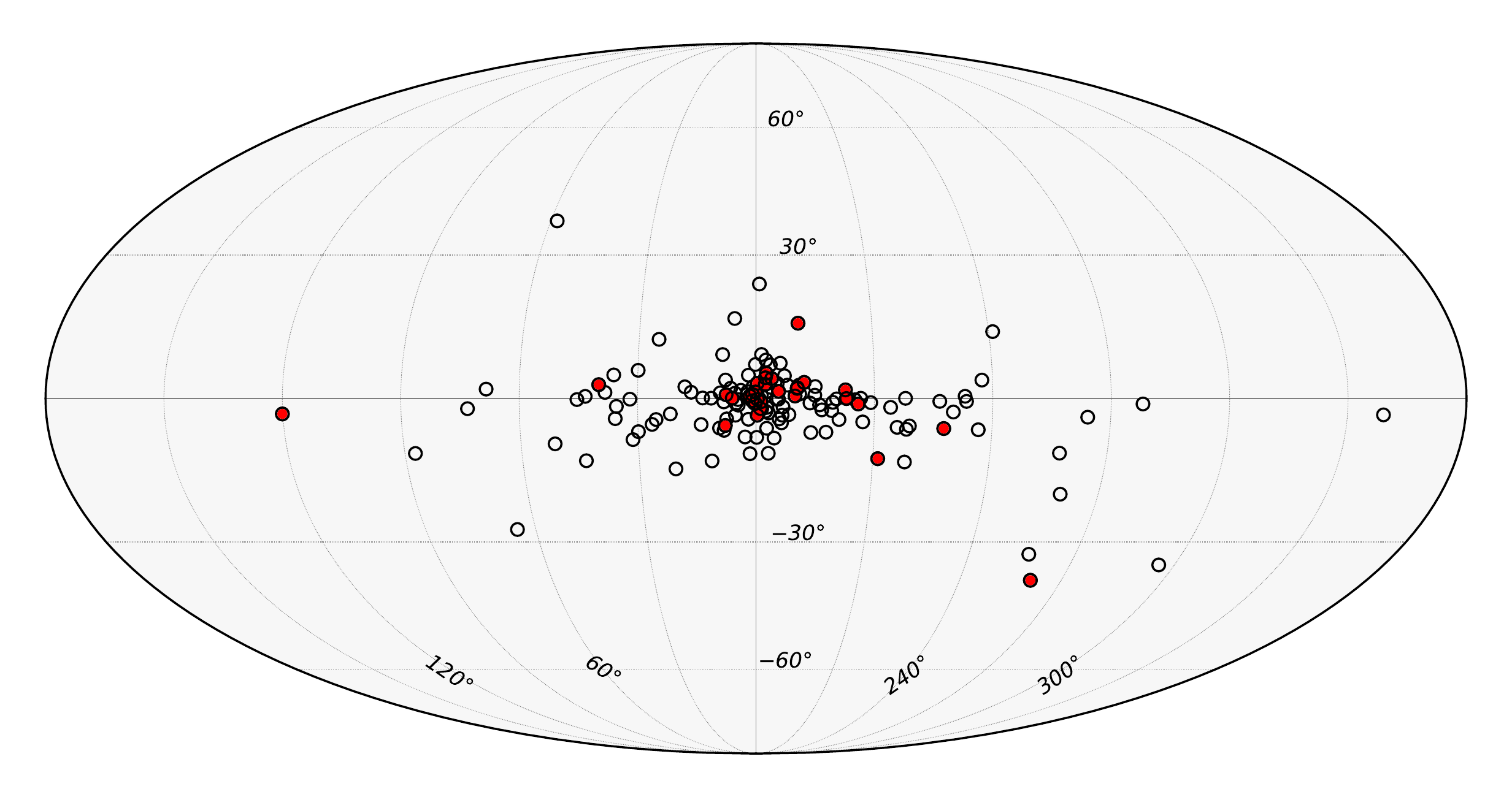}
\includegraphics[width=\textwidth]{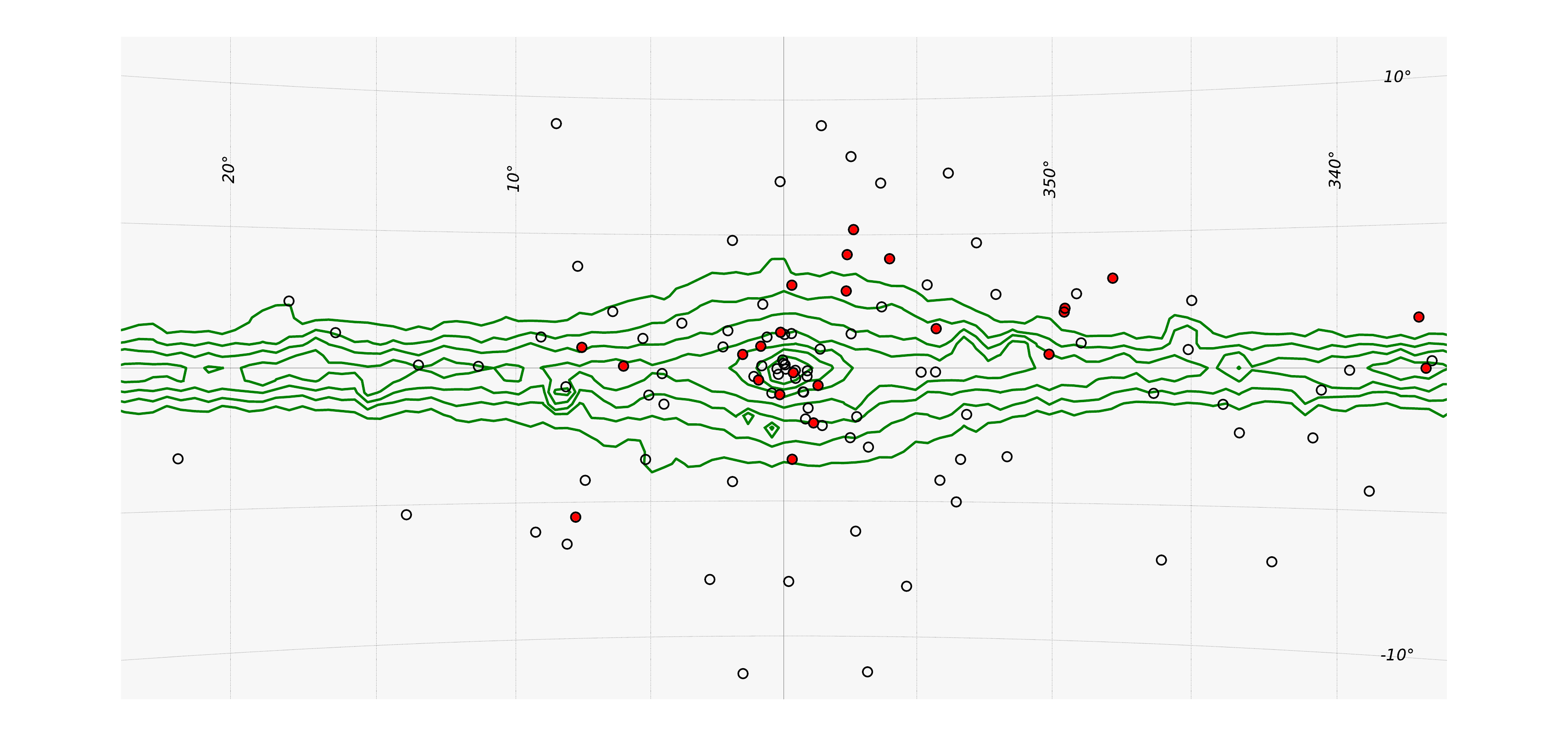}
\caption{\emph{Top:} Positions of the LMXBs detected by \INTEGRAL\ on the sky in Galactic coordinates. The sources discovered by \INTEGRAL\ are shown by filled symbols (in red). \emph{Bottom:} Zoom on the central region. The contours show the 3.5\,$\mu$m NIR surface brightness as measured by \emph{COBE}/DIRBE, which traces the stellar distribution in the Galaxy.}
\label{fig:map}
\end{figure*}

The positions of the sources listed in Table~\ref{table:lmxb} on the sky are shown in Fig.~\ref{fig:map}. Their spatial distribution is discussed in the next Section.

\begin{table*}
\caption{List of LMXBs detected by \INTEGRAL}
\label{table:lmxb}

\begin{tabular}{r|c|r|r|c|r}
\hline
\multicolumn{1}{c}{Id} & 
\multicolumn{1}{c}{Source} & 
\multicolumn{1}{c}{RA} & 
\multicolumn{1}{c}{Dec} &
\multicolumn{1}{c}{Type, comments} &
\multicolumn{1}{c}{D (kpc)}\\
\hline
1&{\bf IGR J00291+5934}  (1)& 7.263 & 59.572 & AMXP (599~Hz) & 4.2 (D1)\\
2&{\bf IGR J04288$-$6702}  (2)& 67.194 & $-$67.075 & $\gamma$-ray binary & 2.4 (D2) \\
3&4U 0513$-$40   & 78.527 & $-$40.044 & UCXB, B & 12.1 (D3)\\
4&LMC X-2   & 80.121 & $-$71.965 & in LMC & 50\\
5&H 0614+091   & 94.280 & 9.137 & UCXB, B ($\nu_{\rm spin}=477$ Hz) & 3.2 (D4)\\
6&EXO 0748$-$676   & 117.141 & $-$67.752 & B ($\nu_{\rm spin}=522$ Hz) & 7.4 (D5)\\
7&GS 0836$-$429   & 129.348 & $-$42.900 & B & $<$9.2 (D6)\\
8&SWIFT J0911.9$-$6452   & 138.010 & $-$64.868 & AMXP (334~Hz)& 9.5 (D7) \\
9&4U 0919$-$54   & 140.11229 & $-$55.20686 & UCXB, B & 4.1--5.4 (D8)\\
10&XSS J12270$-$4859   &  186.994 & $-$48.895 & AMXP (593~Hz) & 1.4 (D9)\\
11&3A 1246$-$588   & 192.41504 & $-$59.08703 & UCXB, B & 4.3 (D10)\\
12&1H 1254$-$690   & 194.405 & $-$69.28917 & B & 7.6 (D11) \\
13&SAX J1324.3$-$6313   & 201.161 & $-$63.228 & B & $<$6.2 (D12)\\
14&4U 1323$-$62   & 201.650 & $-$62.136 & B & 15 (D13)\\
15&GS 1354$-$645   & 209.541 & $-$64.735 & BH & $\ge$25 (D14)\\
16&MAXI J1421$-$613   & 215.408 & $-$61.607 & B & $<$7 (D15)\\
17&{\bf IGR J14298$-$6715}  (3)& 217.499 & $-$67.246 &  & $\sim$10 (D16)\\
18&Cir X-1   & 230.170 & $-$57.167 & B & 9.4 (D17)\\
19&SWIFT J1539.2$-$6227   & 234.799 & $-$62.467 & BHC  & \\
20&4U 1543$-$624   & 236.977 & $-$62.568 & UCXB & 7 (D18)\\
21&XTE J1550$-$564   & 237.744 & $-$56.476 & BH, $\mu$QSO & 4.4 (D19)\\
22&4U 1556$-$60   & 240.259 & $-$60.738 &  & 4 (D20)\\
23&4U 1608$-$522   & 243.179 & $-$52.423 & B ($\nu_{\rm spin}=619$ Hz) & 3.6 (D5)\\
24&{\bf IGR J16194$-$2810}  (4)& 244.888 & $-$28.127 & SyXB & $\le$3.7 (D21)\\
25&Sco X-1   & 244.979 & $-$15.640 & & 2.8 (D22)\\
26&AX J1620.1$-$5002   & 245.091 & $-$50.019 & B & 8.4 (D23)\\
27&4U 1624$-$49   & 247.011 & $-$49.198 & & 15 (D24)\\
28&{\bf IGR J16287$-$5021}  (5)& 247.111 & -50.377 & & \\
29&{\bf IGR J16293$-$4603}  (6)& 247.303 & $-$46.047 & & \\
30&4U 1626$-$67   & 248.070 & $-$67.461 & UCXB, P (0.13~Hz) & 5--13 (D25) \\
31&4U 1630$-$47   & 248.501 & $-$47.394 & BHC & $<$11.8 (D26)\\
32&{\bf IGR J16358$-$4726}  (7)& 248.974 & $-$47.428 & SyXB & $\lesssim$13 (D27)\\
33&4U 1636$-$53   & 250.231 & $-$53.751  & B ($\nu_{\rm spin}=582$ Hz) &  \\
34&GX 340+0   & 251.448 & $-$45.611 & & \\
35&XTE J1652$-$453   & 253.084 & $-$45.344 & BHC & \\
36&GRO J1655$-$40   & 253.500 & $-$39.845& BH, $\mu$QSO & 3.2 (D28)\\
37&Her X-1   & 254.457 & 35.342 & P ($\nu_{\rm spin}=0.8$~Hz) & 6.1 (D29)\\
38&MAXI J1659$-$152   & 254.761 & $-$15.256 & BHC & \\
39&{\bf IGR J16597$-$3704}  (8)& 254.887 & $-$37.120 & UCXB, AMXP (105~Hz) & 9.1 (D30)\\
40&XTE J1701$-$462   & 255.243 & $-$46.185 & & 8.8 (D31)\\
41&XTE J1701$-$407   & 255.434 & $-$40.858 & B & 5.0 (D32)\\
42&H 1658$-$298   & 255.526 & $-$29.945 & B & 9 (D5)\\
43&GX 339$-$4   & 255.706 & $-$48.789 & BH & $>$6 (D33)\\
44&GX 349+2   & 256.435 & $-$36.423 & & 9.2 (D34)\\
45&H 1702$-$429   & 256.563 & $-$43.035 & B & 5.5 (D35)\\

\end{tabular}
\end{table*}

\begin{table*}
\begin{tabular}{r|c|r|r|c|r}
46&{\bf IGR J17062$-$6143}  (9)& 256.569 & $-$61.711 & UCXB, AMXP (163 Hz), B & 7.3 (D36)\\
47& H 1705$-$250 & 257.060 & $-$25.091 & BHC & 8.6 (D37) \\
48&4U 1705$-$32   & 257.226 & $-$32.332 & UCXB, B & 13 (D38)\\
49&H 1705$-$440   & 257.226 & $-$44.102 & B & 8.4 (D37)\\
50&{\bf IGR J17091$-$3624}  (10)& 257.281 & $-$36.407 & BHC, $\mu$QSO &  \\
51&XTE J1709$-$267   & 257.376 & $-$26.655 & B & 8.5 (D39)\\
52&{\bf IGR J17098$-$3628}  (11)& 257.441 & $-$36.466 & BHC & 10.5 (D40)\\
53&XTE J1710$-$281   & 257.551 & $-$28.131 & B & 17.3 (D37)\\
54&4U 1708$-$40   & 258.099 & $-$40.842& B & $<$16 (D41)\\
55&SAX J1712.6$-$3739   & 258.141 & $-$37.643  & B & 7 (D42)\\
56&2S 1711$-$339   & 258.582 & $-$34.046 & B & 7.5 (D43)\\
57&{\bf IGR J17177$-$3656}  (12)& 259.427 & $-$36.934 & BHC & \\
58&{\bf IGR J17191$-$2821}  (13)& 259.813 & $-$28.299 & B ($\nu_{\rm spin}=295$ Hz) & $<$8.6 (D44)\\
59&GRS 1716$-$249   & 259.903 & $-$25.017 & BHC & 2.4 (D45)\\
60&{\bf IGR J17197$-$3010}  (14)& 259.965 & $-$30.033 & SyXB & \\
61&XTE J1720$-$318   & 259.975 & $-$31.748 & BHC & \\
62&{\bf IGR J17233$-$2837}  (15)& 260.845 & $-$28.631 & AMXP (537 Hz) & 0.9 (D3)\\
63&{\bf IGR J17254$-$3257}  (16)& 261.353 & $-$32.954 & B & $<$14.5 (D46)\\
64&XTE J1726$-$476   & 261.707 & $-$47.639 & BHC & \\
65&4U 1722$-$30   & 261.888 & $-$30.801 & B & 9.5 (D47)\\
66&{\bf IGR J17285$-$2922}  (17)& 262.162 & $-$29.362 & BHC & $>$4 (D48)\\
67&3A 1728$-$169   & 262.934 & $-$16.961 & & \\
68&GX 354$-$0   & 262.989 & $-$33.834 & B, P & 5.3 (D37)\\
69&GX 1+4   & 263.009 & $-$24.745 & SyXB & 4.3 (D49)\\
70&{\bf IGR J17329$-$2731}  (18)& 263.227 & $-$27.530 & SyXB & 2.7 (D50)\\
71&4U 1730$-$335   & 263.350 & $-$33.387 & B & 8.8 (D47)\\
72&SWIFT J1734.5$-$3027   & 263.602 & $-$30.398 & B & 7.2 (D51)\\
73&{ \bf IGR J17353$-$3539}  (19)& 263.847 & $-$35.670 & B & $\le$9.5 (D52)\\
74&{ \bf IGR J17379$-$3747}  (5)& 264.495 & $-$37.772 & B, P (468 Hz) & \\
75&SLX 1735$-$269   & 264.571 & $-$26.994 & B & $<$13 (D53)\\
76&4U 1735$-$444   & 264.742 & $-$44.450 & B & 9.4 (D37)\\
77&GRS 1736$-$297   & 264.887 & $-$29.723 & & \\
78&XTE J1739$-$285   & 264.974 & $-$28.496 & B ($\nu_{\rm spin}=1122$ Hz) & $<$10.6 (D54)\\
79&SLX 1737$-$282   & 265.162 & $-$28.296 & B & 7.3 (D55)\\
80&{\bf IGR J17407$-$2808}  (20)& 265.175 & $-$28.124 & & \\
81&GRS 1739$-$278   & 265.654 & $-$27.783 & BHC, T & \\
82&XTE J1743$-$363   & 265.755 & $-$36.372 & SyXB & 5 (D56)\\
83&1E 1740.7$-$2942   & 265.978 & $-$29.745 & BHC, $\mu$QSO & \\
84&{\bf IGR J17445$-$2747}  (21)& 266.126 & $-$27.766 & B & $<$12.3 (D57)\\
85&GRO J1744$-$28   & 266.137 & $-$28.738 & P (2.1 Hz) & $\sim$8 (D58) \\
86&XMMU J174445.5$-$295044   & 266.189 & $-$29.845 & SyXB & 3.1 (D59)\\
87&KS 1741$-$293   & 266.233 & $-$29.351 & B & 6.2 (D60)\\
88&XMM J174457$-$2850.3   & 266.243 & $-$28.837 & B & 6.5 (D61)\\
89&GRS 1741.9$-$2853   & 266.259 & $-$28.913 & B & 7 (D62)\\
90&{\bf IGR J17451$-$3022}  (22)& 266.278 & $-$30.378 & & \\ 
\end{tabular}
\end{table*}

\begin{table*}
\begin{tabular}{r|c|r|r|c|r}
91&SWIFT J174510.8$-$262411   & 266.294 & $-$26.404 & & $<$7 (D63)\\ 
92&{\bf IGR J17454$-$2919}  (23)& 266.366 & $-$29.331 & & \\
93&1A 1742$-$289   & 266.404 & $-$29.018 & B & $\ge$8 (D64)\\
94&1A 1742$-$294   & 266.522 & $-$29.515 & B &  8.1 (D37)\\
95&H 1743$-$322   & 266.566 & $-$32.233 & BHC, $\mu$QSO & 8.5 (D65) \\
96&1A 1743$-$288   & 266.760 & $-$28.883 & B & 7.5 (D66)\\
97&1E 1743.1$-$2843   & 266.587 & $-$28.728 & &  \\
98&XMMU J174716.1$-$281048   & 266.817 & $-$28.180 & B & 8.4 (D67)\\
99&{\bf IGR J17473$-$2721}  (24)& 266.825 & $-$27.344 & B & 6.4 (D68)\\
100&SLX 1744$-$300   & 266.855 & $-$30.044 & B & \\
101&SLX 1744$-$299   & 266.858 & $-$30.020 & B & \\
102&GX 3+1   & 266.983 & $-$26.563 & B & $<$6.5 (D5)\\
103&EXO 1745$-$248   & 267.0216 & $-$24.780 & B & 5.5 (D69)\\
104&1A 1744$-$361   & 267.080 & $-$36.121 & B ($\nu_{\rm spin}=530$ Hz) & $<$9 (D70)\\
105&{ \bf IGR J17494$-$3030}  (25)& 267.349 & $-$30.499 & & \\
106&H 1745$-$203   & 267.222 & $-$20.367 & B & 8.4 (D47)\\
107&SWIFT J1749.4$-$2807   & 267.382 & $-$28.134 & B, AMXP (518 Hz) & 6.7 (D71)\\
108&{\bf IGR J17497$-$2821}  (26)& 267.408 & $-$28.354 & BHC & \\
109&SLX 1746$-$331   & 267.460 & $-$33.198 & BHC & \\
110&{\bf IGR J17498$-$2921}  (27)& 267.480 & $-$29.322 & B, AMXP (401 Hz) & 7.6 (D72)\\
111&1E 1746.7$-$3224   & 267.516 & $-$32.430 & & \\
112&1H 1746$-$370   & 267.552 & $-$37.052  & B & 11 (D47)\\
113&SAX J1750.8$-$2900   & 267.601 & $-$29.038 & B & 6.8 (D5)\\
114&GRS 1747$-$312   & 267.689 & $-$31.292 & B & 11 (D73)\\
115&{\bf IGR J17511$-$3057}  (28)& 267.786 & $-$30.961 & B, AMXP (245 Hz) & $<$6.9 (D74)\\
116&XTE J1751$-$305   & 267.816 & $-$30.625 & AMXP (435 Hz) & \\
117&XTE J1752$-$223   & 268.0628 & $-$22.3423 & BHC & $>$5 (D75)\\
118&SWIFT J1753.5$-$0127   & 268.367 & $-$1.451 & BHC & \\
119&SAX J1753.5$-$2349   & 268.383 & $-$23.820 & B & \\
120&SWIFT J1756.9$-$2508   & 269.238 & $-$25.107 & AMXP (182 Hz) & \\
121&{\bf IGR J17585$-$3057}  (29)& 269.637 & $-$30.956  & & \\
122&{\bf IGR J17591$-$2342}  (30)& 269.761 & $-$23.718 & AMXP (527 Hz) & \\
123&{\bf IGR J17597$-$2201}  (31)& 269.940 & -22.027 & B &\\
124&GX 5$-$1   & 270.284 & $-$25.079 &  & 7.2 (D75)\\
125&GRS 1758$-$258   & 270.302 & $-$25.740 & BHC, $\mu$QSO & $<$12 (D77)\\
126&GX 9+1   & 270.384 & $-$20.528 & & 4.4 (D78)\\
127&1RXS J180408.9$-$342058   & 271.033 & $-$34.347 & B, UCXB(?) & 10 (D79)\\
128&SAX J1806.5$-$2215   & 271.634 & $-$22.238 & B & $<$8 (D80)\\
129&XTE J1807$-$294   & 271.749 & $-$29.408 & AMXP (190 Hz), B & \\
130&SAX J1808.4$-$3658   & 272.114 & $-$36.978 & AMXP (401 Hz), B & 3.5 (D81)\\
131&XTE J1810$-$189   & 272.586 & $-$19.069 & B & 3.5--8.7 (D82)\\
132&SAX J1810.8$-$2609   & 272.685 & $-$26.150 & B & 4.9 (D83)\\
133&GX 13+1   & 273.631 & $-$17.157 & B & 7 (D84)\\
134&4U 1812$-$12   & 273.775 & $-$12.096 & B & 4.1 (D37)\\
135&GX 17+2   & 274.005 & $-$14.0363 & B & $<$13 (D5)\\
 
\end{tabular}
\end{table*}

\begin{table*}
\begin{tabular}{r|c|r|r|c|r}
136&XTE J1817$-$330   & 274.431 & $-$33.018 & BHC & \\
137&XTE J1818$-$245   & 274.603 & $-$24.537 & BHC  & 3.5 (D85)\\
138&4U 1820$-$30   & 275.918 & $-$30.361 & UCXB, B & 8.4 (D86)\\
139&{\bf IGR J18245$-$2452}  (32)& 276.110 & $-$24.857 & AMXP (254 Hz), B & 5.5 (D87)\\
140&4U 1822$-$00   & 276.341 & $-$0.011 & B & 13 (D88) \\
141&3A 1822$-$371   & 276.445 & $-$37.105 & P (1.7 Hz) & 2--2.5 (D89)\\
142&MAXI J1828$-$249   & 277.242 & $-$25.030 & BHC & \\
143&Ginga 1826$-$24   & 277.367 & $-$23.796 & B & 5.7 (D90)\\
144&MAXI J1836$-$194   & 278.930 & $-$19.320 & BHC & 4--10 (D91)\\
145&XB 1832$-$330   & 278.933 & $-$32.981 & B & 9.2 (D92)\\
146&Ser X-1   & 279.989 & 5.035 & B & 7.7 (D5)\\
147&SWIFT J1842.5$-$1124   & 280.572 & $-$11.416 & BHC & \\
148&SWIFT J185003.2$-$005627   & 282.537& $-$0.940 & B & $\le$3.7 (D93)  \\
149&3A 1850$-$087   & 283.270 & $-$8.705 & UCXB, B & 6.8 (D87)\\
150&{\bf IGR J18539+0727}  (33)& 283.491 & 7.468 & BHC & \\
151&XTE J1856+053   & 284.179 & 5.309 & BHC & \\
152&Swift J1858.6$-$0814   & 284.645 & $-$8.237  & &  \\
153&HETE J1900.1$-$2455   & 285.038 & $-$24.920  & AMXP (377 Hz), B & 4.7 (D37) \\
154&XTE J1901+014   & 285.417 & 1.440 & & \\
155&XTE J1908+094   & 287.221 & 9.384 & BHC, $\mu$QSO & \\
156&SWIFT J1910.2$-$0546   & 287.575 & $-$5.785 & BHC & \\
157&Aql X-1   & 287.816 & 0.584  & B ($\nu_{\rm spin}=550$ Hz) & 5.2 (D94)\\
158&GRS 1915+105   & 288.798 & 10.945 & BH, $\mu$QSO & 11 (D95)\\
159&4U 1916$-$053   & 289.699 & $-$5.236 & B & 8.8 (D38)\\
160&SWIFT J1922.7$-$1716   & 290.654 & $-$17.284 & B & $\le$4.8 (D93)\\
161&3A 1954+319   & 298.926 & 32.096 & SyXB & 1.7 (D96)\\
162&MAXI J1957+032   & 299.161 & 3.44 & &  \\
163&4U 1957+115   & 299.850 & 11.708 & BHC, P & \\
164&V404 Cyg   & 306.015 & +33.867 & BHC, $\mu$QSO & 2.4 (D97)\\
165&4U 2129+12   & 322.492 & 12.167 & UCXB, B & 10.4 (D87)\\
166&Cyg X-2   & 326.171 & 38.321 & B & 13.4 (D37)\\
\hline
\end{tabular}

{\bf Note:} The discovery references of \lq IGR' sources are given below, while the ones relative to the source distances (\lq Dn') are listed at the end of the paper.

{\bf Acronyms:} AMXP -- accreting millisecond X-ray pulsar, B -- type I burster, BH -- black hole, BHC -- black hole candidate, $\mu$QSO -- microquasar, P -- pulsar, SyXB -- symbiotic X-ray binary, UCXB -- ultracompact X-ray binary.

{\bf References:} (1) \cite{2004ATel..352....1E}, (2) \cite{2013MNRAS.428...50G}, (3) \cite{2006ATel..810....1K}, 
(4) \cite{2006ApJ...636..765B}, (5) \cite{2007ApJS..170..175B}, (6) \cite{2008ATel.1774....1K}, (7) \cite{2003IAUC.8097....2R}, (8) \cite{2017ATel10880....1B}, (9) \cite{2007A&A...467..529C}, (10) \cite{2003ATel..149....1K}, (11) \cite{2005ATel..444....1G}, (12) \cite{2011ATel.3223....1F}, (13) \cite{2007ATel.1021....1T}, (14) \cite{2010A&A...523A..61K}, (15) \cite{2010A&A...523A..61K}, (16) \cite{2004ATel..229....1W}, (17) \cite{2004ATel..229....1W}, (18)  \cite{2017ATel10644....1P}, (19) \cite{2007A&A...475..775K}, (20) \cite{2004GCN..2793....1G}, (21)  \cite{2006ApJ...636..765B}, (22) \cite{2014ATel.6451....1C}, (23) \cite{2014ATel.6530....1C}, (24) \cite{2005ATel..467....1G}, (25) \cite{2012ATel.3984....1B}, (26) \cite{2006ATel..885....1S}, (27) \cite{2011ATel.3551....1G}, (28) \cite{2009ATel.2196....1B}, (30) \cite{2018ATel11941....1D}, (31) \cite{2003ATel..155....1L}, (32) \cite{2013ATel.4925....1E}, (33) \cite{2003ATel..151....1L}.

\end{table*}

%%%%%%%%
\section{The Galactic LMXB population} 
\label{s:population}
%%%%%%%%

The currently achieved depth of the IBIS maps enables finding hard X-ray sources with luminosity (17--60\,keV) $\Lhx>10^{35}$\,erg\,s$^{-1}$ over most of the Galactic disc, while the bulge is pierced through at even lower luminosities of $\Lhx>5\times 10^{34}$\,erg\,s$^{-1}$ \citep{2017MNRAS.470..512K}. As a result, \INTEGRAL\ provides unique unbiased statistics of both high- and low-luminosity LMXBs. 

%%%%%%%%%%%
\subsection{Spatial distribution} 
\label{s:spatial}

LMXBs belong to the old stellar population and are expected to trace the distribution of stars over the Milky Way (as opposed to HMXBs, which are tracers of recent star formation activity) and, in particular, to concentrate towards the centre of the Galaxy. Although this behaviour was already evident with previous missions \citep{1975IAUS...67..413G,1993A&AS...97..149S,1996rftu.proc..141G,2002A&A...391..923G}, \INTEGRAL, thanks to its unique combination of broad energy band, wide field of view, good angular resolution and more than 15 years of observations, has provided a much richer and sharper picture of the Galactic LMXB population. 

The anticipated concentration of HMXBs to spiral arm tangents and LMXBs towards the Galactic Centre has been becoming more evident as the \INTEGRAL\ (IBIS) catalogues have been increasing in size (e.g. \citealt{2005A&A...443..485D,2007A&A...467..585B,2015MNRAS.448.3766K}). Moreover, the concentration of a substantial fraction of the unidentified sources towards the Galactic Centre strongly suggested their LMXB origin, which has been verified by follow-up observations for a number of objects.

\cite{2008A&A...491..209R} have performed a thorough statistical analysis of the LMXBs in the Galactic bulge. To this end, they (i) selected sources detected by IBIS after the first 4 years of observations \citep{2007A&A...475..775K} in the elliptical region around the Galactic Centre with axes $|l|<10.7^\circ$, $|b|<5.1^\circ$ and having fluxes higher than 0.64\,mCrab in the 17--60\,keV energy band, and (ii) excluded all known HMXBs, CVs and AGN. There remained a number of unidentified sources, most of which are likely LMXBs based on statistical arguments and they were therefore added to the sample of known LMXBs. The resulting sample consisted of 22 persistent and 16 transient sources. The choice of the relatively compact Galactic Centre region guaranteed that most of the studied objects lie at nearly the same distance (8\,kpc) so that the uncertainty in the X-ray luminosity does not exceed a factor of 2 for 80\% of the objects. 

\begin{figure}
\centering
\includegraphics[width=\columnwidth]{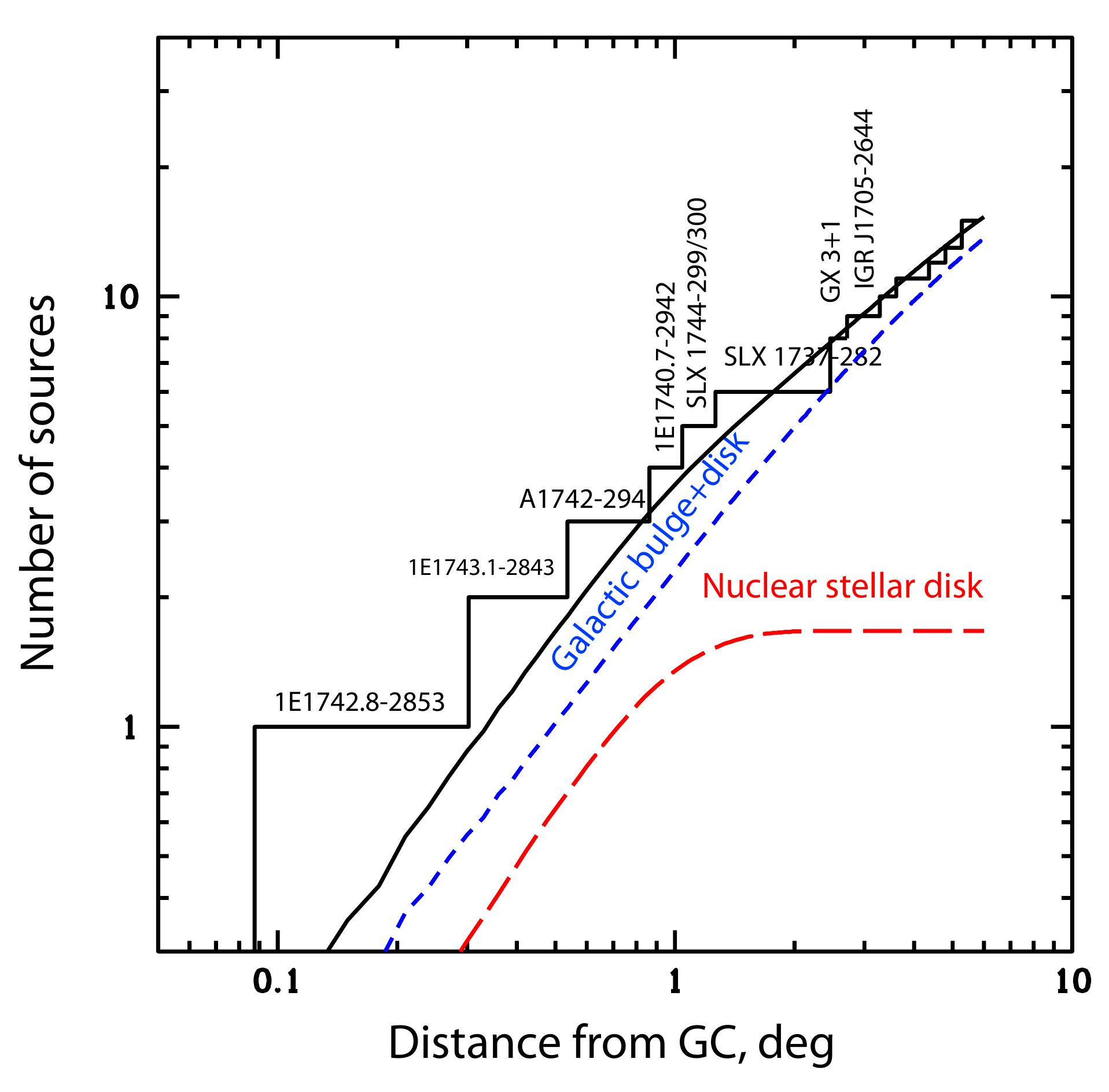}
\caption{Total number of persistent LMXBs within a given projected radius of the Galactic Centre, as seen by \INTEGRAL/IBIS, in comparison with a prediction based on the Galaxy mass model and the LMXB LF model of \cite{2004MNRAS.349..146G} multiplied by 0.4. (From \citealt{2008A&A...491..209R})}
\label{fig:revnivtsev_cumnum}
\end{figure}

As found by \cite{2008A&A...491..209R}, the distribution of persistent LMXBs over the angular distance from the Galactic Centre closely follows the distribution of stellar mass in the bulge within $\sim$10$^\circ$ (see Fig.~\ref{fig:revnivtsev_cumnum}), except in the innermost 1.5$^\circ$ where the observed number (3) of persistent LMXBs is higher (albeit with low statistical significance) than expected for the bulge alone ($\sim$1 object). This central excess is, however, consistent with the presence of the so-called nuclear stellar disc in the central $\sim$200~pc of the Galaxy \citep{2002A&A...384..112L}. The angular distribution of transient LMXBs proves to be slightly more concentrated towards the Galactic Centre compared to the persistent sources, possibly indicating an additional dynamical production of such objects in the dense environment of the Galactic Centre, like in the central region of M31 \citep{2007MNRAS.380.1685V}. However, the statistical significance of this result is also low. 

%%%%%%%%%%%
\subsection{Hard X-ray luminosity function} 
\label{s:lf}

\begin{figure}
\centering
\includegraphics[width=\columnwidth]{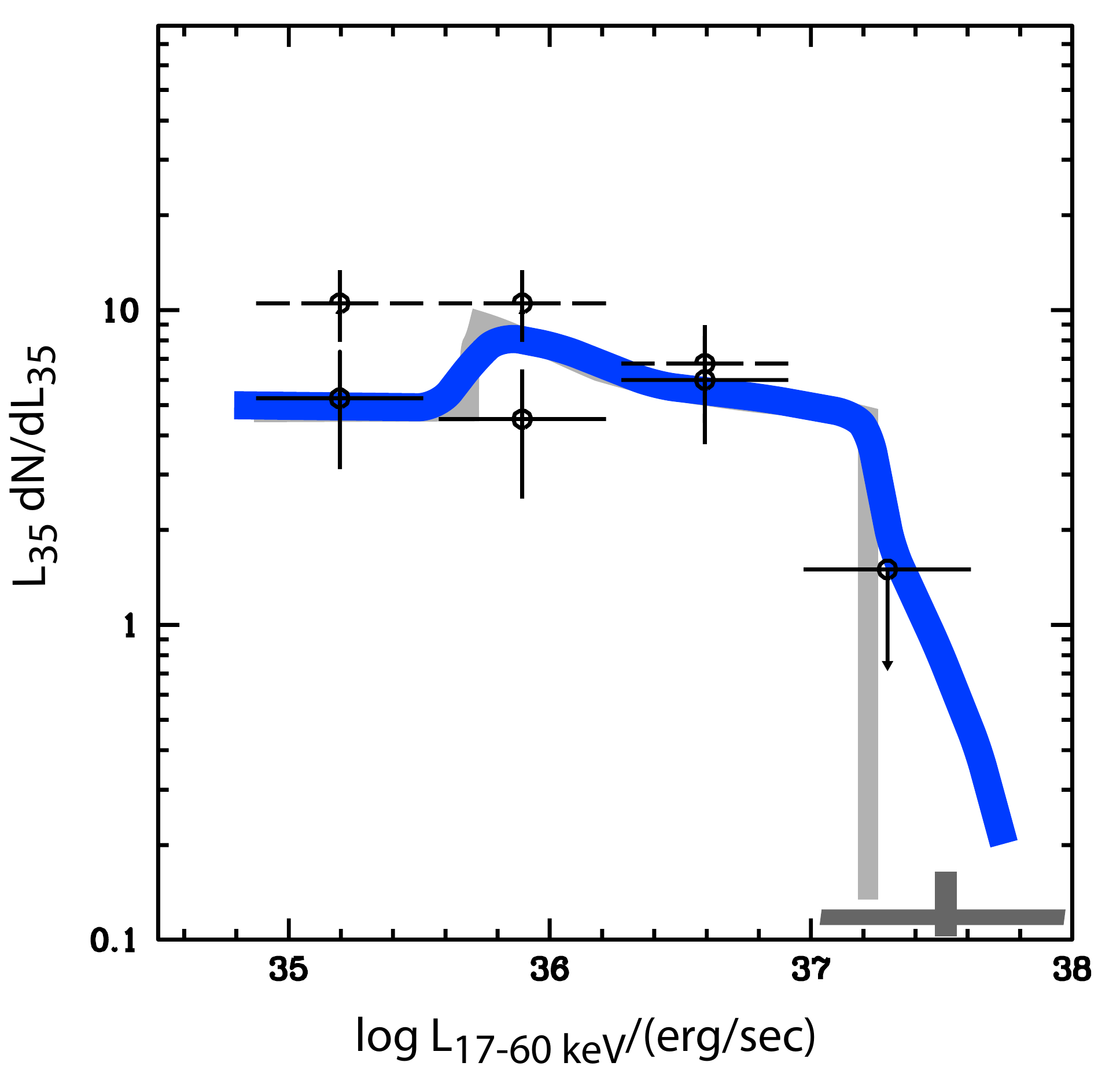}
\caption{Hard X-ray LF of LMXBs detected by IBIS in the bulge (solid crosses -- persistent sources, dashed crosses -- all sources). The thick gray cross is an estimate of the number density of LMXBs with extremely high luminosities based on the only such source in the Galaxy, GRS~1915+105. The gray line is the prediction based on the LMXB LF model of \cite{2004MNRAS.349..146G} multiplied by 0.4; the blue line is similar, taking into account the uncertain distances to the sources. (From \citealt{2008A&A...491..209R})}
\label{fig:revnivtsev_lf}
\end{figure}

The main result of \cite{2008A&A...491..209R} was the measurement of the hard X-ray luminosity function (LF) of Galactic LMXB down to luminosities (17--60\,keV) $\Lhx\sim 10^{35}$\,erg\,s$^{-1}$ and the demonstration that the LF flattens below $\Lhx\sim 10^{37}$\,erg\,s$^{-1}$ (when plotted in units $LdN/dL$), see Fig.~\ref{fig:revnivtsev_lf}. Previously, a qualitatively similar behaviour was reported for the LMXB LF in the standard X-ray band (2--10\,keV), both for the Milky Way \citep{2002A&A...391..923G} and for nearby galaxies \citep{2004MNRAS.349..146G}, but those analyses effectively reached down to just $\Lx\sim 10^{36}$\,erg\,s$^{-1}$ (2--10\,keV) due to the limited sensitivity of the all-sky monitor (ASM) aboard the {\it RXTE} observatory in the former case and the large distances ($\gtrsim$1\,Mpc) to the sources (detected by {\it Chandra}) in the latter. It is thanks to \INTEGRAL's capability of probing most of the Milky Way down to $\Lhx\sim 10^{35}$\,erg\,s$^{-1}$ that the LMXB LF has now been extended down to such low luminosities. 

As can be seen from Fig.~\ref{fig:revnivtsev_lf}, the shape of the hard X-ray LF of Galactic LMXBs is in good agreement with the analytic model of \cite{2004MNRAS.349..146G} for the LF of LMXBs in nearby galaxies. To carry out  this comparison, the model had to be converted from 2--10\,keV to 17--60\,keV: as demonstrated by \cite{2008A&A...491..209R}, the 2--10\,keV/17--60\,keV flux ratio strongly depends on the LMXB luminosity, being of the order of unity at $\Lx<2\times 10^{37}$~erg~s$^{-1}$ and $\sim$40 at higher luminosities. This reflects the fact that high accretion rate sources produce mostly thermal radiation below 10\,keV, while at low accretion rates there is a strong contribution of hard radiation, presumably generated via Comptonization in a hot corona (see \S\ref{s:weakfield} below). 

Interestingly, there are no LMXBs with $\Lhx\gtrsim 10^{37}$\,erg\,s$^{-1}$ in the bulge. This is again related to the fact that high-accretion rate objects have soft spectra. In reality, there is a single LMXB in the Milky Way (located outside of the Galactic Centre region), GRS~1915+105, with $\Lhx\sim 5\times 10^{37}$\,erg\,s$^{-1}$, which is peculiar in many respects (and is addressed in the review by Motta et al. in this volume).

As regards the LF normalisation (the number of LMXBs per stellar mass), a reduction factor of $\sim$0.4--0.7 (depending on whether transient sources are taken into account or not, see Fig.~\ref{fig:revnivtsev_lf}) is required to fit the \cite{2004MNRAS.349..146G} model to the IBIS data points. This suggests that LMXBs may be somewhat underabundant in the Milky Way compared to typical galaxies in the local Universe, as previously  noticed by \cite{2004MNRAS.349..146G}. 

\begin{figure}
\centering
\includegraphics[width=\columnwidth]{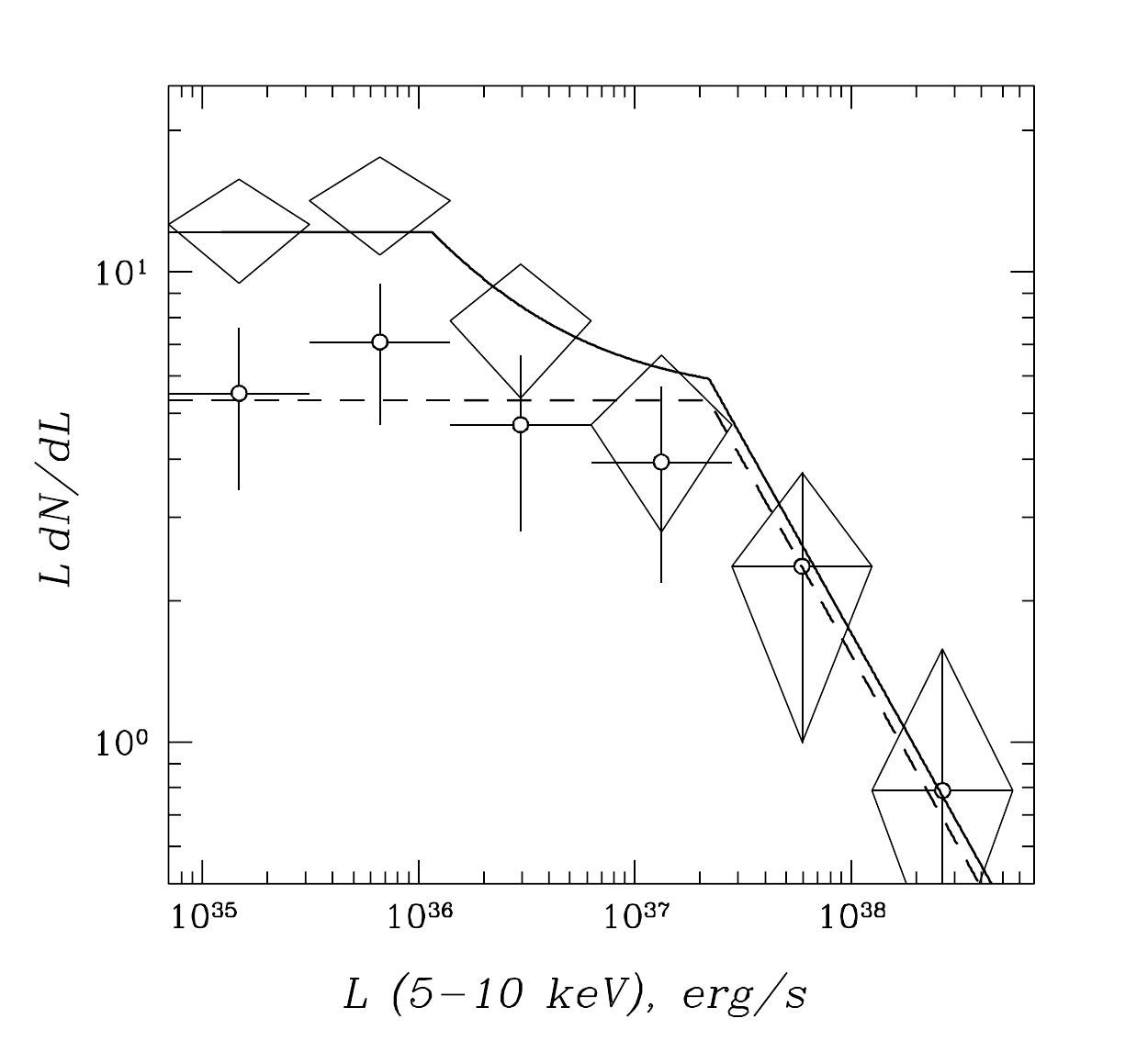}
\caption{X-ray (5–-10\,keV) LF of LMXBs detected by JEM-X in the Galactic Centre region (circles -- persistent sources, diamonds -- all sources). (From \citealt{2015AstL...41..765G})}
\label{fig:grebenev_lf}
\end{figure}

A similar analysis has been done for the softer energy band (5--10\,keV) using the JEM-X survey of the Galactic Centre region \citep{2015AstL...41..765G}. These authors came to similar conclusions regarding (i) the angular distribution of LMXBs within the central $\sim$10$^\circ$ (namely that they closely follow the distribution of stellar mass), and (ii) the LF of LMXBs. Since the JEM-X survey, similarly to that by IBIS, is deep enough to detect all sources with luminosities higher than $10^{35}$\,erg\,s$^{-1}$ in the bulge, selection effects are of minor importance, allowing one to confidently reconstruct the LMXB LF down to $\sim 10^{35}$\,erg\,s$^{-1}$. The resulting LF is shown in Fig.~\ref{fig:grebenev_lf} and shows a pronounced break at $\sim10^{37}$\,erg\,$^{-1}$, similar to that seen in the hard X-ray LF (Fig.~\ref{fig:revnivtsev_lf}). \cite{2015AstL...41..765G} point out that while the shape of the LF of persistent sources is flat below the break, this is no longer true when transient sources are also taken into account: the combined LF of persistent and transient LMXBs slowly rises towards low luminosities. A similar behaviour is actually manifest also in the hard X-ray LF measured by IBIS. This finding should however be regarded with caution since there is no strict distinction between persistent and transient sources, and it is not clear if the 15-year span over which \INTEGRAL\ has been monitoring the Galactic LMXBs is sufficiently long to make a full account of LMXB transient activity. 

A number of explanations have been put forward for the origin of the observed break in the LMXB LF. \cite{2011A&A...526A..94R} pointed out that the majority of LMXBs with luminosities below a few $10^{37}$\,erg\,s$^{-1}$ have unevolved secondary companions (except for those with white dwarf donors), while systems with higher luminosities predominantly harbor giant donors. The duration of the mass transfer phase in the latter systems is expected to be significantly shorter than in systems with main-sequence donors \citep{1983ApJ...270..678W} and this is the most likely cause of the observed steepening of the LMXB LF above a few $10^{37}$\,erg\,s$^{-1}$. 

% May be useful: \cite{2017JCAP...05..056H} -- LMXBs in the inner Galaxy: implications for millisecond pulsars and the GeV excess.

%%%%%%%%%%%
\subsection{LMC and SMC}

For completeness, we should also mention {\it INTEGRAL} deep surveys of the main satellites of our Galaxy, the Large Magellanic Cloud (LMC) and the Small Magellanic Cloud (SMC). 

\cite{2013MNRAS.428...50G} analyzed the IBIS and JEM-X data for the LMC available by mid-2012, with a total exposure of $\sim$7\,Ms. Nine objects physically belonging to the LMC were detected in the 20--60\,keV energy band by IBIS and one additional source (LMC X-3) in the 3--20\,keV energy band by JEM-X. Of these 10 sources, only one is a LMXB (LMC~X-2). In addition, 4 new hard X-ray sources were found. Of these, two have subsequently proved to be of extragalactic origin and another two (IGR~J04288--6702 and IGR~J05099--6913) remain unidentified, although it is possible \citep{2017ATel10412....1M} that IGR~J04288--6702 is associated with the new gamma-ray loud, eclipsing Galactic LMXB 3FGL~J0427.9--6704 discovered by {\it Fermi} \citep{2016ApJ...831...89S}. 

Given the achieved sensitivity of the IBIS survey in the LMC field of 0.5\,mCrab (i.e. $\sim 6\times 10^{-12}$\,erg\,s$^{-1}$\,cm$^{-2}$) and the distance to the LMC of $\sim 50$\,kpc, all objects with 20--60\,keV luminosity higher than $\sim 2\times 10^{36}$\,erg\,s$^{-1}$ should have been detected by IBIS. We can then compare the actual number of such objects found in the LMC (between 1 and 3, depending on the origin of the unidentified objects) with the corresponding number of LMXBs of the same luminosity detected by IBIS in the Galactic Centre region \citep{2008A&A...491..209R}, which is $\sim$6--10 (allowing for some uncertainty in the flux thresholds and distances). On the other hand, the total stellar mass of the LMC is $\sim 3\times 10^9$\,$M_\odot$ \citep{2009IAUS..256...81V}, while the total stellar mass contained in the Galactic Centre region studied by \cite{2008A&A...491..209R} is $\sim 1.7\times 10^{10}$\,$M_\odot$, i.e. about 5 times higher. Therefore, the specific abundance of LMXBs in the LMC appears to be similar to that in the Milky Way. It is worth noting in this connection that given the nearly flat shape of the LMXB LF at low luminosities, even an order of magnitude deeper hard X-ray survey of the LMC (reaching down to $\Lhx\sim 10^{35}$\,erg\,s$^{-1}$) would probably find just a few additional LMXBs in this satellite galaxy. 

\cite{2010MNRAS.406.2533C} analyzed the data of IBIS and JEM-X observations of the SMC in 2008--2009, with a total exposure of $\sim$1\,Ms, and detected seven sources in this dwarf galaxy, including four previously unknown. All these objects are believed to be HMXBs. Taking into account that the stellar mass of the SMC is just $\sim$1/10th of that of the LMC \citep{2009IAUS..256...81V} and that the \INTEGRAL\ survey is somewhat shallower in the former case, it is not surprising that no LMXBs have been found. 

%%%%%%%%
\section{Weakly magnetized neutron stars}
\label{s:weakfield}
%%%%%%%%

The X-ray properties of an X-ray binary are largely determined by the mass transfer rate from the donor, the nature (BH or NS) of the compact object and the strength of the magnetic field in the NS case. Typically LMXBs have a few orders of magnitude weaker magnetic fields than HMXBs (see \citealt{2015SSRv..191..293R} for a review), presumably due to their much older age and the associated field decay induced by accretion (e.g. \citealt{1986ApJ...305..235T,2006RPPh...69.2631H}). Nevertheless, in a minority of LMXBs the magnetic field is strong enough ($B\gtrsim 10^9$\,G) to disrupt the Keplerian flow of accreting matter and to channel it onto the NS magnetic poles, causing X-ray pulsations. Such objects will be considered in \S\ref{s:pulsars}. The present section is devoted to a more numerous subclass of LMXBs where the NS magnetic field is too weak to strongly affect the accretion flow.  

%%%%%%%%%%%
\subsection{Atoll and Z sources}
\label{s:ZAtoll}

Historically, weakly magnetized NS LMXBs have been broadly classified according to their timing and spectral properties \citep{1989A&A...225...79H}. This study, based on \emph{EXOSAT} archival data of sixteen bright NS LMXBs, divided the sample in Z and Atoll sources based on the shape of their track along the X-ray colour--colour (CC) diagram and on the different spectral and timing behaviour that correlates with the position on the tracks. The three branches building up the Z were called Horizontal Branch (HB), Normal Branch (NB) and Flaring Branch (FB), while Atolls have two branches: the island state (IS) 
%in which the sources generally display no significant state change on time scales of hours to days 
and the banana branch (divided in lower branch, LB, and upper branch, UB). Z sources and Atolls move along the branches of their pattern in a smooth way and do not jump from one part of the branch to the other. The position of the source on the pattern was suggested to be linked to a physical quantity that evolves in a continuous way, most likely the accretion rate increasing from HB to FB for Z sources and from the IS to the UB branch for Atolls. 

Atoll sources have typical luminosities of 0.01--0.3\,$L_{\rm Edd}$ (where $L_{\rm Edd}$ is the Eddington luminosity), mostly evolve from the island to the banana state and can be considered the counterparts of BH LMXBs that evolve from the low-hard state to the high/soft states. Z sources are usually brighter ($\sim L_{\rm Edd}$), with softer spectra and weaker variability, in analogy to the high/soft states of BH LMXBs.

In 2002, \emph{RXTE}-based studies \citep{2002ApJ...568L..35M, 2002MNRAS.331L..47G} 
suggested that the clear Z/Atoll distinction on the CC diagram is an artifact due to incomplete sampling: Atoll sources, if observed long enough (years), do exhibit a Z shape as well. Furthermore, in 2006 using \emph{RXTE}, \cite{2006ATel..696....1R} discovered a very important source: \mbox{XTE~J1701--462}. At first classified as a new Z source, it soon proved to be a very peculiar object: not only was it a transient Z (unlike the other Z sources that were persistently bright), but it transformed from a typical Z source into an Atoll source along the outburst \citep{2009AAS...21360303L}. 
Notwithstanding these long-term surveys and results that seem to weaken the historical classification, many differences still remain. Indeed, hard X-ray missions like \emph{GRANAT} and \emph{CGRO} or the more recent \emph{RXTE} and \emph{BeppoSAX} have given a huge boost to our knowledge of these sources, broadening the sample number and observed properties: Atoll sources are more numerous than Z \citep[25 versus 8,][]{2007A&A...469..807L}, they are generally fainter in radio and X-rays, display harder X-ray spectra, complete the (newly discovered) Z pattern along the CC diagram on longer time scales (years versus hours/days), have a different correlated timing behaviour and tend to have a more regular bursting activity \citep{2002astro.ph..7219D, 2006csxs.book...39V, 2006MNRAS.366...79M, 2007A&A...469..807L}. 

\begin{figure}
\centering
\includegraphics[width=\columnwidth]{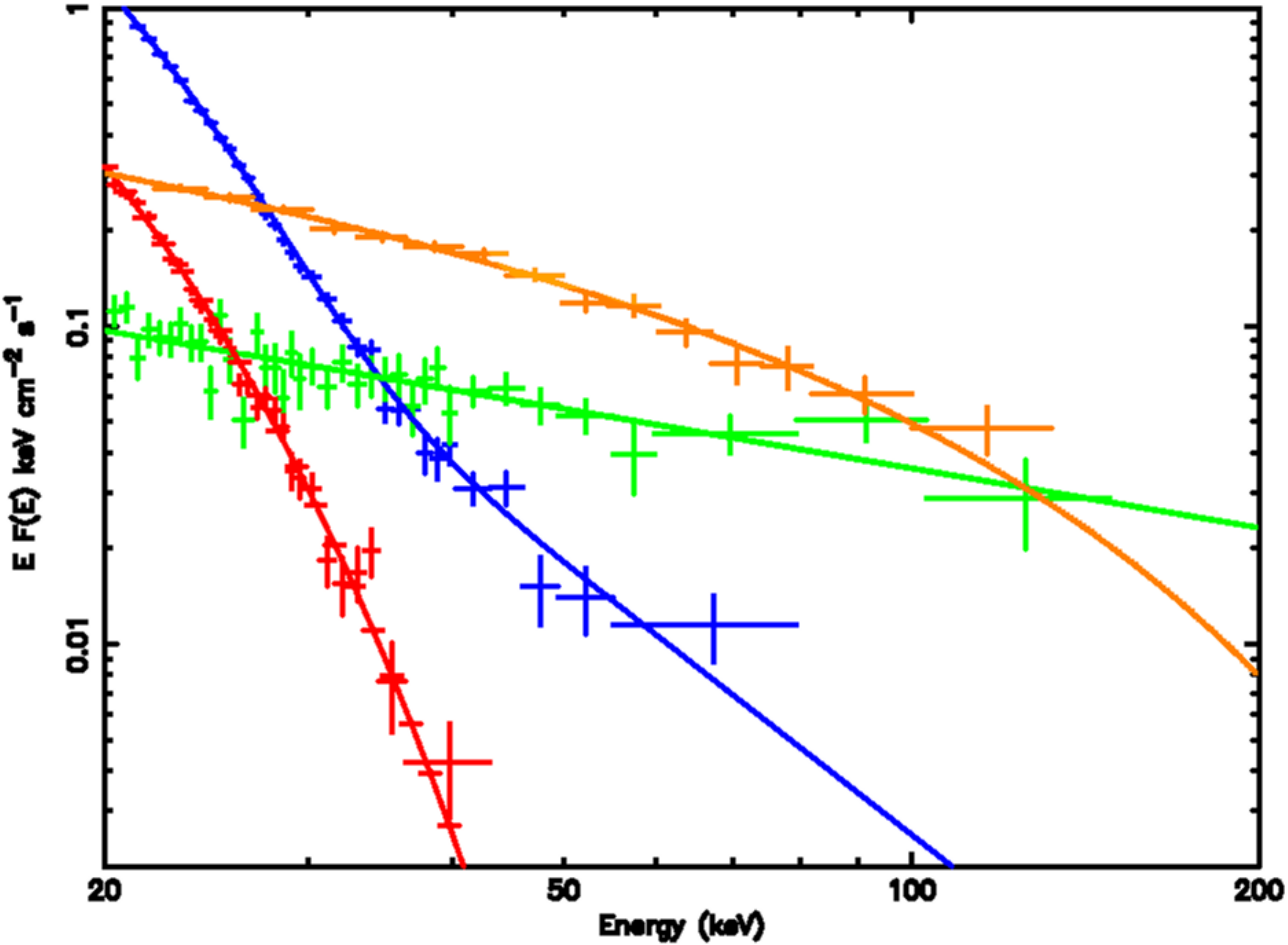}
\caption{Average NS LMXB spectra as seen by IBIS/ISGRI. Red: GX~3$+$1 (bright Atoll), blue: GX~5--1 (Z source), green: H~1608--522 (Atoll), orange: 4U~1728--34 \citep[Atoll, spectrum from][]{2006A&A...458...21F}. (From \citealt{2006A&A...459..187P})
\label{fig:lmavespe}}
\end{figure}

The overall picture is still not understood and detailed long-term studies are essential to grasp the broad physical scenario in which these (apparently similar and yet so different) sources move. The concentration of NS LMXBs towards the Galactic Centre made observations with non-imaging instruments of difficult interpretation. In this respect, \textit{INTEGRAL}, with its imaging capabilities coupled to the best ever available sensitivity to hard X-rays, is a key mission to study the hard X-ray emission of LMXBs as a class. This section will address the non-bursting properties of the spectra of NS LMXBs; burst properties are covered in Section~\ref{s:bursts}.

\subsubsection{Spectral states}

The X-ray spectra of LMXBs hosting a weakly magnetized NS are mainly described as the sum of
a soft and a hard component. The soft/thermal component (e.g., blackbody) is believed to be due to thermal emission by two different sites: the optically-thick, geometrically-thin accretion disc ($kT<2$\,keV) \citep{1973A&A....24..337S} and/or the NS surface ($kT=2$--3\,keV); the hard component (power-law shape with variable cutoff) is mostly interpreted as due to the repeated Compton up-scattering of the soft thermal photons in a hot electron cloud (corona). The final source state is driven by the dominating component and the spectral state transitions are mainly modelled in terms of a gradual change in the electron temperature of the Comptonizing corona. 

These basic ingredients that produce the X-ray continuum are believed to be present in all NS LMXBs but although they would be expected to give similar results through the different sources, important differences arise in the observations.

Figure~\ref{fig:lmavespe} shows the average spectra of four NS LMXBs as obtained with a systematic study with IBIS/ISGRI \citep{2006A&A...459..187P}. The red spectrum is the soft state that is exhibited by Z sources in the FB and NB, by bright \lq GX'  Atoll sources  and typical Atoll sources in the banana UB. The Comptonizing corona temperature obtained is a few keV, resulting in a spectral energy cutoff around 10~keV. The sources that exhibit this spectral state are believed to do so at high accretion rates ($\sim L_{\rm Edd}$), when the corona is efficiently Compton-cooled by abundant thermal disc photons. The green and orange spectra represent hard states that are exhibited only by typical Atoll sources (bright \lq GX' Atoll sources and Z sources have never been observed in such a state). The inferred accretion rate is low (0.01--0.1\,$L_{\rm Edd}$) and it is believed that the Comptonizing corona remains hot (tens of keV) efficiently boosting the soft thermal photons to higher energies. The blue spectrum shows an overall soft spectrum (few keV corona) with a hard tail dominating the spectrum above $\sim$30\,keV. This additional component is simultaneous to the soft spectrum and is
highly variable with  most of the emission remaining soft. Thermal Comptonization of soft photons in a very high temperature plasma is an unlikely interpretation for the hard tails since such a  hot corona would have to co-exist with the cool (3--5\,keV) thermal corona that is clearly observed in the soft states dominating the spectrum below 30\,keV. Possible mechanisms for explaining this transient feature are discussed in the next section.

\subsubsection{Hard tails in Z sources}

Hard X-ray tails (with a photon index $\Gamma\sim$ 2--3) have been detected in the spectra of Z sources when in the HB of their CC diagram, with its intensity dramatically decreasing in the NB and FB: \cite{1994PASJ...46..479A, 2000ApJ...544L.119D,2002astro.ph..7219D, 2005A&A...434...25F, 2005A&A...443..599P, 2006ApJ...649L..91D, 2006A&A...459..187P, 2007ApJ...662.1167F, 2008ApJ...680..602F, 2009A&A...498..509F, 2010A&A...512A..57M, 2014MNRAS.445.1205R}. 

In the case of the Z source GX~5--1, the study of the hard tail has long been contaminated by the nearby (40$^{\prime}$) BH LMXB GRS~1758--258. BH binaries were known to have hard tails extending to hundreds of keV while this was less typical for NSs. Although known to be brighter than GRS~1758--258 below 20\,keV, GX~5--1 was not detected at higher energies by \emph{GRANAT}/SIGMA \citep{1993ApJ...418..844G}. The first detection of a high energy tail in GX~5--1 was done by \cite{1994PASJ...46..479A} using \emph{GINGA} data (1.5--38\,keV). The tail could be approximated by a power-law of photon index 1.8. The contamination from GRS~1758--258 was included in the analysis as a power-law spectrum with the same slope ($\Gamma=1.8$) \citep[as measured by][]{1993ApJ...418..844G}. The presence of the hard tail in the spectrum of GX~5--1 and its decrease from the HB to the FB was confirmed by \INTEGRAL, which unambiguously resolved the two sources in broad-band spectral studies (5--150\,keV) and unveiled the variability of GX~5--1 at energies above 20\,keV on the single-pointing timescale ($\sim$2000\,s)  \citep{2005A&A...443..599P, 2006A&A...459..187P, 2010A&A...512A..57M}.

Upon discovery, hard tails were mostly fit with phenomenological models (power-law) in excess of the thermal Comptonization curvature. Although apt to detect the high energy excess, such a description did not unveil the physical origin of this transient component, and furthermore the lower end of the power-law dominated the soft part of the X-ray spectrum. 

Attempts to describe the hard X-ray tails in terms of physically self-consistent models resulted in different scenarii, mainly Comptonization by a hybrid thermal/non-thermal population of electrons in the corona \citep{1998PhST...77...57P, 1999ASPC..161..375C}, non-thermal electrons coming from the base of a jet \citep{2005ApJ...635.1203M}, jet/disc bulk outflow Comptonization \citep{2000ApJ...544L.119D, 2016A&A...591A..24R, 2018JApA...39...13K}, and bulk inflow Comptonization \citep{1981MNRAS.194.1033B, 1996A&AS..120C.171T, 2008ApJ...680..602F}. 

The long-term broad-band coverage of the Galactic plane and Galactic Centre with \INTEGRAL\ made it possible to investigate in a systematic way the spectral variability of a sample of NS LMXBs. In the frame of a new Comptonization model for weakly magnetized NS LMXBs, \cite{2006A&A...459..187P}, together with \cite{2008ApJ...680..602F, 2010A&A...512A..57M, 2010A&A...509A...2C}, proposed a unified physical scenario to explain the spectral evolution of these sources, including the peculiar transient hard tail: the interplay of thermal and bulk Comptonization  proved very effective in modeling the variety of NS-LMXBs spectral states and the state transitions in individual sources. In this scenario, bulk motion Comptonization (due to first-order process in photon energy gain) is believed to operate in the inner part of the system between the Keplerian disc and the NS surface, producing the hard tails. 

The presence of bulk motion at the origin of the high energy spectra of binaries, originally developed for BH binaries, has also been suggested, modelled and successfully applied to highly magnetized NS, i.e. accretion powered X-ray pulsars \citep{2009A&A...498..825F, 2012A&A...538A..67F, 2007ApJ...654..435B, 2016A&A...591A..29F}.

\begin{figure}
\centering
\includegraphics[width=\columnwidth]{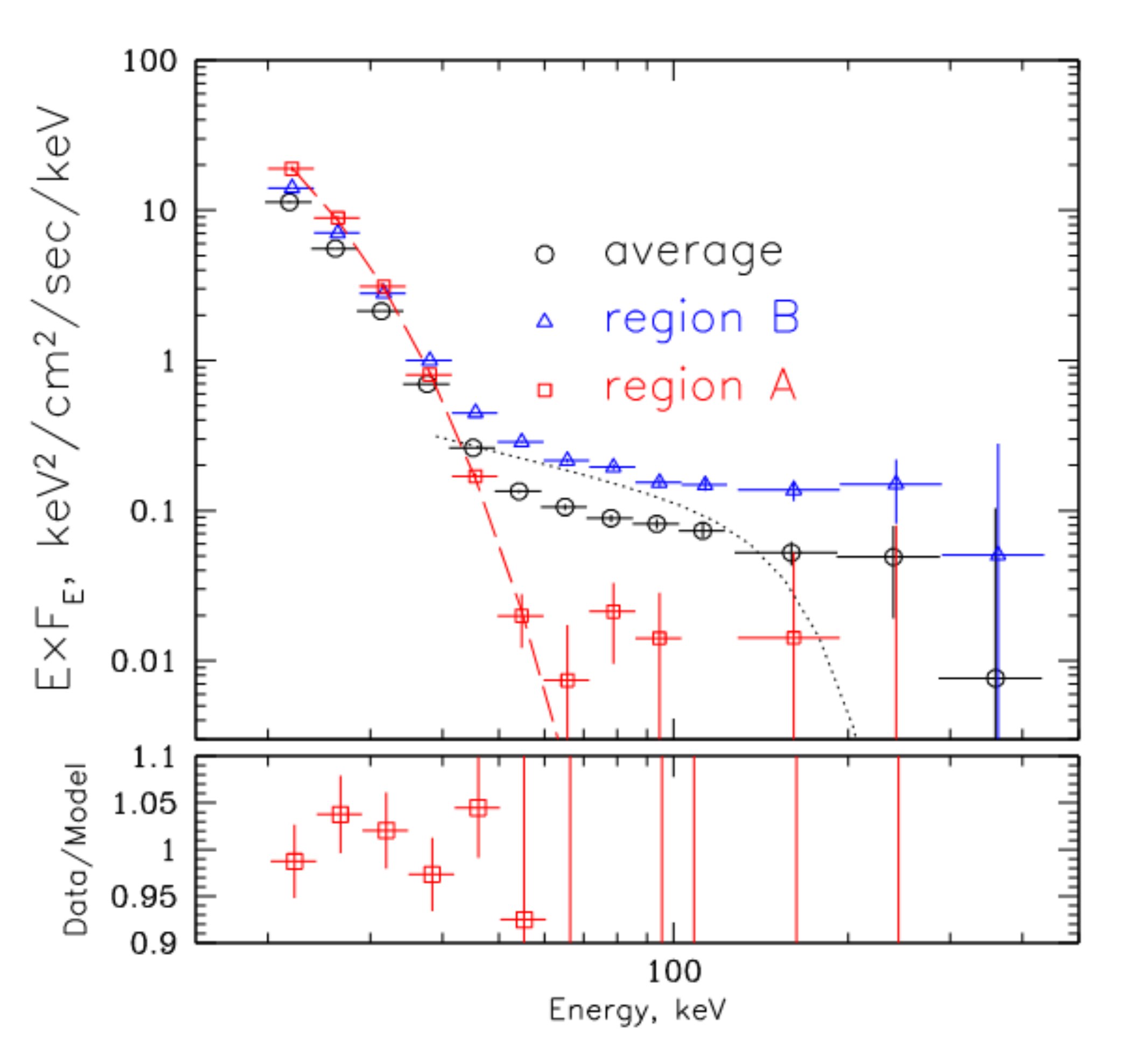} 
\caption{Average 4\,Ms \INTEGRAL\ spectrum of Sco~X--1 (in black) and its spectra from separate regions of the CC diagram (red, FB; blue, HB). (From \citealt{2014MNRAS.445.1205R}) 
\label{fig:ScoX1_MR}} 
\end{figure}

\cite{2014MNRAS.445.1205R} performed a deep study of the available \INTEGRAL\ data of the brightest accreting NS, Sco~X--1. The source, belonging to the Z class, has been known to have a bright hard X-ray tail extending above 30\,keV \citep{2001AdSpR..28..389D}. Figure~\ref{fig:ScoX1_MR} shows the average spectrum (in black, 4\,Ms of dead-time corrected exposure) and two spectra from separate regions of the CC diagram (red, FB; blue, HB). The hard tail is strongest on the HB, as for the remaining Z sources, with statistically significant spectral data points up to about 200\,keV. The black dotted curve represents the analytic approximation  of Monte Carlo simulated bulk-motion Comptonization spectrum following the receipt from \cite{2001ApJ...554L..45Z} for an accretion onto a BH, where bulk motion velocities of matter are expected to be higher than in accreting NSs. The authors conclude that the hard X-ray tail emission cannot be due to bulk motion and instead originates as a Compton up-scattering of soft seed photons on electrons with intrinsic non-thermal distribution. 

Comptonization in hybrid thermal/non thermal plasmas to explain the hard X-ray tails was firstly suggested for BHs \citep{1998PhST...77...57P, 1999ASPC..161..375C} and with the first detection of hard tails also in NS LMXBs was successfully applied by many authors to NSs as well: e.g., \cite{2006ApJ...649L..91D, 2007ApJ...667..411D, 2005A&A...434...25F}. In this scenario, the presence or absence of a hard tail is explained by changing the power injected to accelerate some fraction of electrons over the Maxwellian energy distribution. The mechanism responsible for such injection could be the acceleration in a jet. Indeed radio emission has been detected in the HB of Z sources where also the hard tails appear (\cite{2006MNRAS.366...79M} and \cite{2006A&A...459..187P} for a radio to hard-tail $L_{40-100~{\rm keV}}$ flux correlation in Z and bright \lq GX' Atolls). This could indicate that either the same mechanism is at the origin of both radio and hard-tail emission, i.e. the jet itself, or that both occurrences (radio and hard tail) are triggered by the same source configuration.

\begin{figure*}
\centering
\includegraphics[width=\columnwidth]{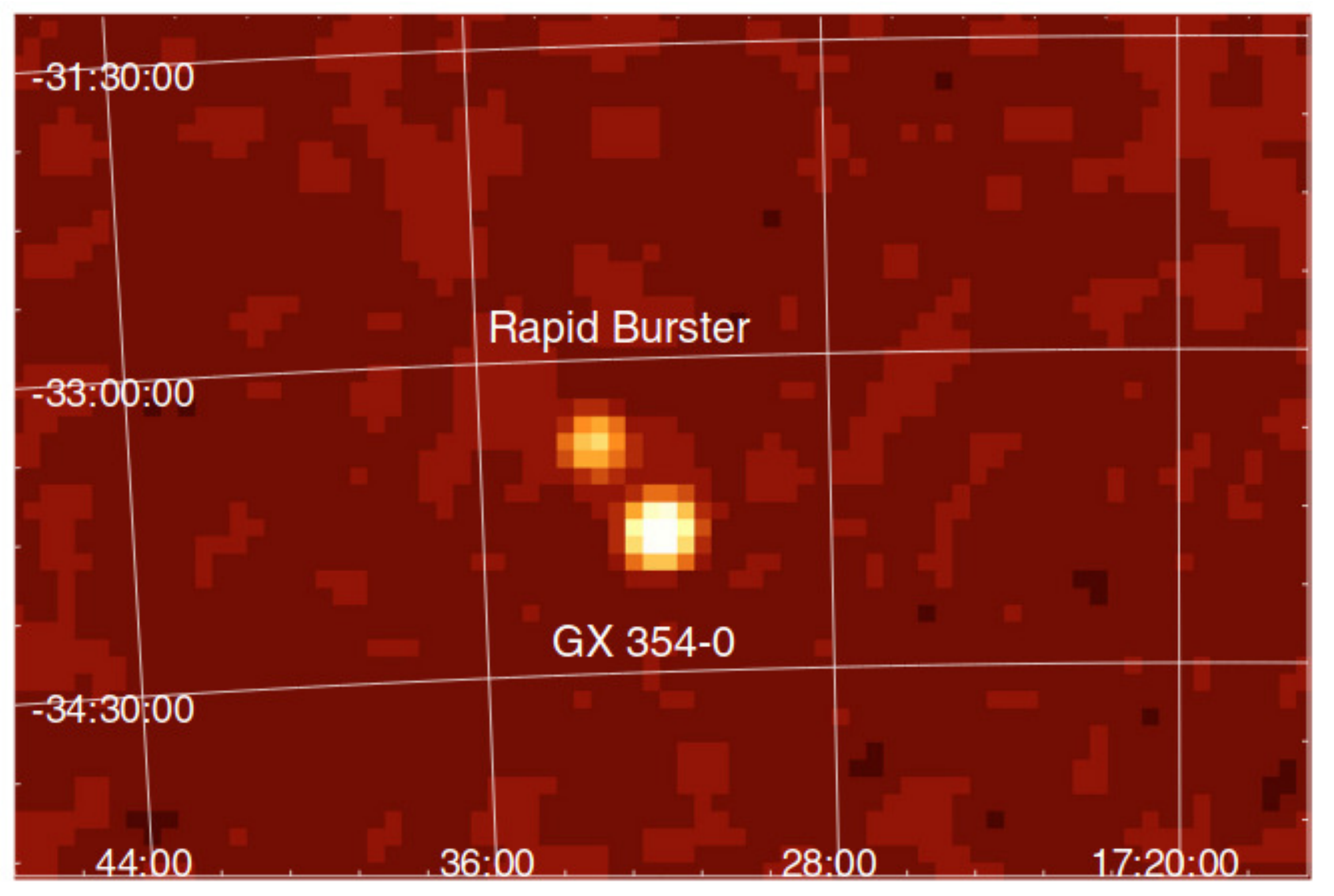}
\includegraphics[width=\columnwidth]{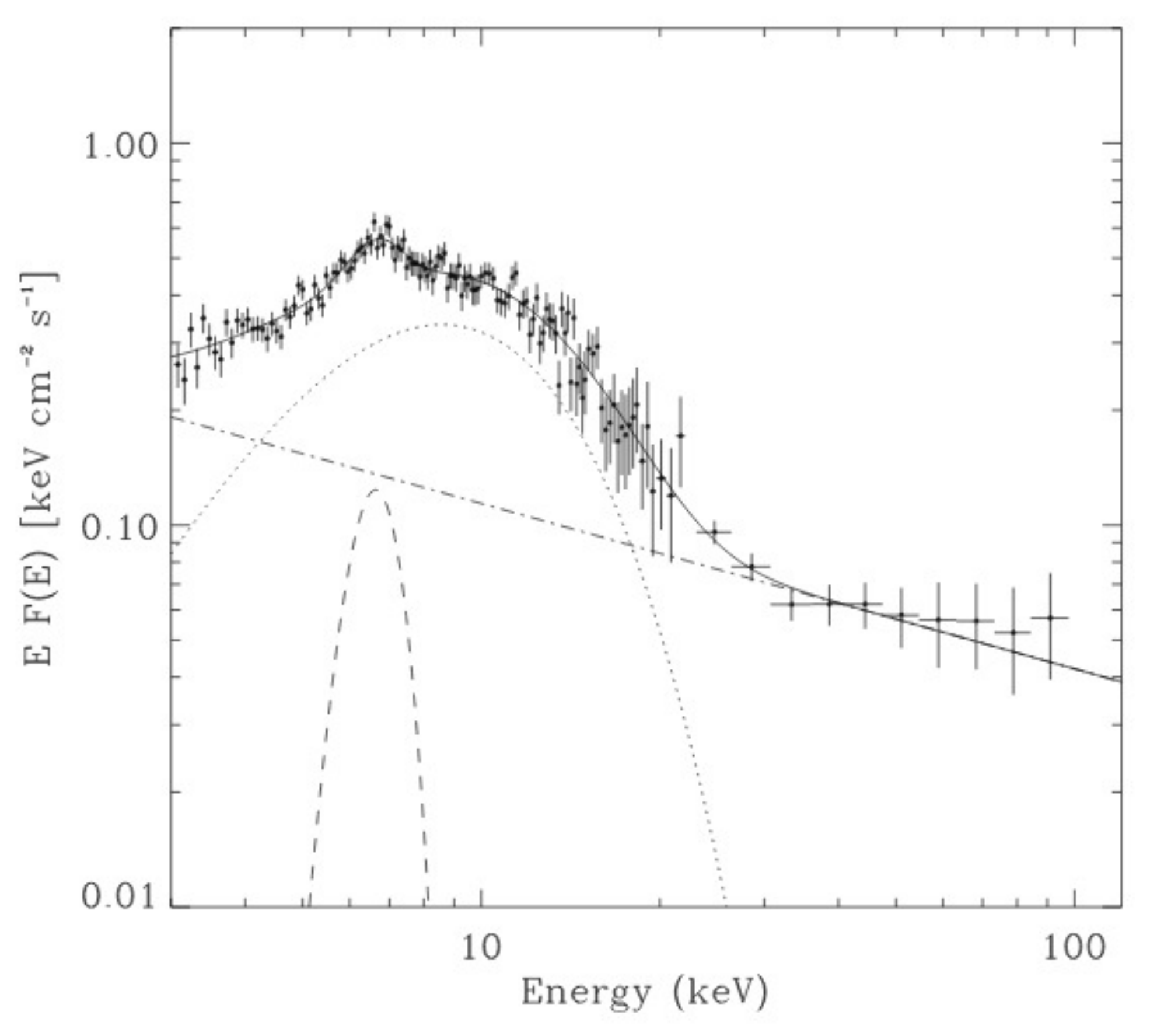} 
\caption{\emph{Left panel}: IBIS/ISGRI 20--40\,keV image of the Rapid Burster, for the first time resolved from the nearby 4U~1728--34=GX~354--0. \emph{Right panel}: JEM-X and IBIS/ISGRI spectrum of the source. The hard tail in a soft spectral state is clearly visible. (From \citealt{2004A&A...426..979F})
\label{fig:falangaRB}} 
\end{figure*}

\subsubsection{Not only Z sources}

Long-term observations thanks to the \INTEGRAL\ deep monitoring of the Galactic plane and Galactic Centre contributed to the discovery of hard tails also in the soft spectra of Atoll sources (similarly to the hard tails in Z sources): GX~13$+$1 \citep{2006A&A...459..187P, 2010A&A...512A..57M}, GX~3$+$1 (albeit non-simultaneous \emph{XMM-Newton}--\INTEGRAL\ coverage, \citealt{2015MNRAS.450.2016P}),  and 4U~1636--53 \citep{2006ApJ...651..416F}. GX~13$+$1 has always been considered a hybrid Atoll/Z source and the hard tail detection and luminosity observed by \INTEGRAL\  together with its radio emission \citep{2006MNRAS.366...79M} confirms its Z-similarity, regardless of its original classification as an Atoll. However, this is not the case for the other sources that unveil a new behaviour of soft states in Atoll sources.

Furthermore, \INTEGRAL\ has provided the first unbiased measurement of the high energy X-ray emission from the Rapid Burster, MXB~1730--335 \citep{2004A&A...426..979F}, without contamination from the nearby (30$^{\prime}$) Atoll source 4U~1728--34 (Fig.~\ref{fig:falangaRB}, left panel). The broad-band JEM--X and IBIS/ISGRI spectrum revealed a very bright hard tail on top of a soft spectrum (Fig.~\ref{fig:falangaRB}, right panel), similarly to the Z and Atoll soft-state spectra previously discussed.

Besides GX~5--1 and the Rapid Burster, \INTEGRAL\ has made it possible to exclude hard X-ray contamination in past observations of the Atoll Ser~X--1 (4U~1837$+$04). In this case however, the source (observed in its bright banana state in the early \INTEGRAL\ observations) unveiled no hard tail, with the source detected up to 30\,keV \citep{2004A&A...423..651M}. 

\begin{figure}
\centering
\includegraphics[width=0.9\columnwidth]{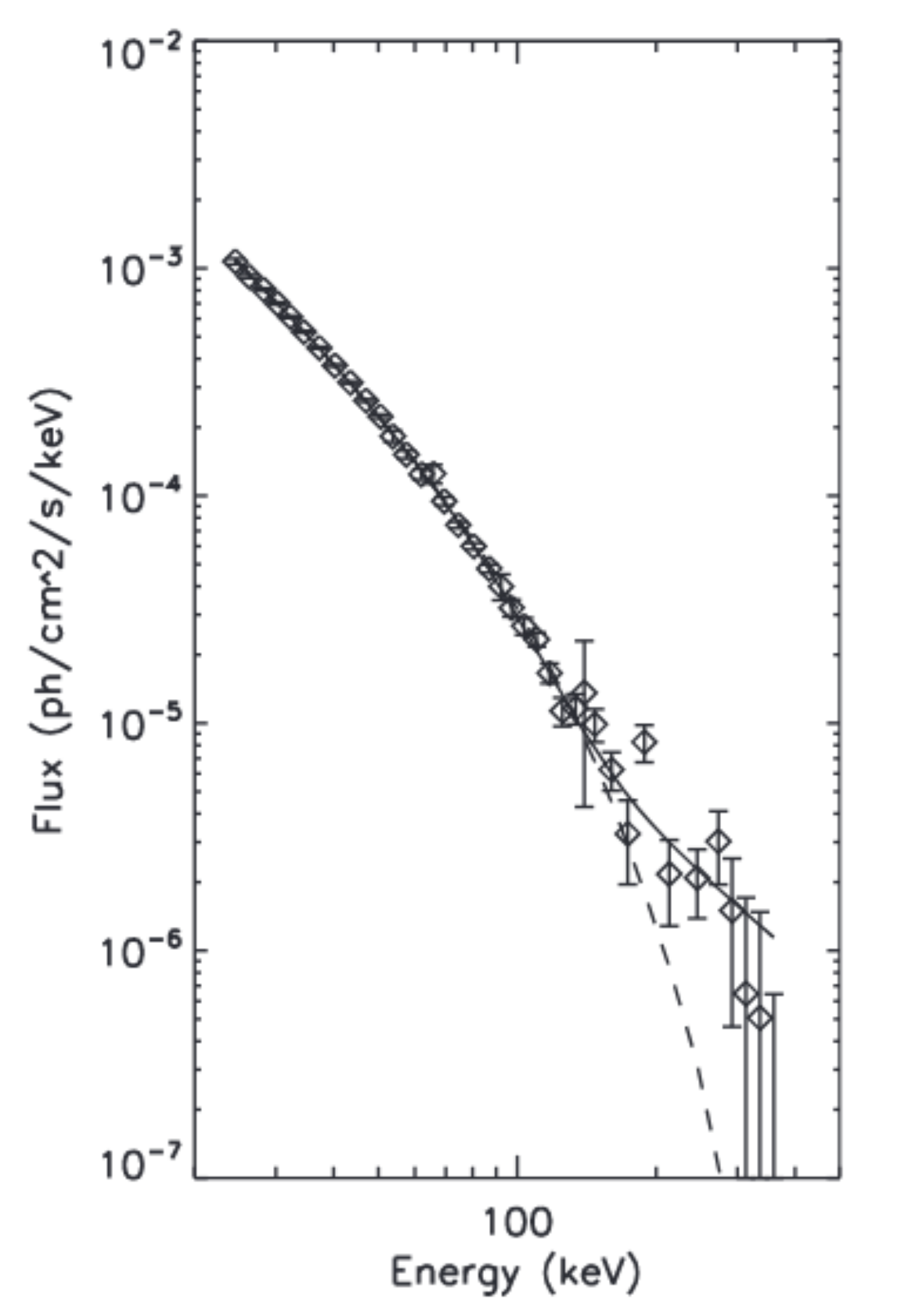} 
\caption{Hard X-ray tail discovered in the Atoll GS~1826--238. The average \INTEGRAL/SPI 25--370\,keV spectrum is shown: the hard X-ray excess on top of the Comptonization hard-state spectrum is clearly visible. (From \citealt{2016ApJ...817..101R})
\label{fig:rodispe}} 
\end{figure}

\INTEGRAL\ deep observations have added yet another important piece to the puzzle: hard tails have been detected in the hard state of Atoll sources (unlike for the cases described above where hard tails dominated the otherwise soft spectrum above 30\,keV). This is the case for GS~1826--238 \citep{2016ApJ...817..101R} where the tail ($\Gamma\sim 1.8$) dominates beyond 150\,keV (Fig.~\ref{fig:rodispe}), 4U~1820--30 \citep[$\Gamma\sim 2.4$,][]{2007ApJ...654..494T} and 4U~1728--34 \citep[$\Gamma\sim 2$,][]{2011MNRAS.416..873T}. Correlations between radio and X-ray fluxes have been reported for 4U~1728--34 \citep{2003MNRAS.342L..67M}, and 4U~1820--30 \citep{2004MNRAS.351..186M} while no detection of GS~1826--238 has been reported in radio so far. However, since radio emission is estimated (and observed) to be at least 5--10 fainter than in the soft state, it could be that hard tails (and radio emission) are there but weak and/or, in the case of X-ray spectra, hidden under the main thermal Comptonization bump at low energies. This hard tail in excess of an otherwise hard-state spectrum could be transient in nature (indeed using \INTEGRAL\ data \citealt{2006A&A...458...21F} detected none in 4U~1728--34, similarly to \citealt{2010A&A...509A...2C} for GS~1826--238) and/or extremely weak, requiring very deep observations such as for GS~1826--238 (10\,Ms SPI spectrum).

The detection of a hard tail in the hard spectra of NS LMXBs is very interesting because this behaviour is similar to BH LMXBs (whose hard states are associated with jet formation), building yet another interesting bridge between NS and BH LMXBs. Indeed, similarly to BHs, the complexity of NS LMXB spectra and radio behaviour confirms that the whole phenomenology observed cannot be driven by a single parameter, originally thought to be the accretion rate \citep[see e.g.,][for NS hysteresis]{2014MNRAS.443.3270M}.

\subsubsection{Discrete features and Compton bump}

The spectra of LMXBs usually show discrete features such as absorption edges and emission lines, interpreted as fluorescence lines from iron (6.4--6.97\,keV) at different ionization states, as well as from other abundant elements (e.g. silicon, argon and calcium) present in the disc. In some cases, an excess of emission in the 20--40\,keV energy range is observed, interpreted as Compton scattering (reflection) of the primary continuum spectrum by the electrons of the disc. The strength of reflection is expressed as the solid angle subtended by the reflector as seen from the Comptonizing corona, in units of 2$\pi$. Even though they are often described independently, these spectral features (lines, edges and Compton bump) are believed to be closely related; they have also been successfully modelled with self-consistent reflection models \citep[][and references therein]{2007ApJ...657..448F, 2015MNRAS.450.2016P, 2016A&A...596A..21I}. Although \INTEGRAL\ is not suited for the detailed analysis of the discrete features that are below or at the lower end of JEM-X energy range, broad-band ($<6$\,keV and \INTEGRAL) quasi-simultaneous studies have been very useful in mapping the presence of the Compton hump related to the discrete features.

In the case of GX~3$+$1, quasi-simultaneous \emph{XMM-Newton} and \INTEGRAL\ observations \citep{2015MNRAS.450.2016P} revealed four asymmetrically broadened emission lines (iron, calcium, argon and sulphur, 2--7\,keV) interpreted as reflection of hard photons from the inner regions of the disc, where Doppler and relativistic effects are stronger. The authors remark that the spectra are consistent with reflection produced at $\sim 10$ gravitational radii by an accretion disc with an ionization parameter $\log(\xi)\sim 2.8$, solid angle $\Omega/2\pi\sim 0.22$\,sr, viewed under an inclination angle of $\sim35^{\circ}$. 

\emph{XMM-Newton} together with simultaneous \INTEGRAL\ data were used also to constrain the Compton reflection component in the Atoll source 4U~1702--429 \citep{2016A&A...596A..21I}. A broad emission line (6.7\,keV) and an absorption edge (8.82\,keV) were detected (associated with Fe XXV), together with an edge at 0.87\,keV (O VIII). A self-consistent reflection model fit the combined 0.3--60\,keV spectra well, implying the reflection being produced in a region with an ionization parameter $\log(\xi)\sim 2.7$ ($\sim$1.9 for the oxygen feature), viewed at $\sim44^{\circ}$, and solid angle $\Omega/2\pi\sim0.07$\,sr. The small solid angle subtended by the reflector as seen from the corona could be due to a compact corona with little superposition with the disc, or a patchy corona above the inner part of the disc.

\begin{figure}
\centering
\includegraphics[width=\columnwidth]{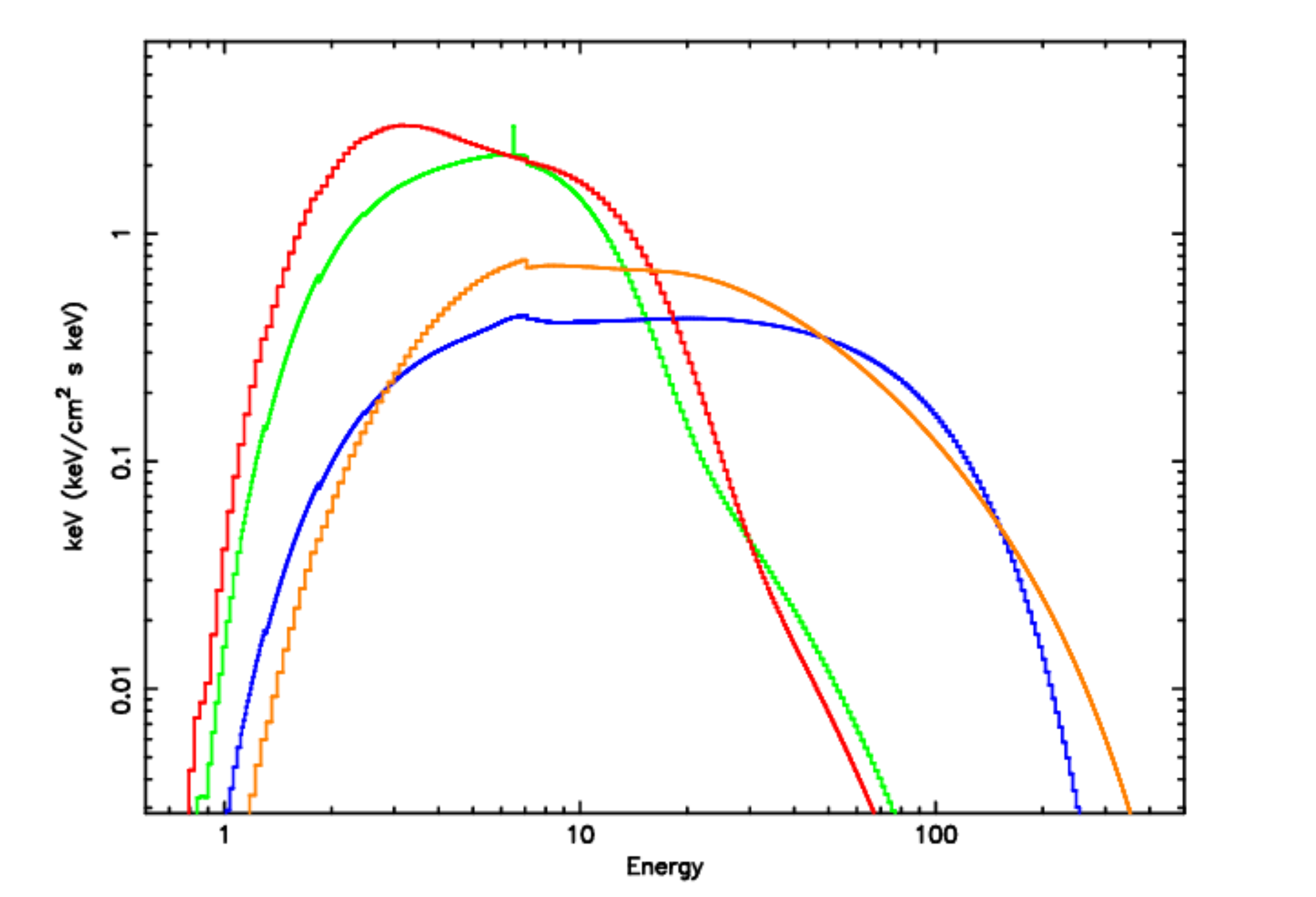} 
\caption{Comparison of the different models describing the spectral states of the Atoll 4U~1705--44. (From \citealt{2007ApJ...657..448F})
\label{fig:fiocchispe}} 
\end{figure}

Using \emph{BeppoSAX} and \INTEGRAL\ data, \cite{2007ApJ...657..448F} detected, for the first time, Compton reflection in the spectrum of the Atoll 4U~1705--44. Unlike the previous cases, the study of the evolution of this feature led to the observation of a strong iron line when the reflection was weak with respect to other epochs, suggesting that these two features, line and Compton hump, are not always positively correlated. The spectral transitions exhibited by 4U~1705--44 are shown in Fig.~\ref{fig:fiocchispe}.

\subsubsection{A comprehensive scenario}

A commonly agreed scenario to explain all the observed spectral phenomenology in weakly magnetized NS LMXBs is yet to come. It is not clear why sources like GX~9$+$1 and GX~5--1, with similar X-ray spectra, exhibit a completely different behaviour regarding the transient hard tail, never detected in GX~9$+$1 notwithstanding the extensive \INTEGRAL\ search \citep{2007ESASP.622..421V, 2009MNRAS.393..569S}. On the other hand, sources like GX~5--1, that do have the transient hard tail, never seem to reach very low accretion rates that are characterized by the low hard state as the ones seen e.g. in 4U~1812--12 \citep{2006A&A...448..335T}, 4U~1728--34 \citep{2006A&A...458...21F}, 4U~1608--522 \citep{2008ApJ...688.1295T}, SAXJ~1810.8--2609 \citep{2009ApJ...693..333F}, GS~1826--238 \citep{2010A&A...509A...2C}, IGR~J17473--2721 \citep{2011A&A...534A.101C} and EXO~1745--248 \citep{2017A&A...603A..39M} for which plasma temperatures $kT_e$ in excess of 20\,keV have been obtained with \INTEGRAL. 

\begin{figure}
\centering
\includegraphics[width=\columnwidth]{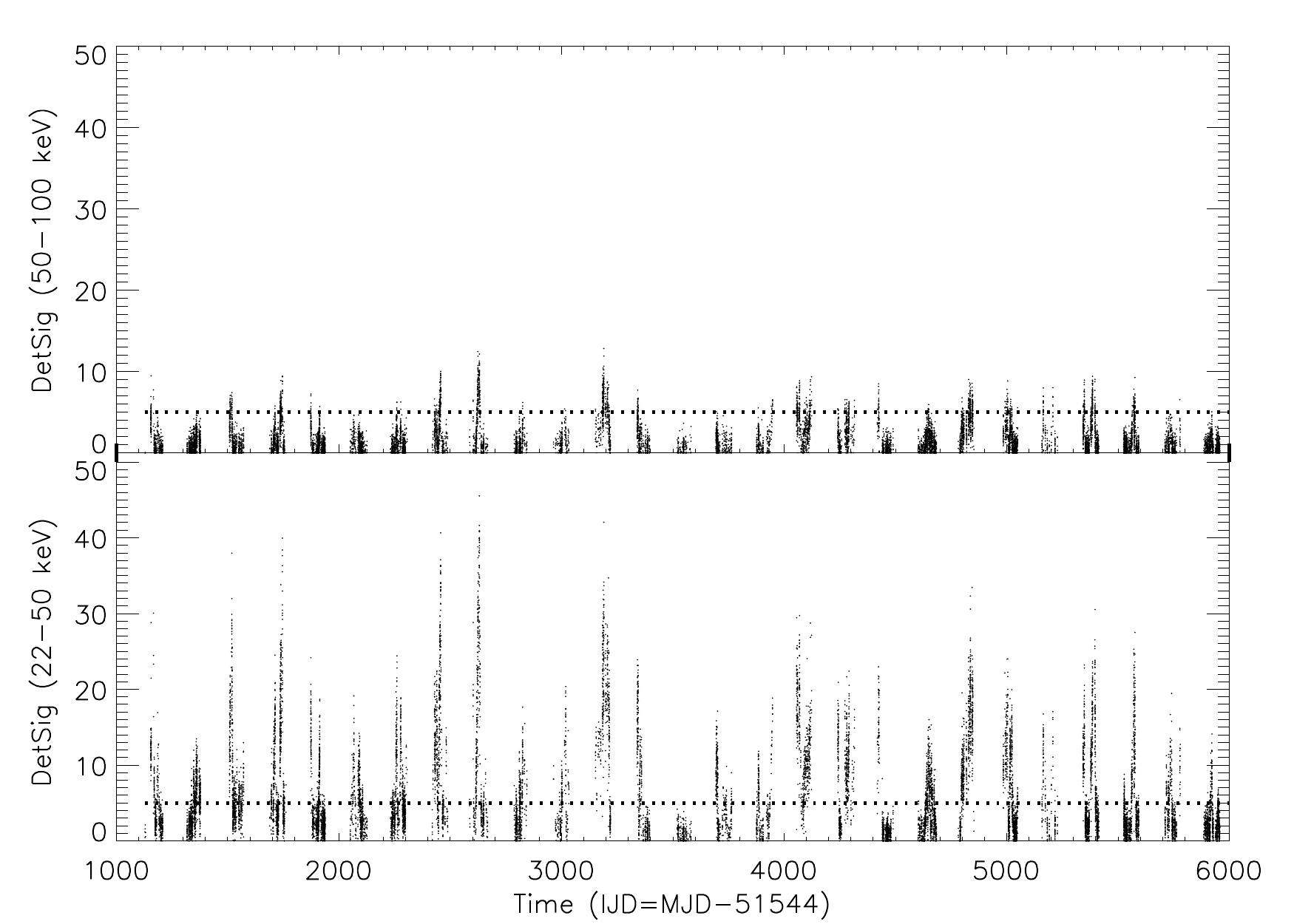}
\caption{IBIS/ISGRI detection significance curve of GX~354--0 ($l,b=354.30, -0.15$) as seen in a $\sim$14 year data span. Each dot corresponds to a single pointing ($\sim$2\,ks, total of $\sim$20 thousand pointings). The horizontal line marks the 5\,sigma detection threshold  \citep[data from][]{2016int..workE..17P}.
\label{fig:gx354m0}} 
\end{figure}

The long and deep coverage of the Galactic plane and Galactic Centre with \INTEGRAL\ (e.g. see Fig.~\ref{fig:gx354m0}) has changed the way weakly magnetized NS LMXBs are perceived, unveiling that what was firstly believed to be prerogative of BH and then Z sources, is actually shared by Atolls as well. With \INTEGRAL, a new step towards understanding these sources is being undertaken. Broad-band and deep coverage is the key: it smooths out the edges of source classifications, enhancing the need to plunge into a physical self-consistent scenario the vast \INTEGRAL\ archive at our disposal, and the one to come.

\subsection{Thermonuclear bursts} \label{s:bursts}

Initial reports of cosmic X-ray bursts date back to 1969--1971 \citep{1972ApJ...171L..87B,1975SvAL....1...32B} and their active studies began in the mid-1970s \citep{1976ApJ...206L.135B,1976ApJ...205L.127G,1976Natur.261..562H}. The bursts were characterized by a peculiar time profile with a fast ($\sim$1\,s) rise and a slow (from seconds to dozens of seconds) quasi-exponential decay, and typically emitted $10^{39}$--$10^{40}$\,erg. Such events were named type I X-ray bursts; their sources (bursters) were identified with NS LMXBs, while the bursts themselves were explained by thermonuclear explosions of hydrogen and/or helium accreted onto the surface of a NS with a relatively weak ($\sim 10^8$\,G) magnetic field \citep{1976Natur.263..101W,1977xbco.conf..127M}. This, in particular, is the case for millisecond X-ray pulsars, from which type I X-ray bursts have been observed with various missions including \INTEGRAL\ (e.g. \citealt{2011A&A...525A..48F}). Type II bursts, which have a different profile shape and are believed to be associated with instabilities in the accretion flow onto the NS are also observed from a number of sources (the archetypal one being called the \lq rapid burster'). For general reviews of bursters see \cite{1993SSRv...62..223L,2006csxs.book..113S,2017arXiv171206227G}.

Type I X-ray bursts have a soft blackbody-like (Wien) spectrum with a temperature $\sim$2--3\,keV, with the bulk of the eneregy released below 10--15\,keV. Thus JEM-X is the primary instrument on board \INTEGRAL\ for studying such events. Yet the IBIS/ISGRI detector has allowed several hundreds of type I X-ray bursts to be detected in the 15--25\,keV energy range \citep{2011AstL...37..597C}. 

In this section we review the main results concerning X-ray bursters that have been obtained using \INTEGRAL\ data.

\subsubsection{Discovery of new X-ray bursters}
\label{s:newbursters}

Over 110 type I X-ray bursters are currently known\footnote{https://personal.sron.nl/$\sim$jeanz/bursterlist.html}. Given the development of the  near-Earth group of astrophysical satellites with wide-field X-ray monitors on board over the last thirty years, it can be asserted that most of the sufficiently bright persistent and more or less regularly flaring bursters in the Galaxy have already been discovered. And this is not surprising:  modern instruments are capable of detecting thermonuclear bursts from the most remote regions of the Galaxy. 

\INTEGRAL\ has discovered or helped to identify a number of new X-ray bursters. Most of them may be called \lq hidden' bursters or \lq burst-only' sources. In such obejcts, thermonuclear explosions occur very rarely, with a characteristic recurrence time of several months to tens of years (\citealt{2006csxs.book..113S}). The rarity of explosions most likely reflects the slowness of accumulation of a critical mass of matter needed for its ignition on the NS surface. The bursts associated with such rare explosions are easy to miss, while in the intervals between them such systems may be observed during deep point observations as faint persistent X-ray sources of unknown origin or remain undetected.
%\emph{Until we obtain a deep all sky survey with X-ray mirror system telescopes (such as a 4-year survey by eROSITA and ART-XC telescopes on board SRG mission, starting in December 2019) such a class of sources will remain.} 

As an example, IGR~J17254--3257 was discovered in 2003 during the first \INTEGRAL\ Galactic Centre survey in the 20--60\,keV energy band \citep{2004ATel..229....1W}. Later, in 2004, a type I X-ray burst was detected from the source in JEM-X 3--30\,keV data \citep{2006ATel..778....1B} indicating a NS LMXB origin. \INTEGRAL\ observations performed on October 1st, 2006 revealed a 15-min long burst from the system \citep{2007A&A...469L..27C}. The authors concluded that a short burst originated from a weak hydrogen flash that prematurely triggered mixed H/He burning, while the intermediate burst resulted from the ignition of a large helium pile beneath a steady hydrogen burning shell.

Using \INTEGRAL\ archival data, \cite{2007AstL...33..807C} found a type I X-ray burst detected with both JEM-X and IBIS/ISGRI from the weak and poorly known source AX~J1754.2--2754. Based on the spectral evidence of a photospheric radius expansion episode at the beginning of the burst, indicating that the Eddington luminosity was reached at the peak, the authors estimated the distance to the source to be $6.6\pm0.3$ or $9.2\pm0.4$\,kpc assuming a pure H or pure He atmosphere, respectively.

A very faint X-ray transient XMMU~J174716.1--81048 was discovered in 2003 but its nature remained unknown until a type I X-ray burst was detected by \INTEGRAL\ in 2005 \citep{2007A&A...468L..17D}. Noting the double peak structure of the burst, the authors inferred the distance to the source to be $\sim 3$\,kpc.

IGR~J17191--2821 was discovered in 2007 by IBIS/ISGRI during Galactic bulge monitoring \citep{2007A&A...466..595K}. The first type I X-ray burst from the system was observed with \emph{RXTE} \citep{2007ATel.1065....1K}. Further \emph{RXTE} observations have revealed burst oscillations and kHz quasi-periodic oscillations \citep{2010MNRAS.401..223A}. 

As a result of a systematic search for type I X-ray bursts in the \INTEGRAL\ archival data,  \cite{2006AstL...32..456C} (see the next subsection) discovered a burst-only source IGR~J17380--3749  \citep{2010AstL...36..895C}. A subsequent analysis of \emph{RXTE} data showed that the burst was detected during a fairly long flare of the source. Later in 2008--2010, three more flares were detected from the source though no other bursts were found. 

While scanning the Galactic Centre region on April 10, 2017, \INTEGRAL\ detected a type I X-ray burst from a faint unidentified source, IGR~J17445--2747 \citep{2017AstL...43..656M}. Under the assumption that the burst luminosity approached the Eddington limit, the authors estimated the maximum distance to the source to be 12.3\,kpc if pure helium was accreted and 7.7\,kpc for the case of matter of solar chemical composition.

%IGR J00291+5934 \citep{2015ATel.7849....1K}
%IGR J17062--6143 \citep{2012ATel.4219....1D}
%IGR J17498--2921 \citep{2011ATel.3560....1F}
%GR J17511--3057 \citep{2009ATel.2198....1B}
%GR J18245--2452 \citep{2013ATel.4959....1P}

\subsubsection{Burst catalogues}

\cite{2006AstL...32..456C} took advantage of the wide field of view of the IBIS instrument to perform a \lq blind' search for type I X-ray bursts in the whole dataset accumulated by the ISGRI detector in 2003--2004. Although this search was necessarily restricted to hard X-rays (namely, 15--25\,keV), an energy range where only a small fraction of burst emission can be detected, it resulted in the detection (in the detector light curves) of 1077 bursts (of 5--500\,s duration) of various origins, 105 of which were localized and identified as type I X-ray bursts. 

Based on this success, the analysis was subsequently extended to 2003--2009 observations \citep{2011AstL...37..597C} and a total of 834 type I bursts (including 105 bursts localized in the previous work) were detected and localized with IBIS/ISGRI, 239 of which were also  detected by JEM-X in the standard X-ray energy band. Interestingly, 587 of the bursts proved to come from a single source -- the \lq slow burster' (see below).

In the third paper of the series, spanning the 2003--2015 period, \cite{2017AstL...43..781C} focused on JEM-X data, not only using detector light curves but also extracting source light curves for 104 known bursters. 
%The authors also continued their analysis of ISGRI detector light curves but it yielded no positive results owing to the degradation of ISGRI sensitivity near the low energy threshold (below $\sim 20$~keV). 
The final joint catalogue of JEM-X and ISGRI bursts comprises 2201 events. Based on this highly representative sample, the authors analyzed the dependence of mean burst rate on persistent luminosity. Also, several multiple bursts were detected and later studied in detail (see Section~\ref{ss:bursts_theory} below).

Another endeavor in cataloguing type I X-ray bursts is the Multi-INstrument Burst ARchive (MINBAR)\footnote{https://burst.sci.monash.edu/minbar/}, which has been developed since 2007 by the world's top researchers in this area. Currently, MINBAR contains information on over 7,000 type I X-ray bursts observed by \emph{RXTE}, \emph{BeppoSAX} and \INTEGRAL. The database is still under preparation at the time of this writing.

\subsubsection{Long bursts}

Most of the several thousand bursts observed by various missions so far are so-called normal bursts: they have duration from a few seconds to several minutes and are believed to originate from unstable thermonuclear burning of the accreted hydrogen and helium on the NS surface. Depending on the composition of the accreted fuel and on the local accretion rate, normal bursts can have different profiles and energetics (e.g. \citealt{2006csxs.book..113S}). Yet there are some outstanding bursts. So-called intermediate bursts last for dozens of minutes and are believed to originate from unstable thermonuclear burning of larger piles of He. When the donor in the binary system is an H-poor star so that the NS accretes mainly He (as in the case of ultracompact X-ray binaries) and the accretion rate is low enough to prevent stable He burning, He can pile up in substantial quantities to later burn in runaway thermonuclear reaction giving rise to an intermediate type I X-ray burst. But even for H-rich accretors, at low local accretion rates $\sim$0.3--1\% of $\dot{M}_{\rm Edd}$ (\citealt{2007ApJ...654.1022P}, though the boundary values are sensitive to the heat generated by the NS crust, \citealt{2007ApJ...661..468C}), unstable H burning can accumulate a thick layer of He leading to an intermediate burst (as e.g. in the case of IGR~J17254--3257 discussed in \S\ref{s:newbursters} above). 

Even more spectacular are the representatives of a rare group of \lq superbursts' -- day-long events that emit $10^{41}$--$10^{43}$\,erg and are usually explained by unstable thermonuclear burning of $^{12}$C-rich matter. A substantial amount of accreted H and He is burned in a stable manner between the bursts in bright sources or in outbursts in low persistent emission transient sources at accretion rates close to and over 10\% of $\dot{M}_{\rm Edd}$. This process is believed to be the source of fuel for carbon superbursts. 

Some bursters show all the existing types of bursts, while other systems stick to just one or two. For example, the ultracompact X-ray binary candidate SLX~1737--82 is known to only emit intermediate bursts with an apparent recurrence time of 86~days \citep{2008A&A...484...43F}. \INTEGRAL\ has allowed us to extend the sample of observed long bursts as well as to expand our knowledge of this yet not fully understood phenomenon. 

On September 17, 2003, \INTEGRAL\ detected a peculiar type I X-ray burst from the known burster SLX~1735--269 \citep{2005A&A...434.1069M}. The burst's duration exceeded 2\,ks, putting it into the rare class of intermediate bursts, and its profile demonstrated several notable features. In particular, the rise time from the beginning of the burst to the moment when the luminosity reached the Eddington limit was atypically long, $\sim$100\,s. The authors argue that such a long rise time cannot be explained by an unusual chemical composition of the burning fuel or by a slow propagation of the thermonuclear burning front over the NS surface, and that there possibly occurred a series of (triggered) thermonuclear bursts on the NS surface that led to this unusual X-ray light curve. Considering a $^{12}$C flash scenario for the long burst and calculating the time needed to accumulate the required amount of carbon, the authors concluded that more likely this burst resulted from burning of a large pile of H and He. They also reported the detection of 5 normal type I X-ray bursts from SLX~1735--269, similar in its characteristics to the only such burst from the source known by that time. Based on this sample, they inferred that most of the persistent luminosity of SLX~1735--269 can be explained by almost pure He burning. Finally, from the observed spectral and burst activity changes with small variations in luminosity, they concluded that the system operates at an accretion rate near the border of two thermonuclear burning regimes.

Thanks to regular monitoring of the Galactic Centre region and the capability of broad-band X-ray spectroscopy, \INTEGRAL\ has also proved very useful in follow-up observations of bursters discovered by other missions. In particular, a new transient, Swift~J1734.5--3027, was discovered by \emph{Swift}/BAT in 2013 and a long ($\sim$2\,ks) type I X-ray burst occurred at the apparent beginning of the outburst. However, it turned out that the source had already been detected with \INTEGRAL/IBIS half a day before the \emph{Swift} discovery, which, together with a broad-band spectral analysis, demonstrated that the outburst developed as a result of the accretion disc's instability, which then paved the way for the long X-ray burst rather than the burst triggered the outburst \citep{2015A&A...579A..56B}. 

\begin{figure}
\centering
\includegraphics[angle=270,width=\columnwidth]{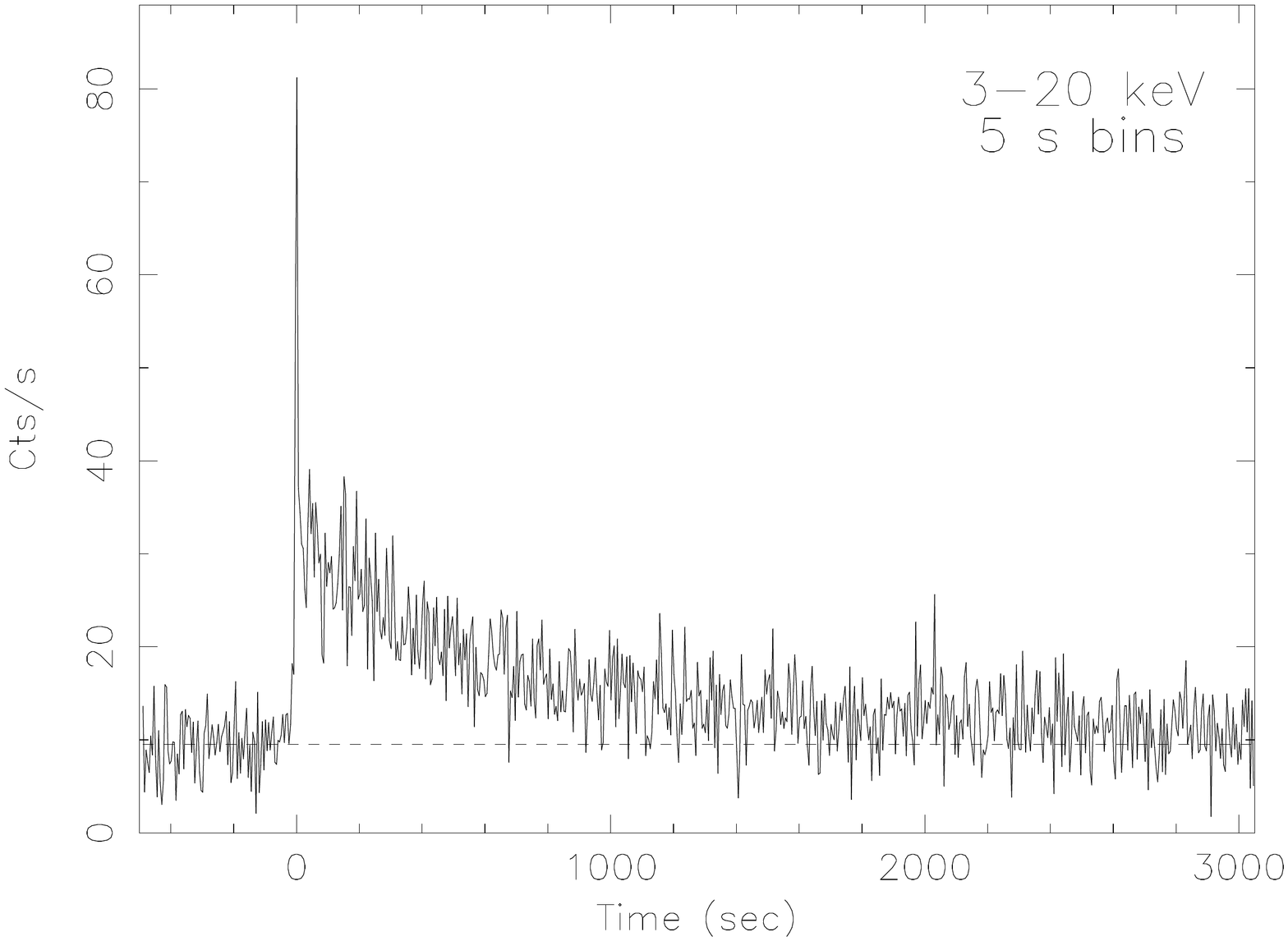}
\caption{Long two-phase X-ray burst from GX 3+1 on August 31, 2004, as observed by JEM-X. (From \citealt{2006A&A...449L...5C})}
\label{fig:chenevez_twophaseburstlc}
\end{figure}

An interesting, more than 2\,ks long, type I X-ray burst was observed by \INTEGRAL\ from GX~3+1  \citep{2006A&A...449L...5C}. It started as a normal burst with a fast 1.3\,s rise and went on to fade away with a decay time of 3\,s, but in the middle of the fall its character changed into a quasi-exponential decay lasting more than 2,000\,s (Fig.~\ref{fig:chenevez_twophaseburstlc}). The rise of the long-lasting emission cannot be distinguished from the initial spike, but the tail is much softer than the initial spike. It remains unclear which of the two events triggered the other one. The authors discuss three possible scenarios of the long burst: (i) unstable burning of a H/He layer involving an unusually large amount of H that slows down the whole process as the rate of hydrogen fusion through rapid proton capture is limited by slow $\beta$ decays, (ii) a flash of a big pile of He and (iii) premature ignition of a carbon layer, eventually favouring the first one.

A unique bursting behaviour was observed by \INTEGRAL\ from SAX~J1747--2853 on February 13, 2011 (\citealt{2011ATel.3183....1C}). Only the beginning of the burst was reported, but rough extrapolation of the burst profile suggests a duration of 4 hours, referring the event to the class of superbursts. This event is unique because an intermediate long burst immediately preceded the superburst \citep{Chenevez_etal_2019_inprep}.

Challenges to our understanding of superburst physics were highlighted with observation of the first superburst from the LMXB transient 4U~1608--522 \citep{2008A&A...479..177K}. The event started 55 days after the beginning of an accretion outburst and resembled previously observed superbursts from 4U~1254--690 and KS~1731--260. The authors analyzed the superburst as well as the long-term accretion and bursting behaviour of 4U~1608--522, using data from \emph{RXTE}, \INTEGRAL\ and other missions. It is often stated that the $^{12}$C fuel for superbursts is produced during outbursts, while the stable nuclear burning of $^{12}$C and the frequent type-I X-ray bursts lower the carbon abundance. However, the 4U~1608--522 outburst was much too short for accumulating the required pile of carbon (a few $10^{12}$\,g\,cm$^{-2}$) for the superburst. Even more challenging is a superburst detected in 2011 from EXO~1745--248 \citep{2012MNRAS.426..927A}, which started less than half a day from the onset of an accretion outburst. It thus appears that significant amounts of carbon can survive the long periods between the outbursts. Clearly, further investigations, taking into account mixing and sedimentation in the NS envelope, are needed to resolve the challenges faced by current superburst ignition models.

\subsubsection{Normal bursts: case studies}

\paragraph{\underline{Slow burster}}

4U~1728--34 is a well-known He accretor often referred to as the \lq slow burster', whose persistent emission and bursting behaviour have been studied by virtually all major X-ray astronomy missions.

\cite{2006A&A...458...21F} used the 2003--2004 \INTEGRAL\ data to study for the first time the source's  spectra in the wide 3--200\,keV energy band. In these observations, 4U~1728--34 underwent a typical (for Atoll sources) spectral transition from an intermediate/hard to a soft state while its luminosity increased from 2\% to 12\% of Eddington. 
%The persistent spectrum could be described in terms of thermal Comptonization with the electron temperature decreasing from 35 to 3 keV and Thomson optical depth increasing from 0.5 to 5. 
A total of 36 type I X-ray bursts were detected, including two demonstrating photospheric radius expansion. These latter events occurred during the intermediate/hard state when the local accretion rate was relatively low, $\sim 1.7 \times 10^{3}$\,g\,cm$^{-2}$\,s$^{-1}$, whereas the other, weaker bursts occurred during the soft state when the accretion rate was high, (2.4--9.4)$\times 10^{3}$\,g\,cm$^{-2}$\,s$^{-1}$. Moreover, an anti-correlation was observed between burst peak flux or fluence and accretion rate.

%\cite{2011MNRAS.416..873T} used one-year 4U-1728--34 monitoring campaign data from \emph{RXTE} and %\INTEGRAL\ to reveal the presence of an excess flux with respect to the Comptonization model at %energies above 50 keV -- the first detection of non-thermal hard component for this burster similar to %the feature detected in few other Atoll sources. 

\cite{2010ApJ...718..947M} reported on their analysis of type I X-ray bursts from 4U~1728--34 based on \emph{RXTE}, \emph{Chandra} and \INTEGRAL\ data. The authors compared the properties of 38 bursts with the predictions of ignition models taking into account the heating and cooling in the crust. The estimated burst ignition column depths proved to be significantly lower than expected. As possible explanations, the authors suggested shear-triggered mixing of the accreted helium to larger depths and fractional covering of the NS surface by the accreted fuel.

\cite{2017A&A...599A..89K} continued studying \INTEGRAL\ data for 4U~1728--34 accumulated over 12 years. Stacking spectra of 123 bursts detected when the source was in the hard state, they detected emission up to 80\,keV during the bursts. Interestingly, the emission above 40\,keV was found to drop during the bursts down to a third of the persistent emission level. This suggests that powerful X-ray burst emission causes an additional cooling of the electrons in the hot part of the accretion flow near the NS surface. The authors also detected a high-energy X-ray tail in the persistent emission when the source was in its soft state, presumably due to Comptonization on non-thermal electrons in a hot flow or a corona. 

\paragraph{\underline{Clocked burster}}

%For some 30 years, GS~1826--238 exhibited regular bursting behaviour in the closest agreement with theoretical model predictions among all known bursters. Due to such consistency, the source is known as the \lq clocked burster'. 

GS~1826--238 was discovered by \emph{GINGA} in 1988 \citep{1988IAUC.4653....2M} and tentatively reported as transient. In view of its similarity to Cyg X-1 and GX 339--4, \cite{1989ESASP.296....3T} suggested that the source is a BH candidate. Later on, GS~1826--238 was monitored by the Wide Field Camera on board \emph{BeppoSAX} (1996--1998) and X-ray bursts were detected from it for the first time, strongly suggesting the compact object to be a weakly magnetized NS. A total of 70 bursts were detected, showing a quasi-periodicity of 5.76\,hr in the burst occurrence time \citep{1999ApJ...514L..27U}. Due to such behaviour, the source is known as the \lq clocked burster'. 

\cite{2010A&A...509A...2C} found that between two sets of \INTEGRAL\ observations carried out in 2003 and 2006, the 2--200\,keV persistent intensity of GS~1826--238 dropped by 30\% while the burst recurrence time also decreased, apparently at variance with the typical limit-cycle bursting behaviour of the source. As a possible explanation, the authors suggested that the 2--200\,keV energy band may miss a significant fraction of the bolometric luminosity and the softer X-ray emission (not probed by the \INTEGRAL\ data) should be taken into account in evaluating the total persistent X-ray flux and the corresponding accretion rate. 

In 2014, the source was caught for the first time in the high/soft spectral state and this was thoroughly investigated using observations with \emph{Swift} and \emph{NuSTAR} and archival data from \emph{RXTE} and \INTEGRAL\ (MINBAR, \citealt{2016ApJ...818..135C}). 
%Interestingly, the estimated accretion rate in the soft state was within the range of previous %observations ($\sim 13$\% of $\dot{M}_{Edd}$). 
Several type I X-ray bursts detected during this period proved to be significantly weaker and shorter and their profiles and recurrence times varied significantly compared to the regular bursts observed during the (usual) low/hard state of the source. Prior to 2014, the bursts were consistent with H-rich thermonuclear burning including the rapid proton capture process (rp-process) leading to longer burst decay times, whereas after 2014, the bursts became flash-like in consistence with He-rich burning regime. This demonstrates that the bursting behaviour is sensitive to changes in accretion flow geometry, although the exact physical explanation is still missing.   

\paragraph{\underline{4U 0614+091}}

A comprehensive analysis of the system was performed by \cite{2010A&A...514A..65K} based on data from the \emph{EURECA}, \emph{RXTE}, \emph{BeppoSAX}, \emph{HETE-2}, \emph{INTEGRAL} and \emph{Swift}  missions. Their efforts increased the number of bursts detected from 4U~0614+091 to 33 including 2 intermediate bursts and one superburst. Based on clear signs of a strong photospheric radius expansion within the first seconds of one of the intermediate bursts, the authors estimated the distance to the source to be 3.2\,kpc. Also, two long (a few hours) faint tails were found, one of which being the remainder of the intermediate burst with photospheric radius expansion. 
%Interestingly, no cooling is seen during the tail while the apparent emitting area decreases. 
%This can be explained by a hot underlying NS but it is surprising that the temperature during the tail %is close to the corresponding temperatures in other bursters that are not accreting such an H-poor %%matter and at such a low rate and are thus not lacking hydrogen burning through the CNO cycle in their surface layers and thus predestined to have hotter NSs.
The authors estimated the column depths and energy release per gram in the bursts (i) by comparing the
observed light curves with models, and (ii) by considering the burst energetics, and found both methods to be in good agreement. As argued by the authors, the sensitive dependence of the He ignition depth on temperature allows one to explain the existence of both normal and intermediate bursts by small variations in the accretion rate. 
%But it is not clear how to reconcile the low He fraction ($\lesssim 10 \%$) as inferred from the optical spectra  with a significantly larger amount required to achieve ignition conditions without stably burning away. The ignition depth for the superburst inferred from the light curve was found lowest of the current sample of superbursts and in good agreement with the constraints from the observed normal burst quenching time, yet the explanation of the superburst faces several problems. At such small accretion rates ($\dot{M}\sim 1$\% of $$\dot{M}_{\rm Edd}$), $^{12}$C ignition is not known to be possible for the layer is too cold and even if it is heat up carbon will burn stably. If the burst is fueled with He than the time to accumulate a necessary amount of it is much longer than the observed though this issue is solvable if an appreciable amount of accreted He survives preceding shorter bursts. But the energy released per nucleon as low as $Q_{nuc} \lesssim 0.6 MeV$ remains a puzzle.

\paragraph{\underline{MX 0836--42}}

MX~0836--42 was discovered in 1971 by OSO-7, and in 1992 type I X-ray bursts were observed from the system establishing its nature as a NS LMXB. \cite{2005AstL...31..681C} used \INTEGRAL\ and \emph{RXTE} data to study the broad-band (3--100\,keV) spectrum of the source for the first time. They also reported the detection of 24 type I X-ray bursts in JEM-X data and 15 bursts in PCA data with a recurrence time of over 2~h, confirming previous estimates. The data show a linear correlation between burst fluence and preburst persistent flux, suggesting complete fuel consumption during the bursts. The authors also set an upper limit of 8\,kpc on the distance to MX~0836--42. 

A later reanalysis of the data by \cite{2016A&A...586A.142A} revealed 61 bursts from MX~0836--42. It was demonstrated that most likely, mixed He/H burning triggered by unstable helium ignition takes place in this system. The authors also reported the detection of four series of double bursts with burst recurrence times of less than 20\,min, with the secondary bursts being shorter and less energetic than the primary and typical bursts from the source. A number of explanations have been proposed for such burst behaviour (see \S\ref{ss:bursts_theory} below)

\paragraph{\underline{IGR J17473--2721}}

In 2008, this \INTEGRAL-discovered X-ray transient underwent a 6-month long outburst that unusually started with an X-ray burst. This outburst allowed \cite{2011MNRAS.410..179C} to study the burst activity of this LMXB in a wide range of accretion rates. Using \emph{AGILE}, \emph{Swift}, \emph{RXTE} and \INTEGRAL\ data, they detected 57 thermonuclear bursts from the source. After the persistent emission rose to the level corresponding to about 15\% of the Eddington accretion rate, burst activity vanished and resumed only in a month at a persistent emission level lower then the pre-outburst one. The authors attribute such a behaviour either to thermal response of the NS crust or to a missed superburst (post-superburst bursting activity quenching has been observed in several sources). They also noticed that bursting activity in IGR~J17473--2721 strongly depends on weather the persistent emission is increasing or decreasing and concluded that we are dealing with a NS accreting hydrogen-rich material. Based on the observations of bursts with photospheric radius expansion, the distance to the source was estimated at $5.5\pm 0.8$\,kpc. 

\paragraph{\underline{GRS 1741.9--2853}}

GRS~1741.9--2853 is a faint transient bursting source in a close vicinity of the Galactic Centre. Its weakness and position determined the low amount of information about its properties. Using \INTEGRAL, \emph{XMM-Newton} and \emph{Swift} data, \cite{2009A&A...504..501T} analysed its flux variability and bursting behaviour during 2 long outbursts the source underwent in 2005 and 2007. GRS~1741.9--2853 is known to produce bursts only while in outbursts and the flares under consideration were not an exception: 15 new type I X-ray bursts were detected from the source. The brightest observed burst allowed the authors to put an upper limit to the source's distance of 7\,kpc. The burst properties and the accretion rate suggest that pure He explosions take place in the system.

\paragraph{\underline{KS 1741--293}}

This is another previously poorly studied faint transient burster located in the crowded Galactic Centre region. Long-term monitoring with \INTEGRAL\ confirmed its transient nature \citep{2007MNRAS.380..615D} and allowed its broad-band (5--100\,keV) spectrum to be studied for the first time (with JEM-X and IBIS). The spectrum could be fit with a combination of black-body emission, originating at the accretion disc and/or at the NS surface, and a hard tail resulting from Comptonization of soft photons in a hot plasma around the NS. Two type I X-ray bursts were detected from the source during these observations.

\paragraph{\underline{SAX J1753.5--2349}}

This source, which was previously considered a burst-only one, provided a unique opportunity to study its non-burst spectral properties with \INTEGRAL/IBIS and \emph{Swift}/XRT in a broad (0.3--100\,keV) energy band during its first ever outburst that took place in 2008 and lasted longer than 14 days. The authors estimated the accretion rate during the outburst and the outburst duty cycle, and concluded that SAX J1753.5--2349 is likely a very compact binary system (see a discussion on ultracompact X-ray binary systems in \S\ref{s:ucxb}). 

\subsubsection{New insights into theory} \label{ss:bursts_theory}

\begin{figure}[t]
\centering
\includegraphics[width=\columnwidth]{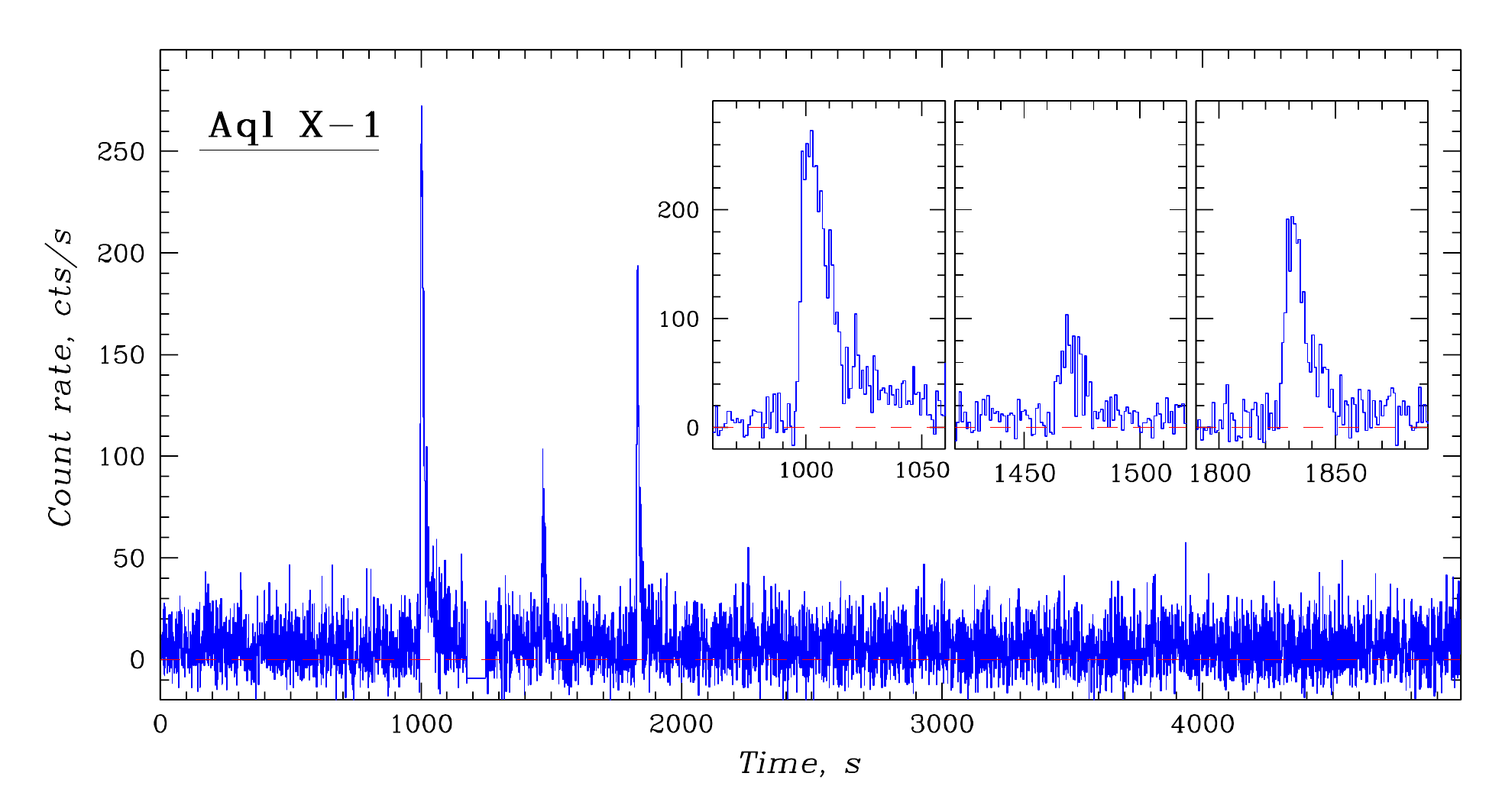}\\
\includegraphics[width=\columnwidth]{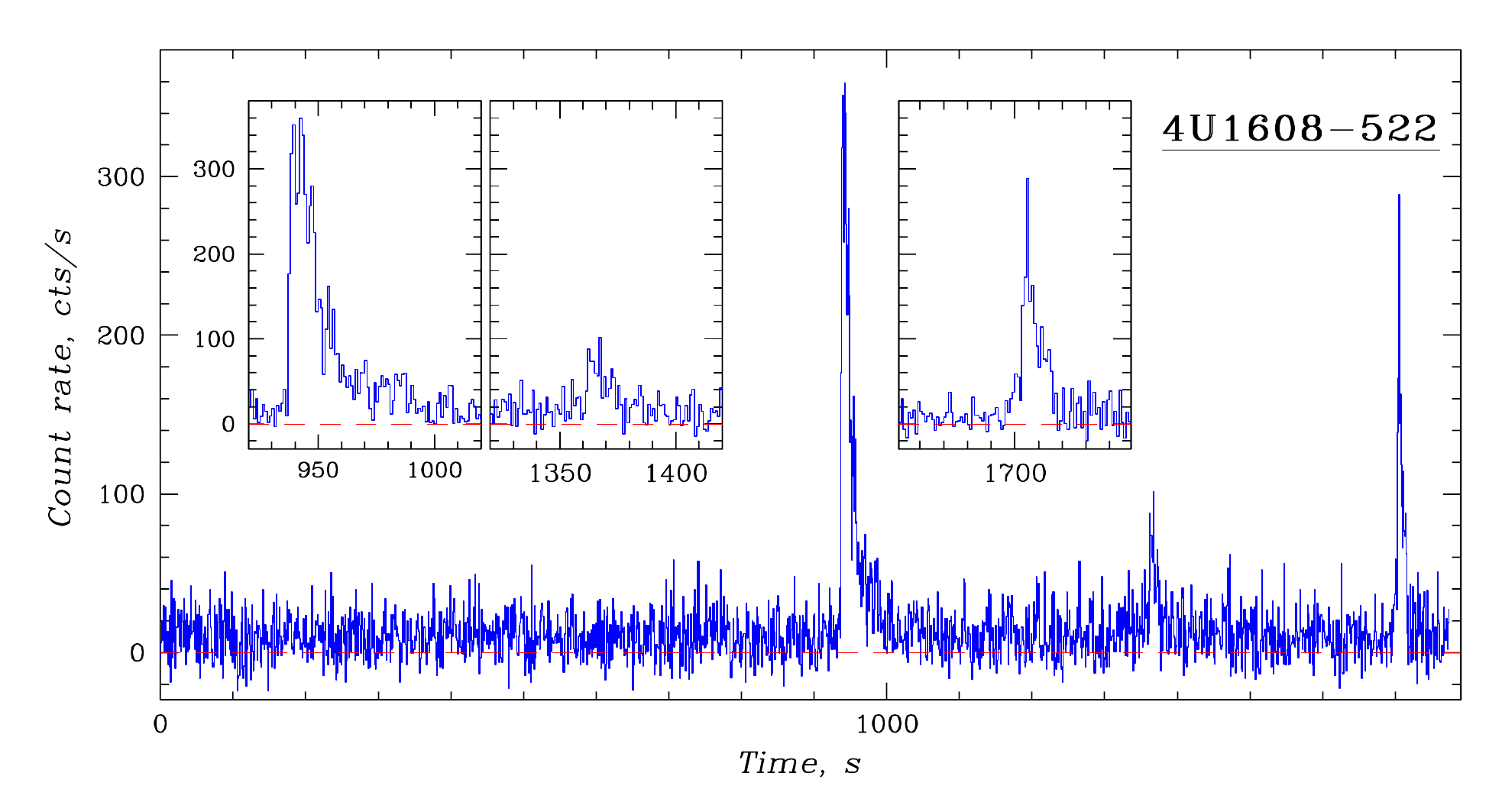}
\caption{JEM-X 3--20\,keV photon count rates obtained on March 24, 2004 (top), and March 28, 2009 (bottom), during which triple thermonuclear X-ray bursts were detected from the bursters Aql~X-1 and 4U~1608--522. (From \citealt{2017AstL...43..583G})}
\label{fig:3bursts}
\end{figure}

Several multiple bursts (sets of 2 or 3 events with recurrence times too short to accumulate enough fuel for the next burst ignition) were discovered (see Fig.~\ref{fig:3bursts}) in preparing the catalogue of type I X-ray bursts detected by \INTEGRAL\ \citep{2017AstL...43..781C}. Analysis of these events allowed \cite{2017AstL...43..583G} to suggest a natural explanation of their origin within the model of a spreading layer of accreted matter on the NS surface \citep{1999AstL...25..269I} in the case of a sufficiently high ($\dot{M} \gtrsim 10^{-9}$\,$M_{\odot}$\,yr$^{-1}$) accretion rate.

According to the proposed scenario, the first burst of the series begins at high latitude on the surface of the NS where the bulk of accreted matter settles down and the ignition conditions are reached. When all the local fuel is burnt out, the flame propagates with a deflagration wave speed $v \simeq 0.01$\,km\,s$^{-1}$ over the stellar surface towards the equator and then towards the opposite stellar pole and the second ring zone. On reaching it, the second burst of the series begins. The existence of triple bursts suggests that a central ring zone might also be important, although in the standard model of a spreading layer no accumulation of matter is believed to occur in this region.

\begin{figure}
\centering
\includegraphics[width=\columnwidth]{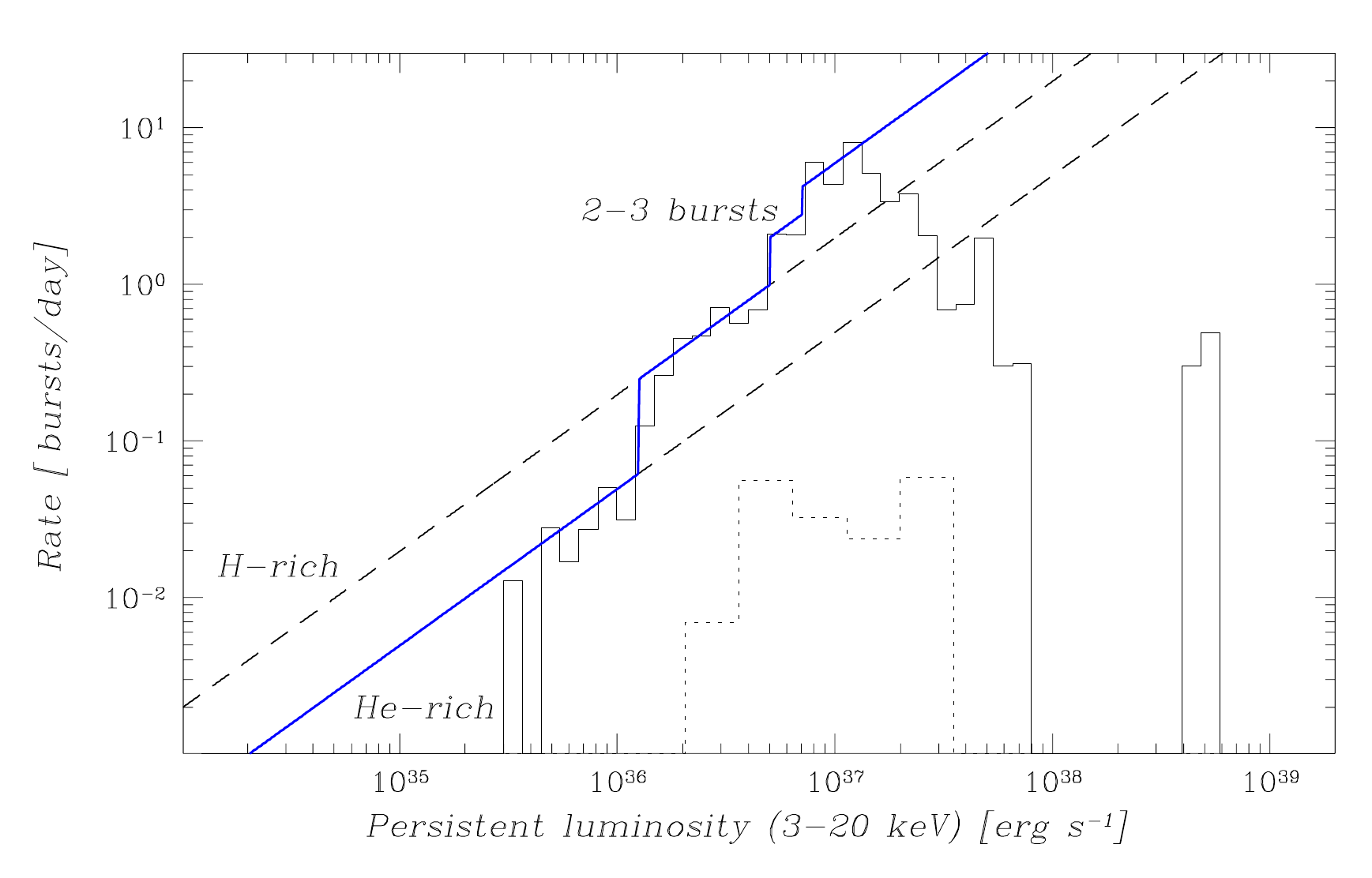}
\caption{Rate of X-ray bursts vs. the persistent luminosity of bursters. The dashed lines show the predictions of the standard model of complete burning in a burst of the matter fallen to the NS surface since the previous burst (with the upper and lower lines corresponding to different assumed helium abundances). The thick solid line shows the prediction taking into account the latitude structure of the  spreading layer. (From \citealt{2018AstL...44..777G})}
\label{fig:brate}
\end{figure}

\cite{2018AstL...44..777G} further suggested that the enhanced column density of the accreted matter in the high-latitude ring zones compared to the rest of the NS surface should facilitate the ignition of matter in these zones, which may explain the observed high rate of type I X-ray bursts in sources with a high persistent luminosity (see Fig.~\ref{fig:brate}). 

There is another possible explanation for short waiting time (SWT) bursts \citep{2010ApJ...718..292K}. \citealt{2017ApJ...842..113K} performed one-dimensional  simulations assuming that some amount of hydrogen fuel is left unburned at low column depth during the burst. As the opacity of the ash layer is anti-correlated with the temperature, it only takes a few minutes of cooling for the ashes to become opaque and to initiate convective mixing. This can bring the unburned hydrogen deep enough for a new burst to start.

%%%%%%%%%%%
\subsection{Dips and eclipses}

Several LMXBs exhibit periodic temporal decreases in the X-ray intensity (dips) as well as (partial or total) eclipses \citep[22 and 13, respectively, as of 2007,][]{2007A&A...469..807L}. It is commonly agreed that these phenomena are associated with high-inclination systems ($i>60^{\circ}$); indeed eclipses are believed to occur when the central source's x-ray emission is blocked by the companion star, while dips are thought to arise due to obscuration by a bulge in the accretion disc, where the matter from the companion impacts on the outer disc \citep{1985SSRv...40..167W,1987A&A...178..137F}. 

The spectral variations during the dipping activity cannot be described by a simple increase in photo-electric absorption  by cold absorbing material with normal abundances, and several scenarii have been originally proposed  \citep[mainly, \lq absorbed plus non absorbed continuum', and \lq progressive covering', respectively:][]{1986ApJ...308..199P, 1999A&A...349..495B}.
With the advent of new-generation X-ray telescopes such as \textit{XMM-Newton} and \textit{Chandra}, an important step forward was performed: systematic and detailed analyses of the complex changes in the 0.6--10\,keV continuum and absorption lines during dips  could be self-consistently explained 
via changes in highly ionized absorbers -- atmospheres or winds -- present above the accretion discs in LMXBs, with no need for unusual abundances or partial covering of extended emission regions \citep[][and references therein]{2005A&A...436..195B,2006A&A...445..179D}. 

Although \INTEGRAL\ is not suited for characterizing the soft X-ray emission where the dipping activity is more pronounced, a broad-band approach has been successfully obtained for the dipper XB~1254--90 \citep{2009A&A...493..145D}. A simultaneous \INTEGRAL/\emph{XMM-Newton} spectral fitting (disc-blackbody and thermal Comptonisation) showed that the continuum responsible for ionizing the warm absorber is the  high-energy (Comptonization) component, sampled by \INTEGRAL\ (JEM-X and IBIS/ISGRI). 

In the case of the transient dipping and eclipsing LMXB IGR~J17451--3022 \citep{2016A&A...589A..42B}, the source remained quite soft during most of the nine-month outburst and was not significantly detected by \INTEGRAL\ albeit a brief time interval (revolution 1458), when the source transited to a harder state with detection up to about 100\,keV (photon index $\Gamma\sim 2$). The \emph{XMM-Newton} coverage revealed the signature of the ionized absorber (multiple absorption features) in the observed dips and eclipses.

%\cite{2009AJ....138...50B} -- A Study of the Low-Mass X-Ray Binary Dip Sources XB 1916 - 053, XB 1323 - 619, X 1624 - 490 and 4U 1746 - 371 Observed with INTEGRAL.
% NOT INCLUDED IN TEXT

%\cite{2009A&A...493..145D} -- Variations in the dip properties of the low-mass X-ray binary XB 1254-690 observed with XMM-Newton and INTEGRAL.

%\cite{2016A&A...589A..42B} -- IGR J17451-3022: A dipping and eclipsing low mass X-ray binary.

%%%%%%%%
\section{X-ray pulsars}
\label{s:pulsars}
%%%%%%%%

%Background information: According to \cite{2007A&A...469..807L}, a dozen or so LMXB pulsars were known in 2007:
%IGR J00291+5934 -- transient, radio-loud, millisecond pulsar, Porb=2.46 h 
%XTE J0929-314 -- transient, radio-loud, millisecond pulsar, Porb=0.73 h (ultra-compact)
%4U 1626-67 -- pulsar, Porb=0.69 h (ultra-compact)
%2A 1655+353=Her X-1 -- intermediate-mass X-ray binary, dipping, eclipsing, Porb=40.8 h {\bf %Kretschmar et al. review?}
%3A 1728-247=GX 1+4 -- dipping, Porb=1160.8 h 
%GRO J1744-28 (bursting pulsar) -- transient, type-II burster, dipping, Porb=284.2 h
%XTE J1751-305 - transient, millisecond pulsar, Porb=0.71 h (ultra-compact)
%XTE J1807-294 -- transient, millisecond pulsar, Porb=0.668 h (ultra-compact) 
%SAX J1808.4-3658 -- transient, millisecond pulsar, burster, radio-loud, Porb=2.014167 
%XTE J1814-338 -- transient, millisecond pulsar, burster, Porb=4.274620 h
%2A 1822-371 -- eclipsing, Porb=5.570000
%HETE J1900.1-2455 -- transient, millisecond pulsar, burster, Porb=1.39 h (ultra-compact)
%This list might be incomplete. 

One of the main achievements of the \emph{Uhuru} observatory was the discovery of X-ray pulsars -- binary systems where accretion occurs onto a strongly magnetized, spinning NS. The first such objects to be discovered turned out to be a high- and an intermediate-mass X-ray binary: Cen X-3 and Her X-1, respectively, and it is currently known that the vast majority of X-ray pulsars are HMXBs. Interestingly though, the first X-ray pulsar was actually tentatively discovered before the launch of {\it Uhuru}, in balloon observations performed in October 1970, and it was GX 1+4 \citep{1971ApJ...169L..17L}, which is a LMXB, more specifically a symbiotic X-ray binary. 

In this section, we discuss the \INTEGRAL\ results on (i) \lq normal' LMXB and IMXB pulsars -- objects where accretion onto the NS proceeds via Roche lobe overflow of the companion, and (ii) symbiotic X-ray binaries, where accretion occurs from the stellar wind of a giant companion. As noted before, accretion-powered millisecond pulsars are reviewed separately by Papitto et al. in this volume. 

%\cite{2010A&A...515A..25D} -- The intriguing nature of the high-energy gamma ray source XSS J12270-4859. \cite{2013A&A...550A..89D} -- X-ray follow-ups of XSS J12270-4859: a low-mass X-ray binary with gamma-ray Fermi-LAT association. \cite{2014MNRAS.438.2105P} -- A propeller scenario for the gamma-ray emission of low-mass X-ray binaries: the case of XSS J12270-4859.

%\cite{2016ApJ...831...89S} -- A New gamma-Ray Loud, Eclipsing Low-mass X-Ray Binary. 3FGL J0427.9–6704.

%\cite{2018A&A...610L...2S} -- Discovery of 105 Hz coherent pulsations in the ultra-compact binary IGR J16597-3704. {\bf Already discussed in the ultracompact binaries subsection}

%\cite{2012MNRAS.423.1178P} -- Spin period evolution and other properties of IGR J17480-2446 in Terzan 5. Possible relation to millisecond pulsars. There is more literature on it. 

%%%%%%%%%%%
\subsection{\lq Normal' X-ray pulsars}

\subsubsection{Her X-1}

Since its discovery by \textit{Uhuru} \citep{1972ApJ...174L.143T}, Her X-1 has been studied and monitored by almost all X-ray missions and therefore deserves special attention. It was the first X-ray pulsar in which the NS magnetic field was estimated directly from measurements of the cyclotron resonance scattering feature (CRSF) in the X-ray spectrum \citep{1978ApJ...219L.105T}. A review of the cyclotron line measurements in magnetized NSs can be found in \citet{2019A&A...622A..61S}. 

Her~X-1 is an IMXB consisting of a 1.8--2.0\,$M_\odot$ evolved sub-giant star and a 1.0--1.5\,$M_\odot$ NS \citep{1972ApJ...174L.143T}. The binary orbital period is 1.7 days, and the NS spin period is 1.24\,s. The optical star HZ Her \citep{1974ApJ...187..345C} fills its Roche lobe and an accretion disc is formed around the NS. Due to the X-ray illumination, the optical flux from HZ Her is strongly modulated with the orbital period, as was first found by the inspection of archival photo-plates \citep{1972IBVS..720....1C}. 

The X-ray light curve of Her X-1 is additionally modulated with an approximately 35 day period \citep{1973ApJ...184..227G}. Most of the 35-d cycles last 20.0, 20.5 or 21.0 orbital periods \citep{1983A&A...117..215S,1998MNRAS.300..992S,Klochkov2006}. The cycle consists of a 7-orbits \lq main-on' state and a 5-orbits \lq short-on' state of lower intensity, separated by 4-orbits intervals during which the X-ray flux vanishes completely. 

\begin{figure}
\includegraphics[width=0.49\textwidth]{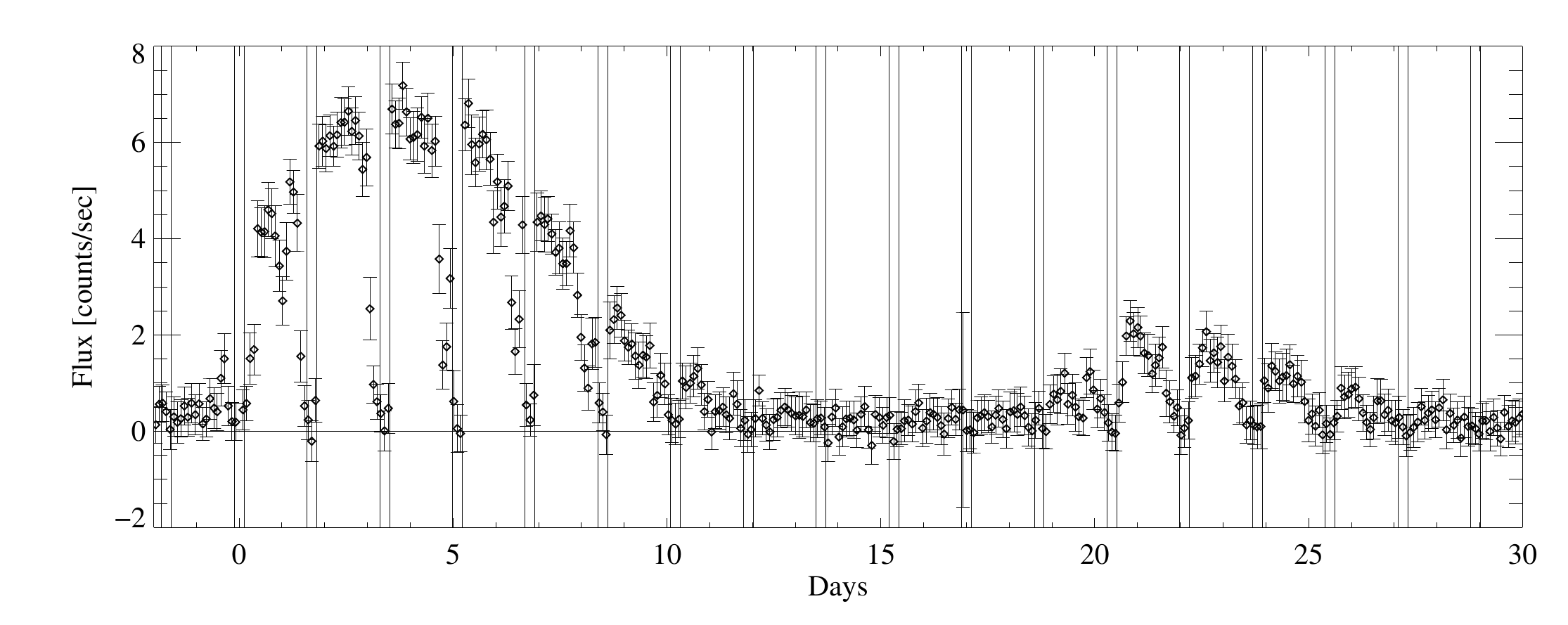}
\includegraphics[width=0.49\textwidth]{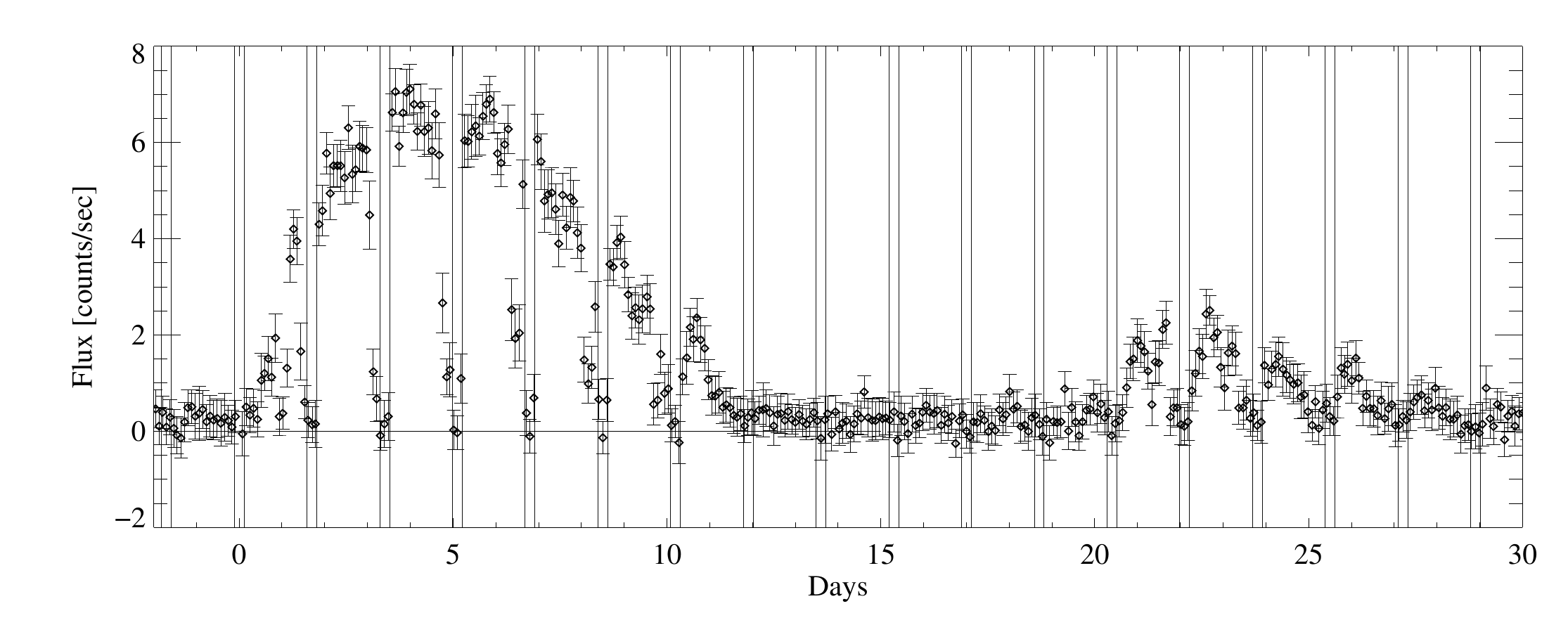}
\caption{35-day \textit{RXTE}/ASM 2--12\,keV mean light curve of Her X-1 constructed for cycles started at the orbital phase $\sim$0.2 and $\sim$0.7 (the upper and lower panel, respectively). Orbital eclipses are shown by vertical lines. (From \citealt{Klochkov2006})}
 \label{fig:HerX1ASM}
\end{figure}

It is now widely recognized that the 35-d cycle of Her X-1 can be explained by the retrograde orbital precession of the accretion disc \citep{1976ApJ...209..562G,1999ApL&C..38..165S}. The 35-d cycle turn-ons most frequently occur at the orbital phases $\sim$0.2 or $\sim$0.7, owing to the tidal nutation of the outer parts of the disc with the double orbital frequency when the viewing angle of the outer parts of the disc changes most rapidly \citep{1973NPhS..246...87K,1982ApJ...262..294L,1978pans.proc.....G}. The orbital and 35-d modulation of the X-ray light curve of Her X-1 as constructed from the \textit{RXTE}/ASM observations is shown in Fig. \ref{fig:HerX1ASM} for the 35-d cycles starting at the orbital phases close to 0.2 and 0.7.

Since the first \textit{Uhuru} observations \citep{1973ApJ...184..227G}, the nearly regular 35-d X-ray light curve behaviour of Her X-1 attracted much attention. It is generally accepted that the shape of the light curve is nicely explained by the precessional motion of the accretion disc. It was then noticed that the shape of the 1.24-s pulse profiles showed variations with the phase of the 35-d modulation, which was tentatively attributed to a possible free precession of the NS \citep{1986ApJ...300L..63T}. This was further investigated by \cite{2009A&A...494.1025S} and \cite{2013MNRAS.435.1147P} using an extensive set of \textit{RXTE} data. \cite{2013A&A...550A.110S}, however, showed that the possibly existing two \lq 35-d clocks', i.e. the precession of the accretion disc and the precession of the NS, are extremely well synchronized -- they show exactly the same irregularity. This requires a strong physical coupling mechanism which could, for example, be provided by the gas-dynamical coupling between the variable X-ray illuminated atmosphere of HZ Her and gas streams forming the outer part of the accretion disc \citep{1999A&A...348..917S,2013A&A...550A.110S}.

\begin{figure}
\includegraphics[width=0.49\textwidth]{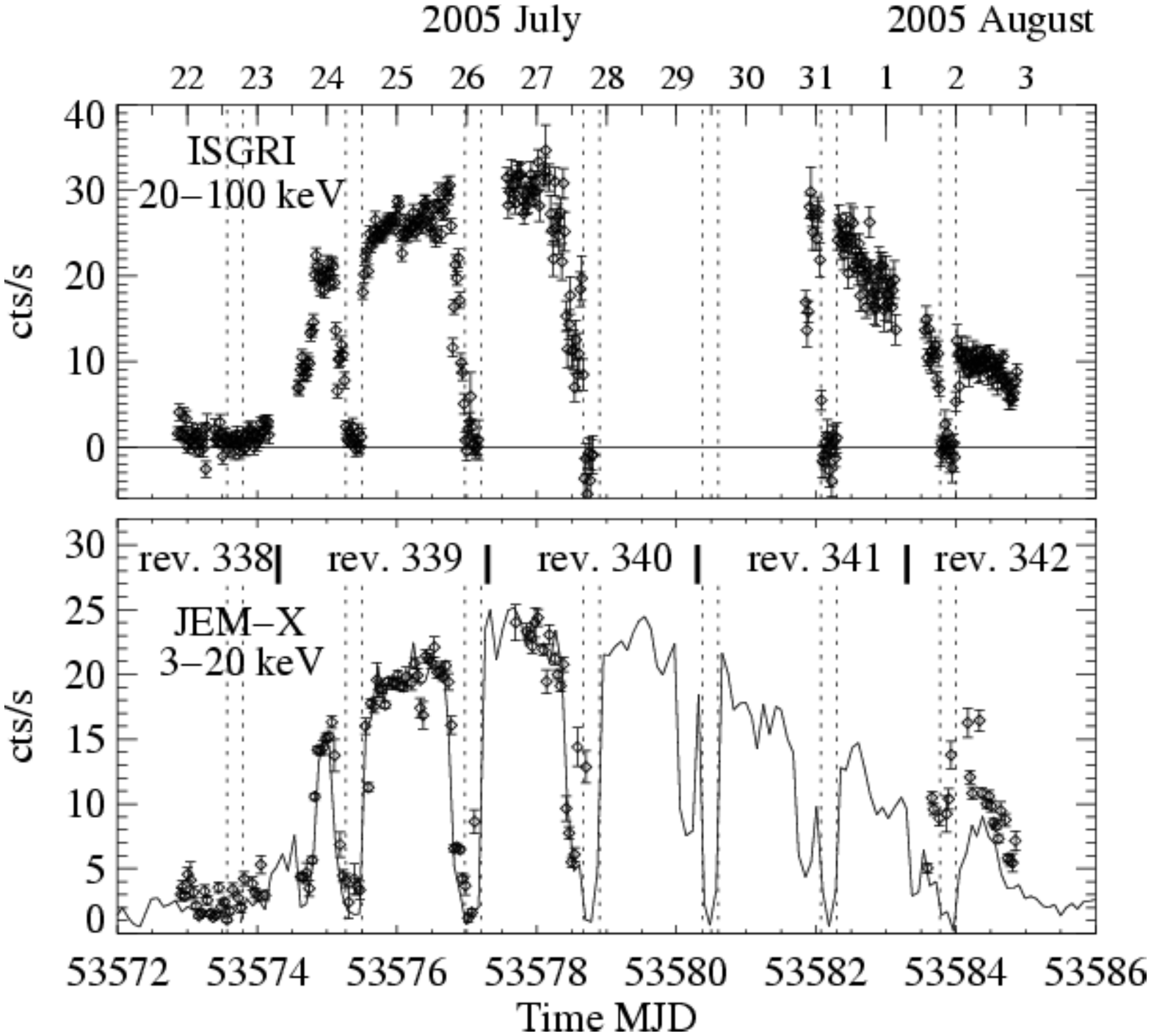}
\caption{The IBIS/ISGRI 20--100\,keV and JEM-X 3--20\,keV light curve of Her X-1 obtained in the first \INTEGRAL\ observations in 2005. The vertical dotted lines show orbital eclipses. The thin solid line in the bottom panel shows the mean \textit{RXTE}/ASM light curve. (From \citealt{2008A&A...482..907K})}
 \label{fig:HerX1INTEGRAL}
\end{figure}

Dedicated \INTEGRAL\, and \textit{RXTE} observations of Her X-1 were carried out covering an almost entire main-on stage \citep{2008A&A...482..907K,2013A&A...550A.110S} (see Fig. \ref{fig:HerX1INTEGRAL}). These observations confirmed a strong pulse profile variation with the 35-d phase and enabled pulse phase resolved spectral measurements. They also revealed a change in the cyclotron line energy with the pulse phase. But most importantly, they triggered a suite of further X-ray observations to monitor the long-term behaviour of the cyclotron line energy (see below).

The timing analysis of  the \textit{RXTE} and \INTEGRAL\, observations using the phase connection technique enabled the orbital ephemeris of Her X-1 to be updated, the orbital period secular decay to be improved and the orbital eccentricity to be measured for the first time \citep{2009A&A...500..883S}. 

%\subsubsection{The neutron star free precession in Her X-1}

\subsubsection{4U 1626--67}

%\textbf{To be written}
This is a very interesting X-ray pulsar with a pulse period of 7.7\,s in an LMXB with an orbital period of 42\,min, discovered from optical observations \citep{1981ApJ...244.1001M,1998ApJ...492..342C}. In this ultracompact X-ray binary (see Section~\ref{s:ucxb} below), the NS acccrets from a very low-mass companion ($\sim$0.04\,$M_\odot$ for the binary inclination angle $i=18^\circ$, \citealt{1988ApJ...327..732L}). In the X-ray spectrum of 4U 1626--67, a CRSF with an energy of $E_{\rm cycl}\sim 37$\,keV is observed (e.g. \citealt{1998ApJ...500L.163O,2002ApJ...580..394C,2012A&A...546A..40C,2019ApJ...878..121I}). 

A prominent feature of this pulsar is abrupt torque reversals observed in 1990 \citep{1997ApJ...481L.101C} and in 2008 \citep{2010ApJ...708.1500C}. Interestingly, a possible CRSF in emission was tentatively detected during a spin-down episode by \textit{Suzaku} \citep{2012ApJ...751...35I}.
In recent \textit{Suzaku} and \textit{NuSTAR} observations carried out during the ongoing spin-up episode of the source started after the torque-reversal in 2008, the pulse period change was found to be $\dot P\approx -3\times 10^{-11}$\,s/s, in apparent agreement with the popular Ghosh and Lamb accretion model \citep{1979ApJ...234..296G} for accreting X-ray pulsars \citep{2016PASJ...68S..13T}.

The source was observed by \INTEGRAL\ in March--October 20013 during the Galactic plane scanning and the deep Galactic center survey \citep{2005AstL...31..729F}. The parameters of the hard X-ray spectrum from the IBIS data were found to be similar to those originally determined by \cite{1998ApJ...500L.163O} from \textit{BeppoSAX} observations. The spectral and timing variability of 4U~1626--67 during the ongoing spin-up episode based on the \textit{Suzaku} and \textit{NuSTAR} observations is discussed in detail by \cite{2019ApJ...878..121I} who also used an original relativistic ray-tracing code to model the phase-resolved pulse profiles in this source. 

\subsubsection{Positive CRSF energy -- luminosity correlation and long-term CRSF energy evolution}

\begin{figure}
\includegraphics[width=0.5\textwidth]{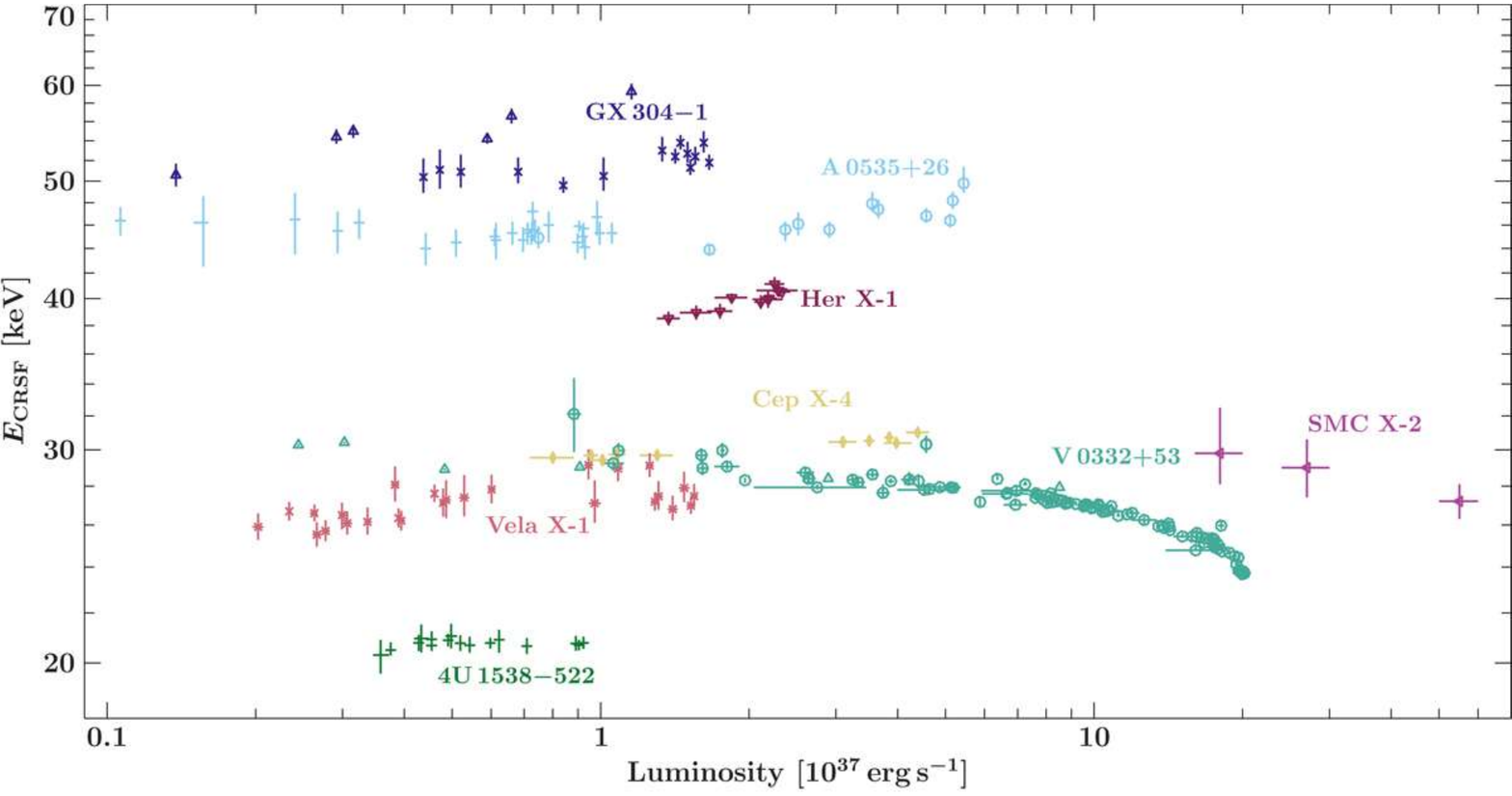}
\caption{CRSF energy -- X-ray luminosity correlation in X-ray pulsars. (From \citealt{2019A&A...622A..61S})}
 \label{fig:EcycLxall}
\end{figure}

Using \textit{RXTE} data, \cite{2007A&A...465L..25S} found for the first time a positive correlation between the CRSF centroid energy $E_{\rm cycl}$ and X-ray flux in Her X-1, with an increase of $\sim$5\% in energy for a factor of two increase in luminosity $\Lx$. This correlation was later confirmed by a thorough analysis of \INTEGRAL, \textit{Swift}/BAT and \textit{NuSTAR} data \citep{2014A&A...572A.119S}. In bright transient X-ray pulsars, a negative $E_{\rm cycl}$--$\Lx$ correlation was known before \citep{1995PhDT.......215M,2006MNRAS.371...19T}. It is generally thought that the negative $E_{\rm cycl}$--$\Lx$ correlation is due to the changing height of a radiative shock that can be present in high-luminosity X-ray pulsars above some critical luminosity  \citep{1976MNRAS.175..395B}. In low-luminosity pulsars, the radiation-supported accretion column cannot be formed, and the accreting matter is halted close to the NS surface \citep{2007A&A...465L..25S}, with the possible formation of a collisionless shock. This model suggested a possible explanation to a number of spectral correlations observed in other low-luminosity X-ray pulsars (HMXBs), including GX 304--1 \citep{2017MNRAS.466.2752R} and Cep X-4 \citep{2017A&A...601A.126V}, and triggered a lot of further studies. Figure~\ref{fig:EcycLxall} shows the measured $E_{\rm cycl}$--$\Lx$ correlations for a number of pulsars \citep{2019A&A...622A..61S}.

\begin{figure}
\includegraphics[width=0.5\textwidth]{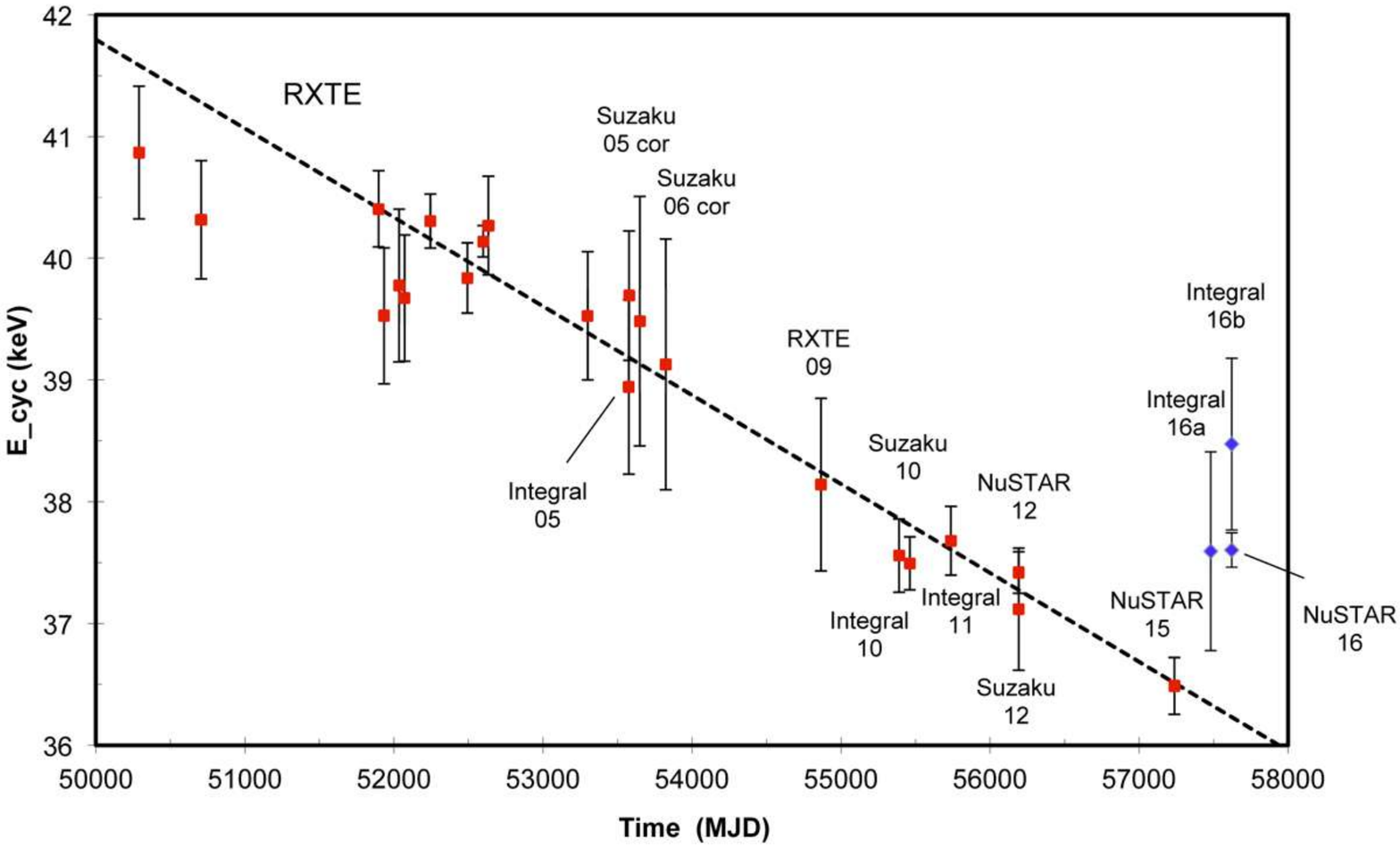}
\caption{Secular decrease of the CRSF energy in Her X-1 observed in 1996--2015 and reversed afterwards. Measurements from different X-ray satellites are shown. (From \citealt{2017A&A...606L..13S})}
 \label{fig:HerX1secular}
\end{figure}

Dedicated measurements of the cyclotron line energy in the X-ray spectrum of Her X-1 with \INTEGRAL\, and other satellites resulted in the discovery of a secular decrease of $E_{\rm cycl}$ on a time-scale of a few tens of years \citep{2014A&A...572A.119S}, which apparently stopped or reversed around 2015 (\citealt{2017A&A...606L..13S} and Staubert et al. 2019, in preparation) (see Fig. \ref{fig:HerX1secular}). The physics of the cyclotron line energy decrease can be related to the spread of accreted matter over the NS surface. This interesting issue definitely deserves further observational and theoretical studies.

The CRSF energy evolution was also searched for in other bright X-ray pulsars (HMXBs) Vela X-1 and Cen X-3 using \textit{Swift}/BAT long-term survey  observations \citep{2019MNRAS.484.3797J}. A secular decrease in the cyclotron line energy on a time-scale similar to that of Her X-1 was detected in Vela X-1 but was not found in Cen X-3. 

%%%%%%%%%%%
\subsection{Symbiotic X-ray binaries}
\label{s:symbiotic}

%Usually have long orbital periods, long spin periods and strong magnetic fields (X-ray pulsars). See Postnov et al. papers %on their formation scenarios.
%
%\paragraph{GX 1+4}
%\cite{2007A&A...462..995F} -- INTEGRAL observation of the accreting pulsar GX 1+4. \cite{2009AstL...35..433L} -- Timing %characteristic of 10 X-ray pulsars, including 9 HMXBs and GX 1+4. \cite{2012A&A...537A..66G} -- Spin period evolution of GX %1+4. \cite{2017A&A...601A.105I} -- Variability of GX 1+4. Enhanced activity near periastron passage.
%
%\paragraph{A newly classified symbiotic X-ray binary 4U 1954+319}
%\cite{2006A&A...460L...1M} -- Source discovered long ago. INTEGRAL helped to classify it. \cite{2011ApJ...742L..11M} -- 
%The 5 hr Pulse Period and Broadband Spectrum of 3A 1954+319. \cite{2014ApJ...786..127E} -- Spectral and Timing Nature of 4U %1954+319: The Slowest Rotating Neutron Star in an X-Ray Binary System. 
%
%\cite{2007A&A...470..331M} -- IGR J16194-2810: a new symbiotic X-ray binary. Multi-wavelength study. No X-ray pulsations so %far? {\bf Overlap with the Identification section.}
%
%\cite{2015AstL...41..394K} -- IGR J17463-2854, a possible symbiotic binary system in the galactic center region. The %compact object is likely a white dwarf, so probably not a LMXB.
%
%\cite{2018A&A...613A..22B} -- IGR J17329-2731: The birth of a symbiotic X-ray binary. X-ray pulsar.

Symbiotic stars demonstrate peculiar optical spectra implying the presence of a red giant and a companion hot enough to sustain HeII (or higher ionization) emission lines \citep{1984PASAu...5..369A}. These are
binary systems at an advanced evolutionary stage in which accretion onto a compact object (a white dwarf or a neutron star) occurs from a low-mass red giant companion (see, e.g., \citealt{2019arXiv190901389M} for a recent review).  

Symbiotic X-ray binaries (SyXBs) represent a relatively small subclass of long-period Galactic LMXBs consisting of a late-type giant (K1-M8) and an accreting magnetized NS. Candidate systems CGCS~5926 \citep{2011A&A...534A..89M} and CXOGBS~J173620.2--29333 \citep{2014ApJ...780...11H} possibly host carbon stars. The first SyXB was identified more than 40 years ago \citep{1977ApJ...211..866D}, and since then only a dozen of SyXBs have been discovered by various X-ray missions, including \INTEGRAL. Table~\ref{t:syxbs}, taken from \cite{2019MNRAS.485..851Y}, lists their main observed properties: the NS spin period $P^*$, binary orbital period $P_\mathrm{orb}$, X-ray luminosity $\Lx$ and the distance to the source $d$. 

\begin{table*}
  \caption{Parameters of observed and suspected SyXBs \citep{2019MNRAS.485..851Y}.
}
\label{t:syxbs}
\begin{tabular}{lcccccc}
\hline
\hline
 SyXB&$P^*$  &$P_{\rm orb}$ &$\Lx$ & d %& Sp. type
\\
            & (s)  & (day)       & (erg\,s$^{-1}$) & (kpc) %&
\\
\hline
 GX 1+4 &$\simeq140$$^{(1)}$ &1161$^{(2, 3)}$&$10^{35}$--$10^{36}$$^{(4)}$&4.3$^{(2)}$%&M5 III$^{(1)}$
\\
                     & & $295\pm70^{(5)}$ & & &\\
                     & & 304$^{(6, 7)}$ & & &\\
4U 1954+319 & $\sim 18300$$^{(8)}$ & $\gtrsim400$?$^{(9)}$ & $4\times10^{32}$$^{(9)} $ &1.7$^{(9)}$%&M4-5~III$^{(5)}$
\\
4U~1700+24 &? & $404\pm20^{(6)}$ & $2\times10^{32}$--$10^{34}$$^{\rm(10)}$ & $0.42\pm0.4$$^{(10)}$
%&M2 III$^{(6)}$
\\
                     & & 4391$^{(23)}$ & & &\\
Sct X-1&113$^{(11)}$  & ? & $2\times10^{34}$$^{(11)}$ & $\geq4^{(11)}$\\%&Late K/\\
%& & & & & Early M I-III$^{(7)}$
IGR J16194--2810 &?&?&$\leq 7\times10^{34}$$^{\rm
(12)}$&$\leq3.7^{(12)}$\\
IGR J16358--4726 & 5850$^{(13)}$ & ? & $3\times10^{32}$--$2\times10^{36}$$^{(14)}$ & 5--6; 12--13$^{(15)}$ \\
CGCS 5926 & ? & $\sim151^{(16)}$ & $\leq3\times10^{32(16)}$ & $5.2^{(16)}$\\ 
CXOGBS J173620.2--293338 & ? &  ?& $\sim9\times10^{32}$$^{(17)}$ &?
 \\
XTE J1743--363 &? &? &? &$\sim5^{(18)}$  
\\
%%%%%%%%% new systems
XMMU~J174445.5--295044 & ? & ? & $\gtrsim 4\times10^{34}$$^{(19)}$& $3.1^{+1.8}_{-1.1}$$^{(19)}$ \\
3XMM~J181923.7--170616 & $407.9$$^{(20)}$ & ? &  $2.78\times10^{34}d_{10}^2$$^{(20)}$ & ?\\
IGR~J17329--2731 & $6680\pm3$$^{(21)}$& ? & ?& $2.7^{+3.4}_{-1.2}$$^{(21)}$ \\
IGR~J17197--3010 &                      &   & $\lesssim1.6\times10^{35}$$^{(22)}$ & 6.3--16.6$^{(22)}$\\
\hline
\end{tabular}
\vskip 0.05cm
%REFS:
\begin{flushleft}
1 --  \citet{2007A&A...462..995F},
2 -- \citet{2006ApJ...641..479H},
3 -- \citet{2017A&A...601A.105I},
4 -- \citet{GonzalezGalan2011}, %gonzal
5 -- \citet{2016pas..conf..133M},
6 -- \citet{2002ApJ...580.1065G}, %gal2002
7 -- \citet{1999ApJ...526L.105P},
8 -- \citet{2008ApJ...675.1424C},
9 -- \citet{2006A&A...453..295M}, %mas2006
10 -- \citet{2002A&A...382..104M},  %mas2002
11 -- \citet{2007ApJ...661..437K}, %% kaplan
12 -- \citet{2007A&A...470..331M}, % mas 2007
13 -- \citet{2004ApJ...602L..45P}, %patel 2004
14 -- \citet{2007ApJ...657..994P}, %patel 2007
15 -- \citet{2005A&A...444..821L}, \\%lut
16 -- \citet{2011A&A...534A..89M}, %mas. 2011
17 -- \citet{2017AAS...23031704H}, %hynes
18 -- \citet{2012MNRAS.422.2661S}, %smith
19 -- \citet{2014MNRAS.441..640B}, %bahram
20 -- \citet{2017ApJ...847...44Q},  %qui
21 -- \citet{2018A&A...613A..22B}, %bozzo 
22 -- \citet{2012A&A...538A.123M}, % mas
23 -- \citet{2018arXiv181208811H}. % hinkle 4U
\end{flushleft}
\end{table*}

The observed properties of SyXBs can be explained by quasi-spherical accretion from slow stellar winds of the red giant companions. The \INTEGRAL\ monitoring of the steady spin-down in GX 1+4 \citep{2012A&A...537A..66G} triggered a deep study of wind accretion onto a slowly rotating magnetized NS, which resulted in the construction of a new theory of quasi-spherical settling accretion \citep{2012MNRAS.420..216S,2018ASSL..454..331S}. 

For the Bondi--Hoyle--Lyttleton stellar-wind accretion in a circular binary hosting a NS with mass $M_\mathrm{x}$, the captured mass rate from the stellar wind from a donor star with mass $M_\mathrm{o}$ and the stellar wind mass-loss rate $\dot M_\mathrm{o}$ with velocity $v_\mathrm{w}$ is roughly $\dot M_\mathrm{B}\sim (1/4) \dot M_\mathrm{o}(R_B/a)^2$, where $R_\mathrm{B}=2GM_\mathrm{x}/(v_\mathrm{w}^2+v_\mathrm{orb}^2)$ is the Bondi radius ($v_\mathrm{orb}$ is the orbital velocity of the NS). The fate of the captured matter depends on the ability of the NS magnetosphere to absorb the accreting plasma, which is determined by the plasma cooling mechanism. If the plasma cools rapidly, $t_\mathrm{cool}<t_\mathrm{ff}$ (here $t_\mathrm{ff}$ is the free-fall time at the magnetospheric boundary and $t_\mathrm{cool}$ is the plasma cooling time), it freely enters the magnetosphere, and the classical Bondi--Hoyle supersonic accretion is realized with $\dot M_\mathrm{x}=\dot M_\mathrm{B}$. In the opposite case of slow plasma cooling, $t_\mathrm{cool}\gg t_\mathrm{ff}$, it can be shown \citep{2012MNRAS.420..216S} that the actual accretion rate onto the NS which determines the observed X-ray luminosity $\Lx$ is $\dot M_\mathrm{x}\approx \dot M_\mathrm{B}(t_\mathrm{ff}/t_\mathrm{cool})^{1/3}$ and can be much lower than the Bondi rate.

The settling accretion regime should take place if the X-ray luminosity of the source is below $\sim 4\times 10^{36}$\,erg\,s$^{-1}$. In this case, a quasi-spherical hot convective shell forms above the magnetosphere and accretion occurs subsonically. Turbulent stresses in the shell mediate the angular momentum transfer to/from the NS magnetosphere, which adequately explains the observed spin-up/spin-down properties of GX 1+4 \citep{2012A&A...537A..66G,2012MNRAS.420..216S} and other slowly rotating X-ray pulsars with moderate X-ray luminosity \citep{2011ApJ...742L..11M,2017MNRAS.469.3056S}.

Recently, the Galactic population of SyXBs was modeled by \cite{2019MNRAS.485..851Y}. The formation of a SyXB follows the classical evolutionary scenario of binary systems with NSs elaborated in the 1970s \citep{1972NPhS..239...67V,1973NInfo..27...70T}. To form a NS, the mass of the primary component of a binary system should exceed $\sim$8\,$M_\odot$. The primary evolves off the main sequence and mass transfer occurs onto the secondary component. For a large mass ratio of the primary (more massive) to the secondary (less massive) binary component, a common envelope (CE) is expected to form during the first mass-transfer episode. After the CE stage, the helium core of the primary evolves to a supernova that leaves behind a NS remnant. 

If the binary survives the supernova explosion, a young NS appears in a binary system accompanied by a low-mass 
main-sequence star. When the low-mass star evolves off the main sequence, accretion onto the NS from the stellar wind of the red giant companion can occur, leading to the formation of a SyXB. 

\begin{figure}
\includegraphics[width=0.49\textwidth]{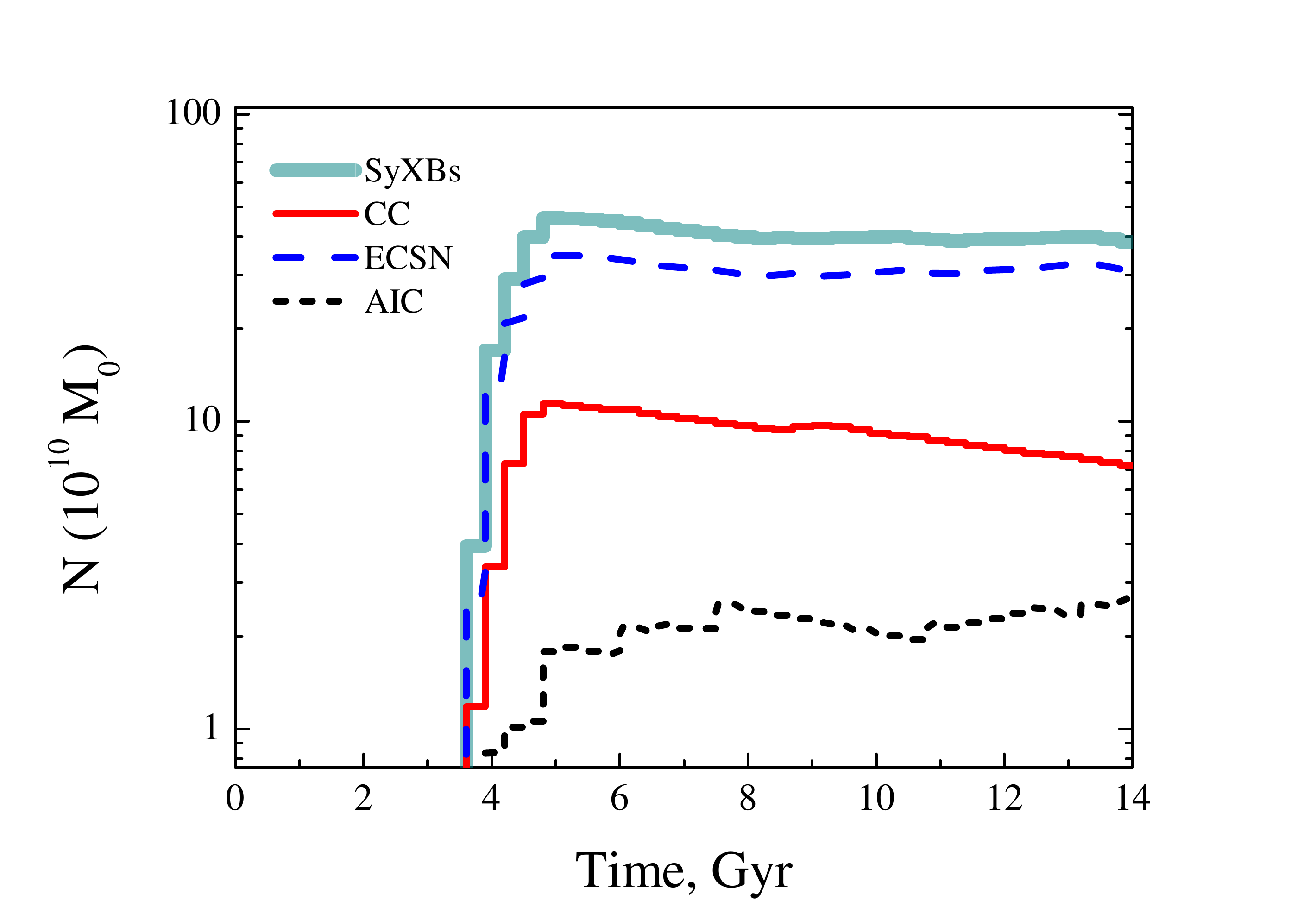}
\caption{Number of SyXBs in the Galaxy as a function of time for SyXBs with the NS-components produced 
via the iron CC, ECSN and AIC channels (see text). (From \citealt{2019MNRAS.485..851Y})
}
 \label{f:SyXBrate}
\end{figure}

During the supernova explosion from the iron core collapse (CC) of a massive star, a newborn NS can acquire high anisotropic (kick) velocity. However, in a narrow range of primary masses close to the lower mass limit for the NS formation, after the first mass exchange in the binary system the evolution of the helium core presumably ends up with the formation of a NS via an electron-capture supernova (ECSN) (\citealt{1980PASJ...32..303M}; see \citealt{2018A&A...614A..99S} for a recent study), which is not accompanied by a large NS kick. However, neither the precise mass range for ECSN nor the kick velocity amplitude have been found to significantly affect the results of the SyXB population synthesis \citep{2019MNRAS.485..851Y}. Another possible NS formation channel, via the accretion-induced collapse (AIC) of an oxygen-neon white dwarf which accumulated mass  close to the Chandrasekhar limit (1.4\,$M_\odot$) in a low-mass binary system \citep{1976A&A....46..229C}, has been shown to be strongly subdominant for the Galactic population of SyXBs  \citep{2019MNRAS.485..851Y}. In Fig.~\ref{f:SyXBrate}, taken from \cite{2019MNRAS.485..851Y}, the number of SyXBs in the Galaxy is shown as a function of time for different formation channels (iron CC, ECSN and AIC). For descendants of AICs, the upper estimate of SyXB  obtained by assuming 100\% efficient accretion is shown. In this calculations, a model of the Galactic star formation rate proposed by \cite{2010A&A...521A..85Y} was assumed. It is seen that only a few dozen SyXBs are expected to exist in the Galaxy, mostly produced via the ECSN channel, with their number remaining nearly constant in the last 10 billion years. The expected properties of the Galactic SyXB population (orbital periods, NS spin periods, X-ray luminosities) prove to be in agreement with observations. 
 
%%%%%%%%
\section{Ultracompact X-ray binaries}
\label{s:ucxb}
%%%%%%%%

Ultracompact X-ray binaries (UCXBs) are a distinct subclass of LMXBs characterized by orbital periods ($P_{\rm orb}$) shorter than $\sim$1\,hr. The short orbital period implies the mass of the donor to be a few hundredths of a solar mass and that it is hydrogen-poor \citep{1986ApJ...304..231N, 1986A&A...155...51S}. Specifically, the donor can be an O/Ne/Mg, C/O or He white dwarf, or a non-degenerate He star accreting onto a NS or a BH. The combination of compact objects, short period variability and peculiar chemical composition makes UCXBs a unique astrophysical laboratory for studying accretion onto NSs under hydrogen-poor conditions and for testing the binary evolution theory. The interest in compact binary evolution has greatly increased recently thanks to the detection of gravitational waves from coalescing binary BHs and NSs. Understanding the physical characteristics of UCXBs is particularly important in connection with ESA/NASA \emph{Laser Interferometer Space Antenna} (\emph{LISA}) future mission, as such objects are expected to be strong sources of gravitational waves within the frequency range ($10^{-4}$--1\,Hz) accessible to \emph{LISA}  \citep{2014LRR....17....3P,2018PhRvL.121m1105T,2018arXiv180701060N}. 

Direct identification of UCXBs is possible through measurement of $P_{\rm orb}$, which is difficult and while the predicted number of such systems in our Galaxy is as large as 10$^6$ \citep{2018arXiv180701060N}, only 38 confirmed (of which 23 have a firm $P_{\rm orb}$) or candidate (suggested by indirect methods) UCXBs are currently known \citep{2007A&A...469.1063I,2013ApJ...768..184H, 2017A&A...598A..34S, 2018ApJ...858L..13S, 2016A&A...587A.102B, 2017MNRAS.467.2199B, 2018A&A...610L...2S, 2019MNRAS.486.4149G}. Thanks to the availability of new data and our better understanding of the physical properties of accreting compact objects, the UCXB sample has been rapidly growing recently; note that only 12 such systems (with measured or suggested orbital periods) were known in 2005 \citep{2010NewAR..54...87N}. 

\begin{figure}
\centering
\includegraphics[angle=-90,width=\columnwidth]{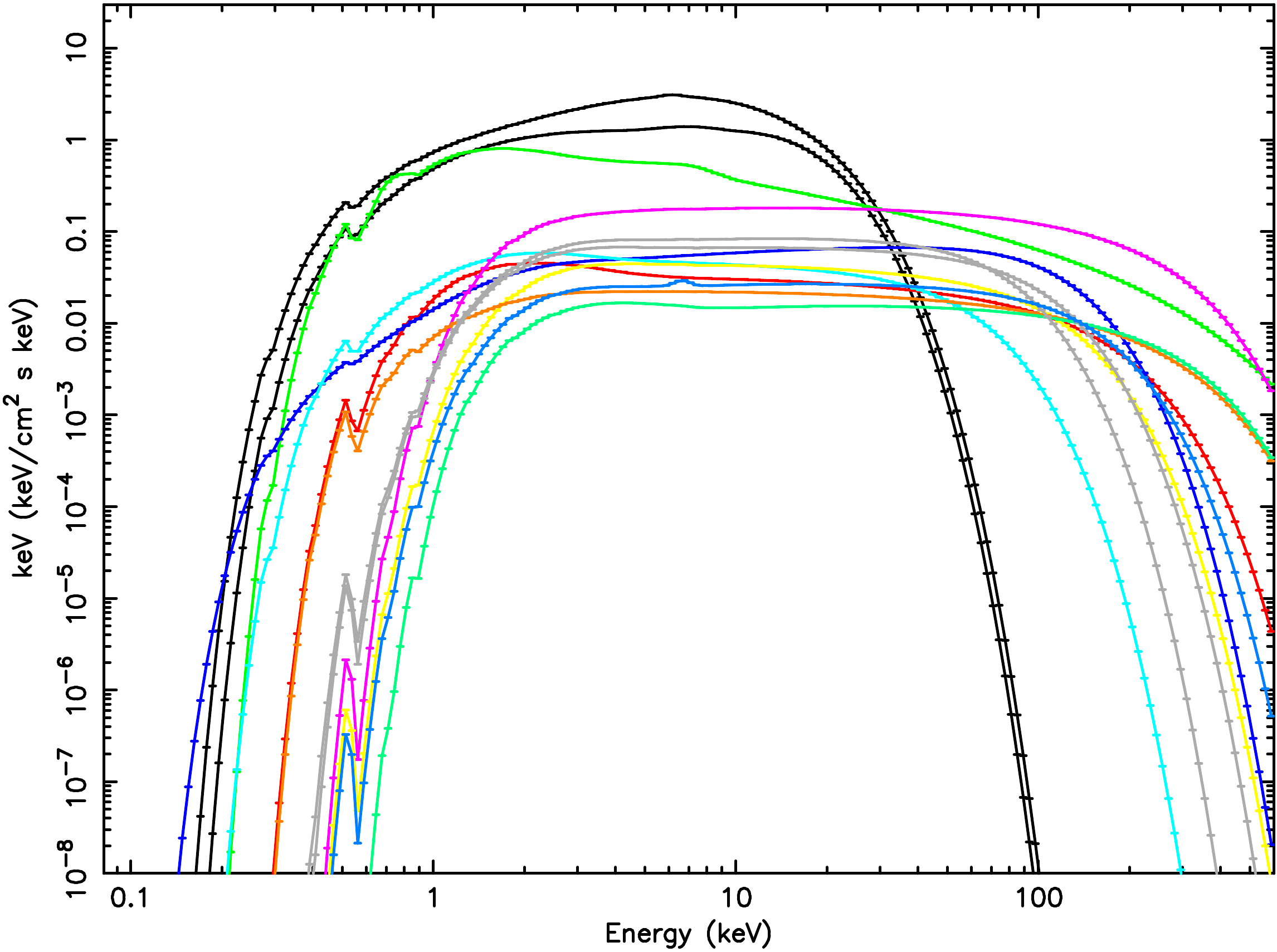}
\caption{Extrapolated models of the spectra of burster UCXBs measured with \INTEGRAL, \emph{BeppoSAX} and \emph{Swift}. (From \citealt{2008A&A...492..557F})}
\label{fig:fiocchi2007}
\end{figure}

Observations of UCXBs with \INTEGRAL\ have provided a wealth of new information on the accretion process, thermonuclear burning (see Section~\ref{s:bursts}) and orbital period variability in these systems. In particular, detailed studies of the broad-band X-ray spectra have been carried out for a number of systems. A systematic analysis of the \INTEGRAL, \emph{BeppoSAX} and \emph{SWIFT} data for a sample of UCXBs \citep{2008A&A...477..239F, 2008A&A...492..557F, 2011MNRAS.414L..41F} showed that these sources spend most of the time in the canonical low/hard state, with the hard Comptonization component dominating the emission (see Fig. \ref{fig:fiocchi2007}). The derived physical parameter distributions show that the luminosities are lower than $\sim7\times 10^{36}$\,erg\,s$^{-1}$, the temperature of the electrons in the corona ($kT_{\rm e}$) is typically greater than 20\,keV and the corona's optical depth $\tau \leq 5$. Thanks to their hard spectra, UCXBs often turn out to be bright at high energies and indeed 13 of them have been detected in the 100--150\,keV sky maps obtained with \INTEGRAL\ \citep{2006ApJ...649L...9B, 2015MNRAS.448.3766K}. 

The UCXB 4U 1850--087 was one of the first to be studied in detail using \INTEGRAL\ data \citep{2006A&A...460..229S}. X-ray emission was detected up to 100\,keV. A broad-band spectrum based on quasi-simultaneous \INTEGRAL\ and \emph{XMM-Newton} observations in September 2003 was modelled with a combination of a disc-blackbody component with an inner temperature of $kT_{\rm in}=0.8$\,keV and a power law with a photon index of 2.1. A similar behaviour has been observed for SLX 1737--282 \citep{2008A&A...484...43F}: the joined JEM-X/IBIS 3--100\,keV broad-band spectrum could be well fit by a power law with a photon index of 2.1. 

Very rarely UCXBs are observed in the classical high/soft state, with the exception of 4U 1820--30 which spends most of the time in this state. \cite{2012A&A...539A..32C} reported on an observation of this peculiar source in a high/soft state. The broad-band continuum was studied with \emph{XMM-Newton} and \INTEGRAL\ up to 40\,keV and could be described by a classical combination of blackbody ($kT_{\rm bb}\sim0.3$\,keV) and Comptonized emission ($kT_{\rm e}\sim 4$\,keV, $\tau\sim 5$), while the absorption features produced by cold matter along the line of sight were studied using \emph{XMM-Newton} and \emph{Chandra}. No evidence of iron emission was detected. The weakness or non-detection of this feature is a common property of UCXBs \citep{2007A&A...469.1063I}. Very similar spectral parameters have been obtained for another UCXB observed (by various X-ray telescopes including \INTEGRAL) in a high/soft state, 4U 0513--40: $kT_{\rm bb}\sim0.3$\,keV, $kT_{\rm e}\sim2$--5\,keV and $\tau\sim 6$ \citep{2011MNRAS.414L..41F}. For comparison, during the typical low/hard state of the source, the Comptonizing plasma is much hotter: $kT_{\rm e}\sim15$--21\,keV, with $\tau\sim2$. In soft X-rays, a 17\,min orbital period was detected whenever the source was in a high/soft state, regardless of its luminosity (which varied between 0.7 to $5.2\times 10^{36}$\,erg\,s$^{-1}$ in the 0.5--50\,keV energy band). This periodicity disappeared when the source was in a low/hard state.  These facts are consistent with the orbital motion modulating the soft thermal emission coming from a small region around the NS but not the Comptonization component generated in a more extended corona surrounding the NS, and further imply a high inclination angle of the system ($>80^\circ$).

An interesting result has been obtained by \cite{2015A&A...582A..32I} for the UCXB XB 1916--053. Based on its light curves in soft and hard X-rays constructed with data from a large set of instruments including \INTEGRAL/JEM-X and spanning 37 years, they found a long-term sinusoidal modulation (in addition to the linear trend) of the orbital period suggesting the presence of a third body with a mass $\sim$0.1\,$M_\odot$ orbiting the X-ray binary with a period $\sim$50 years (assuming conservative mass transfer). 

Most of the known UCXBs show thermonuclear X-ray bursts, implying accretion on to a NS. A number of UCXBs are known to be accreting millisecond X-ray pulsars. Recently, a new member of this class has been discovered thanks to \INTEGRAL: X-ray pulsations with a period of 9.5\,ms, modulated by an orbital period of 46\,min, were detected with \emph{NuSTAR} and \emph{Swift} \citep{2018A&A...610L...2S} from the transient IGR J16597--3704 discovered by \INTEGRAL\ \citep{2017ATel10880....1B} in the globular cluster NGC 6256. The broad-band spectrum based on data collected by \INTEGRAL\ and other observatories can be described by a sum of emission from an accretion disc and a Comptonizing corona with $kT_{\rm in}\sim 1.4$\,keV and $kT_{\rm e}\sim 30$\,keV, respectively, with no evidence for spectral iron lines or reflection humps similarly to other such systems. 

\begin{figure}
\centering
\includegraphics[angle=90,width=\columnwidth]{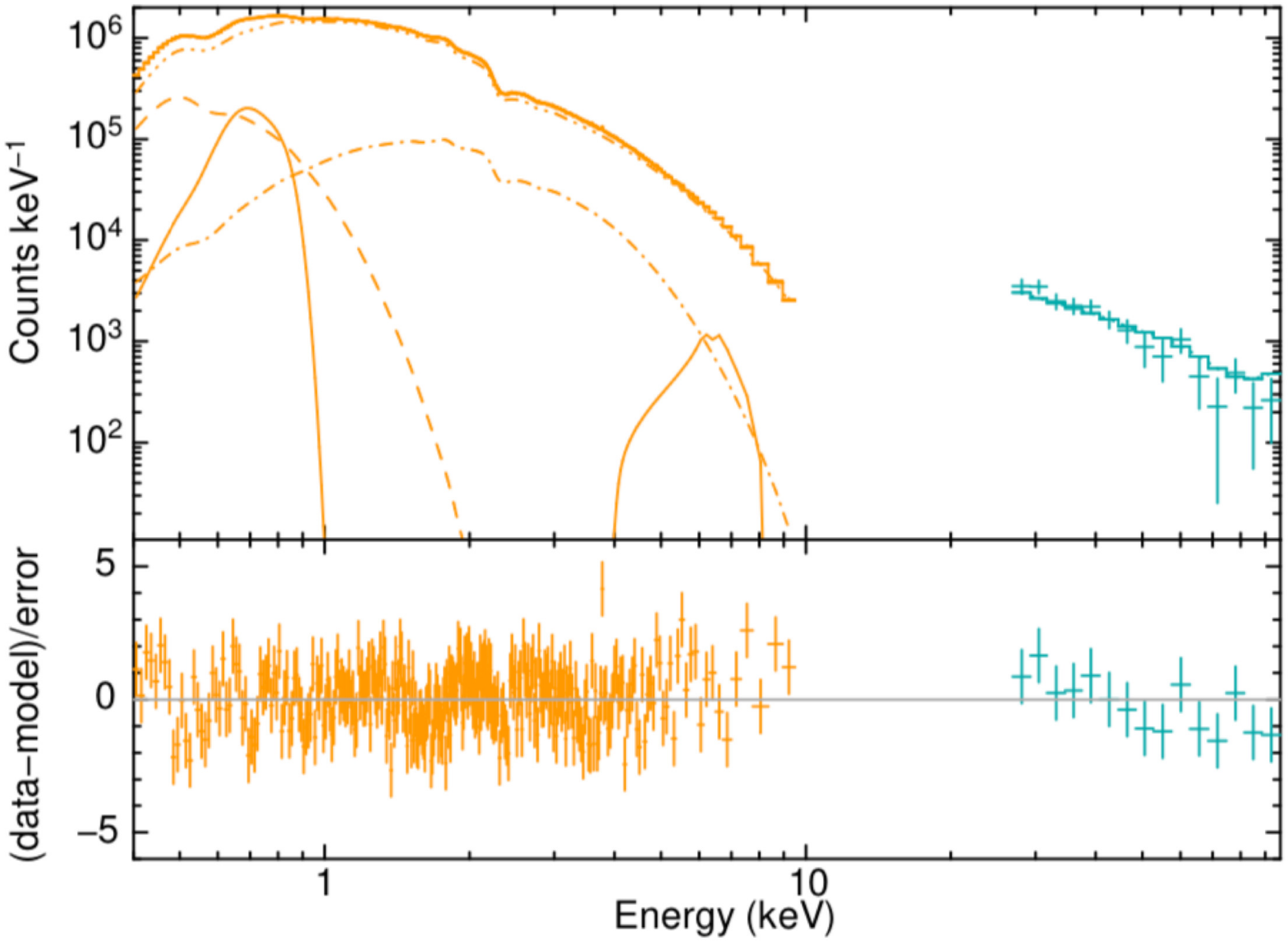}
\caption{Spectral modelling of \emph{NICER} (orange) and \INTEGRAL\ (teal) data for the UCXB 4U~1543--624. (From \citealt{2019arXiv190800539L})}
\label{fig:ludlam2019}
\end{figure}

In a number of UCXBs it remains unclear whether the primary component is a NS or a BH, as e.g. in 4U~1543$-$624. This source is normally stable and exhibits no timing signatures such as millisecond variability or type-I X-ray bursts. However, recently 4U~1543--624 showed an episode of enhanced accretion during \emph{Swift} monitoring \citep{2017ATel10690....1L}, and a 10-day follow-up observational campaign with \emph{NICER}, \emph{Swift} and \INTEGRAL\ in X-rays and ACTA in the radio was performed \citep{2019arXiv190800539L}. During the \INTEGRAL\ observation, the source was clearly detected at 20--80\,keV while it did not show a significant increase in soft X-rays or in the radio \citep{2017ATel10719....1M}. The \emph{NICER} and \INTEGRAL/IBIS data together can be described by the hybrid spectral model for NSs \citep{2007ApJ...667.1073L} consisting of a multi-temperature blackbody for the accretion disc, a single temperature blackbody for a NS boundary layer and a power-law component (see Fig.~\ref{fig:ludlam2019}). In the radio, only a $3\sigma$ upper limit on the flux density could be obtained. When placed on the X-ray--radio luminosity diagram for X-ray binaries, the source proves to lie significantly far from the range expected for BHs, which suggests that the primary compact object is a NS. 

Recently, \cite{2019arXiv191009325F} reported on the source 1RXS~J180408.9$-$342058, an X-ray burster and UCXB candidate. It has been observed in three different spectral states (high/soft, low/very-hard and transitional) using quasi-simultaneous \INTEGRAL, \emph{Swift} and \emph{NuSTAR} observations. The authors show that the source is a new \emph{clocked} X-ray burster, with an accretion rate of $\sim4\times10^{-9}M_\odot$\,yr$^{-1}$ and the decay time of the X-ray bursts longer than $\sim$30\,s. This thermonuclear emission can be explained by a mixed H/He burning triggered by thermally unstable He ignition and strengthens the source's classification as UCXB. During the high/soft state, blackbody emission with $kT_{\rm bb}\sim1.2$\,keV is generated from the accretion disc and the NS surface, which is Comptonized by an optically thick corona with $kT_{\rm e}\sim2.5$\,keV. During the transitional and low/very-hard states, the spectra can be interpreted in terms of a double Comptonizing corona: the seed disc/NS photons ($kT_{\rm bb}\sim1.2$\,keV) are Comptonized by a cool corona ($kT_{\rm e}\sim8$--10\,keV), while another component originates from lower temperature photons  ($kT_{\rm bb}\sim 0.1$\,keV) Comptonized by a hot corona ($kT_{\rm e}\sim35$\,keV).

%In summary, the \INTEGRAL\ broad-band X-ray observations, in combination with other data, have %substantially deepened our understanding of UCXBs and there are bright prospects for further advances %in this field, in particular associated with the upcoming spaceborne gravitational wave astronomy. 

%{\bf *** Bazzano, Fiocchi ***}

%\cite{2006A&A...460..229S} -- The ultra-compact binary 4U 1850-087 observed with INTEGRAL: hard X-ray emission from an X-ray burster. {\bf Overlap with the bursters subsection}

%\cite{2008A&A...484...43F} -- Intermediate long X-ray bursts from the ultra-compact binary candidate SLX 1737-282 {\bf Overlap with the bursters subsection}.

%\cite{2008A&A...492..557F} -- The INTEGRAL long monitoring of persistent ultra compact X-ray bursters.

%\cite{2008A&A...477..239F} -- SAX J1712.6-3739: a persistent hard X-ray source as monitored with INTEGRAL.

%\cite{2011MNRAS.414L..41F} -- The 17-min orbital period in the ultracompact X-ray binary 4U 0513-40

%\cite{2012A&A...539A..32C} -- XMM-Newton observation of 4U 1820-30. Broad band spectrum and the contribution of the cold interstellar medium.

%\cite{2015A&A...582A..32I} -- Signature of the presence of a third body orbiting around XB 1916-053. 

%\cite{2018A&A...610L...2S} -- Discovery of 105 Hz coherent pulsations in the ultra-compact binary IGR J16597-3704. {\bf Strong magnetic field! Overlap with the X-ray pulsars section}

%%%%%%%%
\section{Persistent BH LMXBs}
\label{s:persBH}
%%%%%%%%

Most of the known or candidate BH LMXBs are transients and just three objects (1E~1740.7--2942, GRS~1758--258 and 4U~1957+115) may be considered persistently bright X-ray sources in the sense that they have never been observed to \lq switch' on or off during more than 30 years of their monitoring (see \citealt{2016ApJS..222...15T} for a recent review). It is important to note, however, that there is probably no true dividing line between the transient and persistent classes as even the bona fide persistent BH LMXBs might well be long-term transients. Bearing in mind this uncertainty in the classification, we refer the reader to the review by Motta et al. in this volume for a discussion of \INTEGRAL\ observations of transient BH LMXBs and proceed to discussing \INTEGRAL-based results on persistent BH LMXBs. 

%%%%%%%%%%%
\subsection{1E~1740.7--2942 and GRS~1758--258}

Both sources are located in the bulge of the Galaxy and are known as microquasars due to the discovery of double-sided radio jets \citep{1992Natur.358..215M, 1992ApJ...401L..15R}. Both objects spend most of their time in the low/hard state when they exhibit a flat or inverted radio spectrum with the radio variability correlated with that in hard X-rays. These properties are remarkably different from those associated with outbursts of BH X-ray transients and suggest that the jets in persistent BH LMXBs are intimately related to the Comptonization process \citep{2001MNRAS.322...31F}. 

1E~1740.7--2942, also known as the Great Annihilator, was discovered in 1984 in soft X-rays by \emph{Einstein} \citep{1984ApJ...278..137H} and then reported as a bright hard X-ray source \citep{1987Natur.330..544S}. The observed correlation between its hard X-ray and radio emission \citep{1993A&AS...97..193M} and the X-ray spectral behaviour similar to that of other galactic BH candidates \citep{1991ApJ...383L..49S, 1999ApJ...525..215S} point to a BH origin of the compact object. The source aroused great interest when the \emph{GRANAT} satellite detected a transient broad line around the electron--positron annihilation energy of 511\,keV \citep{1991ApJ...383L..45B, 1991ApJ...383L..49S}, although it was not confirmed by \emph{Compton GRO}/OSSE data \citep{1995A&A...295L..23J}. GRS~1758--258 was discovered during observations of the Galactic Centre region with \emph{GRANAT} \citep{1990IAUC.5032....1M,1991A&A...247L..29S} and later was suggested to be a LMXB \citep{1998A&A...338L..95M, 2002ApJ...569..362S, 2002ApJ...580L..61R}.

The optical/infrared companions of both sources have been difficult to identify owing to the crowded stellar environment and a large column ($\sim 10^{23}$\,cm$^{-2}$) of cold gas in their direction \citep{2002MNRAS.337..869G,2006smqw.confE..80K}. For 1E~1740.7--2942, a tentative NIR counterpart was suggested by \cite{2010ApJ...721L.126M} based on astrometric coincidence with the X-ray and radio position of the microquasar's central core. \cite{2015A&A...584A.122L} excluded an extragalatic origin of 1E 1740.7--2942 by putting an upper limit of 12\,kpc on its distance based on the detection of structural changes in the arcminute radio jets on time scales of about 1 year. Furthermore, these authors revealed a precession of the jets with a period of 1.3 years which allowed them to estimate the distance at 5\,kpc. For GRS 1758--258, recently a reliable counterpart candidate has been finally identified based on astrometric and NIR photometric variability criteria \citep{2010ApJ...721L.126M,2014ApJ...797L...1L}. Moreover, the source's multi-wavelength properties were found to be consistent with its suspected BH LMXB origin. However, \cite{2016A&A...596A..46M} based on optical spectroscopy suggested that GRS 1758--258 hosts a main-sequence star of mid-A spectral type, implying that the system might actually be an intermediate-mass X-ray binary. 

As already mentioned, most of the time these sources have been observed in the low/hard state characterized by an absorbed power law with a photon index of 1.4--1.5 and a high-energy cutoff, as is typical for BH candidates. However, on a number of occasions transitions to a softer state have been recorded. \emph{RXTE} monitored 1E 1740.7--2942 and GRS 1758--258 for many years and revealed an interesting hysteresis effect for changes within the hard state \citep{1999ApJ...525..901M,2002ApJ...569..362S}. Namely, the power-law index ($\Gamma$) was found to anti-correlate with the derivative of the photon flux, i.e. the spectrum is softest when the photon flux is dropping. This is in striking contrast with the behaviour of the BH HMXB Cyg X-1, where soft spectra are observed only when the luminosity is highest. In one observation, another extraordinary event was observed by \emph{RXTE} in 1E 1740.7--2942 \citep{2002ApJ...569..362S}, when its spectrum hardened dramatically while the total photon flux remained unchanged. The same monitoring X-ray data have also been used to tentatively measure the orbital period of 1E~1740.7--2942 at 12.7 days \citep{2002ApJ...569..362S}, implying that if the system accretes by Roche-lobe overflow it must have a red-giant companion. Finally, there is an indication of a very long ($\sim 600$ days) periodicity in this object \citep{2002ApJ...578L.129S}, possibly associated with cyclic transitions between a flat and warped disc, such as seen at somewhat shorter timescales in LMC X-3 and Cyg X-1. For GRS~1758--258, the RXTE monitoring revealed a soft state \citep{2002ApJ...578L.129S}, with the soft component decaying slower than the hard one on a time scale of 28 days.

\begin{figure}
\centering
\includegraphics[width=\columnwidth]{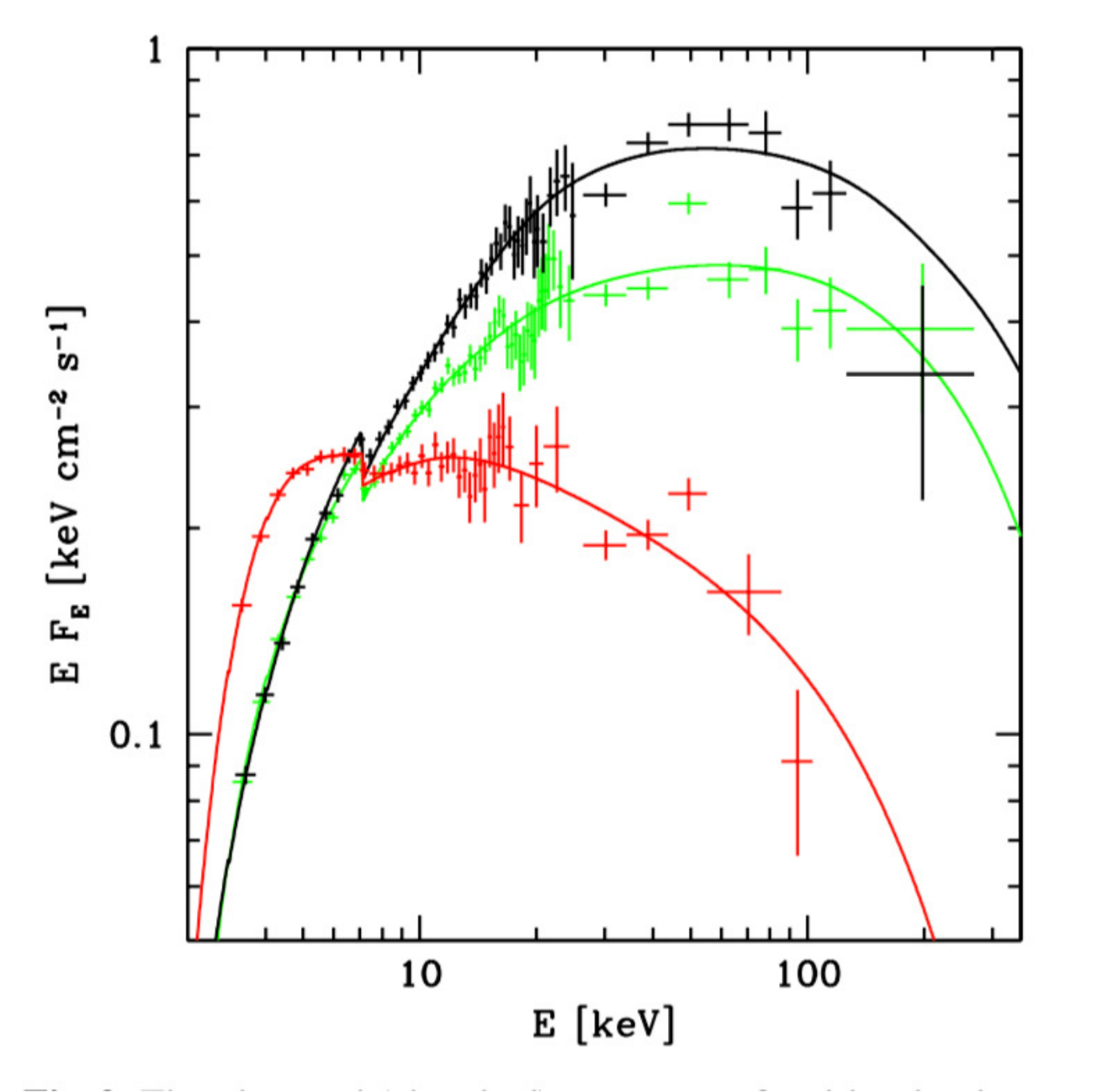}
\caption{Spectra of 1E~1740.7--2942 in the hard (black and green) and soft (red) states obtained with  \emph{RXTE} and \INTEGRAL, fitted by the absorbed thermal Comptonization model. (From \citealt{2005A&A...433..613D})}
\label{fig:1E_delsanto}
\end{figure}

Starting February 2003, \INTEGRAL\ has extensively observed the Galactic Centre region during the visibility periods of 1E~1740.7--2942 and GRS~1758--258. \cite{2005A&A...433..613D} performed a joint spectral and long-term variability analysis of 1E~1740.7--2942 with \INTEGRAL\ and \emph{RXTE}. In the hard state of the source, its spectrum could be measured with IBIS up to a rollover near 100\,keV, which made it possible to estimate the temperature and optical depth of the Comptonizing corona. In September 2003, a spectral transition was detected: the flux at lower energies increased and the slope of the power-law component increased from $\Gamma=1.3$ to 2.3 (see Fig.~\ref{fig:1E_delsanto}). Interestingly, the total flux in this soft state proved somewhat lower than in the hard state. No Compton reflection component was detected in the soft state, which was attributed to a precessing accretion disc being caught nearly edge-on. Recently, \cite{2014ApJ...780...63N} reported on nearly simultaneous \INTEGRAL\ and \emph{NuSTAR} observations of 1E~1740.7--2942 in its low/hard state. The derived spectrum spanning a very broad range of 3--250\,keV could be described by Comptonization on thermal electrons with $kT_{\rm e}\sim 40$\,keV. 

\begin{figure}
\centering
\includegraphics[width=\columnwidth]{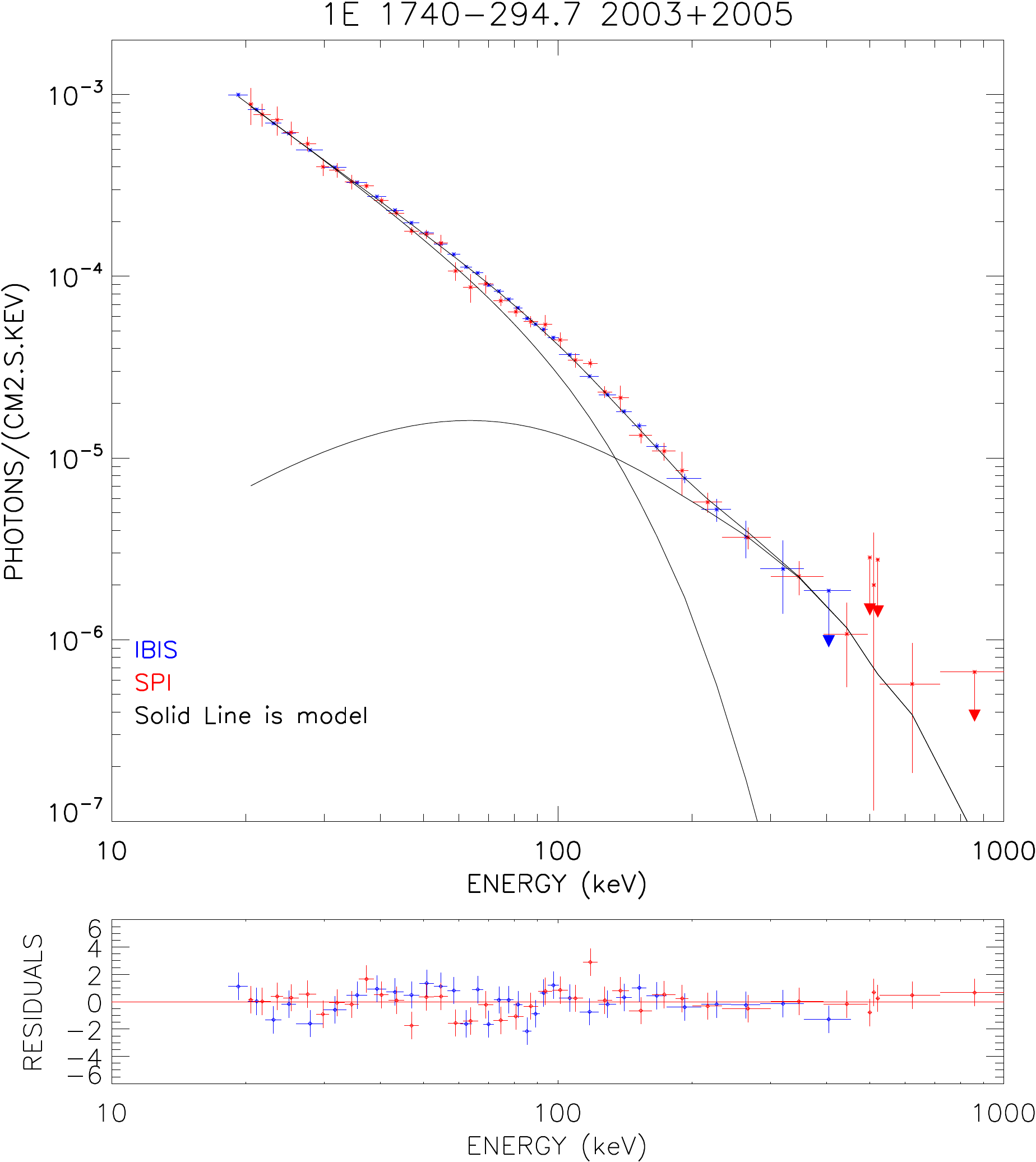}
\caption{1E~1740.7--2942 spectrum measured with IBIS and SPI during the 2003 and 2005 observations. The solid line shows the best-fitting two-temperature Comptonization model. (From \citealt{2009ApJ...693.1871B})}
\label{fig:1E_Bouchet}
\end{figure}

Combining the data from the IBIS and SPI instruments obtained in 2003--2005, \cite{2009ApJ...693.1871B} extended the study of the spectral analysis of 1E~1740.7--2942 into the soft gamma-ray band. Since the angular resolution of SPI is not sufficient to separate 1E 1740.7--2942 from several other nearby sources, information on their relative fluxes provided by IBIS was used in the spectral analysis of the SPI data. Mean spectra of 1E~1740.7--2942 were obtained for the 2003 and 2005 observations, when the source was in the canonical low/hard state (while during 2004, it was very dim in hard X-rays). The hard X-ray continuum was measured up to 600\,keV and revealed excess emission at energies above 200\,keV on top of the thermal Comptonization component (with a high-energy cutoff at $\sim 140$ keV). Alternatively, the broad-band spectrum could be described by a Comptonization model with two populations of hot electrons with $kT_{e1}\sim 30$\,keV and $kT_{e2}\sim 100$\,keV (see Fig.~\ref{fig:1E_Bouchet}), suggesting the presence of two distinct heating regions/mechanisms or a temperature gradient in the Comptonizing plasma. A similar high-energy excess has been found by \INTEGRAL\ during the low/hard state of the BH transient GX~339--4 \citep{2007ApJ...657..400J,2008MNRAS.390..227D} and was suggested to reflect spatial/temporal variations in plasma temperature (see e.g. \citealt{2000A&A...359..843M}). Previously, high-energy components had only been observed in the high/soft state of BH X-ray binaries and were interpreted as Comptonization by a non-thermal electron population (e.g. \citealt{2004PThPS.155...99Z}). 

Finally, no electron-positron annihilation signature has been detected from 1E~1740.7--2942 by \INTEGRAL\ despite its multi-year monitoring of the Galactic Centre region: the $2\sigma$ upper limit obtained with IBIS is $1.6\times10^{-4}$\,ph\,cm$^{-2}$\,s$^{-1}$ on any emission line in the 491--531\,keV range \citep{2011A&A...531A..56D}, which is well below the flux reported from \emph{GRANAT}/SIGMA observations \citep{1991ApJ...383L..45B, 1991ApJ...383L..49S}. This implies that if positron annihilation does occur in 1E~1740.7--2942, it must have a low duty cycle. 

\begin{figure}
\centering
\includegraphics[width=\columnwidth]{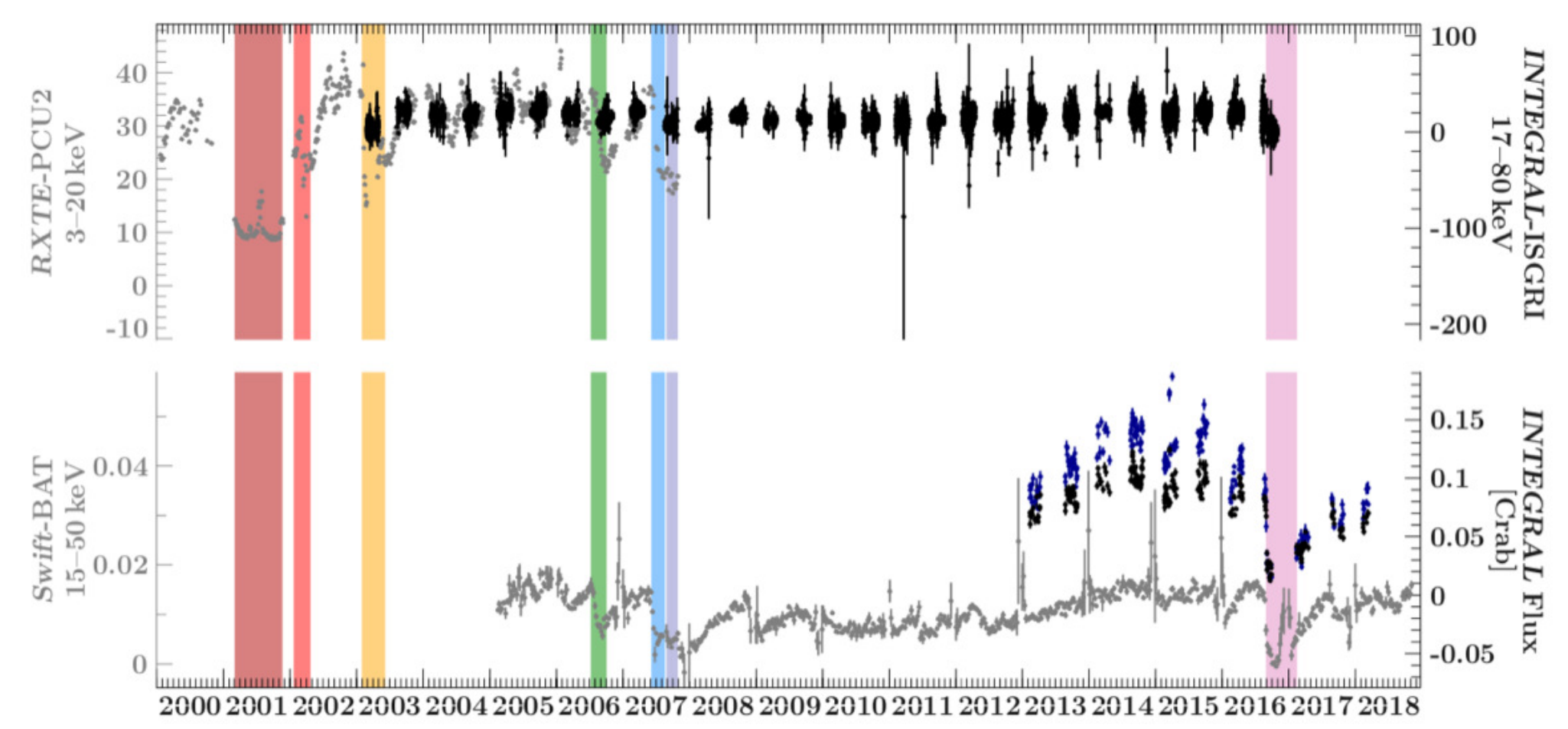}
\caption{Light curve of GRS 1758--258 in various X-ray bands as seen with \emph{RXTE}/PCA (3--20\,keV -- top, grey), \emph{Swift}/BAT (15--50\,keV -- bottom, grey) and \INTEGRAL/IBIS (17--80\,keV -- top, black, 20--40\,keV -- bottom, black and 40--80\,keV -- bottom, blue). (From M. Hirsch's thesis, https://www.sternwarte.uni-erlangen.de/docs/theses/)}
\label{fig:GRS}
\end{figure}

Similar broad-band X-ray studies based on the \INTEGRAL\ and \emph{RXTE} long-term monitoring campaigns have been performed also for GRS~1758--258 (\citealt{2006IAUS..230...93C,2006A&A...452..285P}, see Fig.~\ref{fig:GRS}). The source was found in the hard state most of the time. The derived spectra can be fit by a cutoff powerlaw or a thermal Comptonization model in the 3--300\,keV energy band, with the parameter values typical of the hard state of BH X-ray binaries. \cite{2008int..workE..98P} made an attempt to extend the spectrum of GRS~1758--258 into the soft gamma-ray range using the data from the PICsIT high-energy detector of the IBIS instrument and reported the detection of a positive signal up to $\sim$800\,keV, suggesting the presence of an additional emission component, similarly to the case of 1E~1740.7--2942 in its hard state discussed above. 

On a few occasions, GRS~1758--258 was observed to switch to a soft state \citep{2008int..workE..98P,2016ATel.9625....1P} when its inferred bolometric luminosity remained nearly the same as during the hard state. This \lq dim' soft state, similar to that observed by \INTEGRAL\ in 1E~1740.7--2942 (see above), seems to be different from the canonical soft state observed in persistent BH HMXBs like Cyg X-1 or LMC~X-3, where softening is associated with higher mass accretion rates, and is likely a manifestation of the same hysteresis effect as displayed by BH transients like GX 339--4. This is consistent with persistent BH LMXBs being intrinsically similar to BH transients, as we are probably dealing with Roche lobe overflow from a low-mass companion in both cases in contrast to wind-fed accretion in BH HMXBs.  

%%%%%%%%%%%
\subsection{4U 1957+115} 

The X-ray source 4U~1957+115 and its optical counterpart V1408 Aql have been known for more than 40 years \citep{1974ApJS...27...37G,1978ApJ...221..907M}. Although there are strong indications that the compact object is a BH rather than a NS (X-ray spectral and variability properties as well as a lack of X-ray bursts, e.g. \citealt{2002MNRAS.331...60W,2012ApJ...744..107N}), it has not yet been possible to tightly constrain the parameters of the binary (other than the orbital period of 9.3~hours) and put a lower limit on the mass of the compact object \citealt{2015ApJ...809....9G}). The X-ray properties of 4U~1957+115 are quite unique. It has remained active for more than 40 years and has always been observed in a soft, disc-dominated spectral state with no detectable radio jets \citep{2002MNRAS.331...60W,2011ApJ...739L..19R,2014ApJ...794...85M}. 

Because of its persistently soft spectrum, 4U~1957+115 has remained undetectable in hard X-rays by \INTEGRAL\ in most observations, apart from a two-day interval in mid-October, 2008 when it was detected in the 20--40\,keV energy band \citep{2016ApJS..223...15B}, perhaps indicating a soft-to-hard spectral transition. However, the source is clearly seen with the JEM-X instrument in the 5--10\,keV energy band, with a typical flux $\sim$30\,mCrab.

%%%%%%%%
\section{Sgr A*: persistent emission, flares, and X-ray echoes of the past activity}
\label{s:sgrA}
%%%%%%%%

The supermassive black hole (SMBH) at the centre of our Galaxy, Sgr~A*, is in a very low accretion state at the present epoch, with the estimated accretion rate $\dot{M}\lesssim10^{-7}~M_\odot$\,yr$^{-1}$, i.e. $\lesssim10^{-6}$ of the critical rate for a $\sim4\times10^{6}M_\odot$ black hole \citep[e.g.][]{2015ApJ...809...10Y}. As a result, Sgr~A* is an extremely faint source, with the bolometric-to-Eddington luminosity ratio $L_{\rm bol}/L_{\rm Edd}\sim 10^{-9}$ \citep{2010RvMP...82.3121G} and persistent X-ray luminosity of about $3\times10^{33}$\,erg\,s$^{-1}$\citep{2003ApJ...591..891B}\footnote{Of which a significant fraction may actually be produced by the nuclear cluster of stars \citep{2012MNRAS.420..388S}.}. Detection of such a source with \INTEGRAL\ is hardly feasible, especially taking into account that the Galactic Centre region is densely populated by much brighter (both persistent and transient) hard X-ray sources that cannot be unequivocally separated from Sgr~A* at \INTEGRAL/IBIS's $\sim 12'$ angular resolution \citep[see e.g.][and Section~\ref{s:population} of this review]{2004A&A...425L..49R,2006ApJ...636..275B}.

\begin{figure}
\centering
\includegraphics[width=0.55\columnwidth, angle=-90]{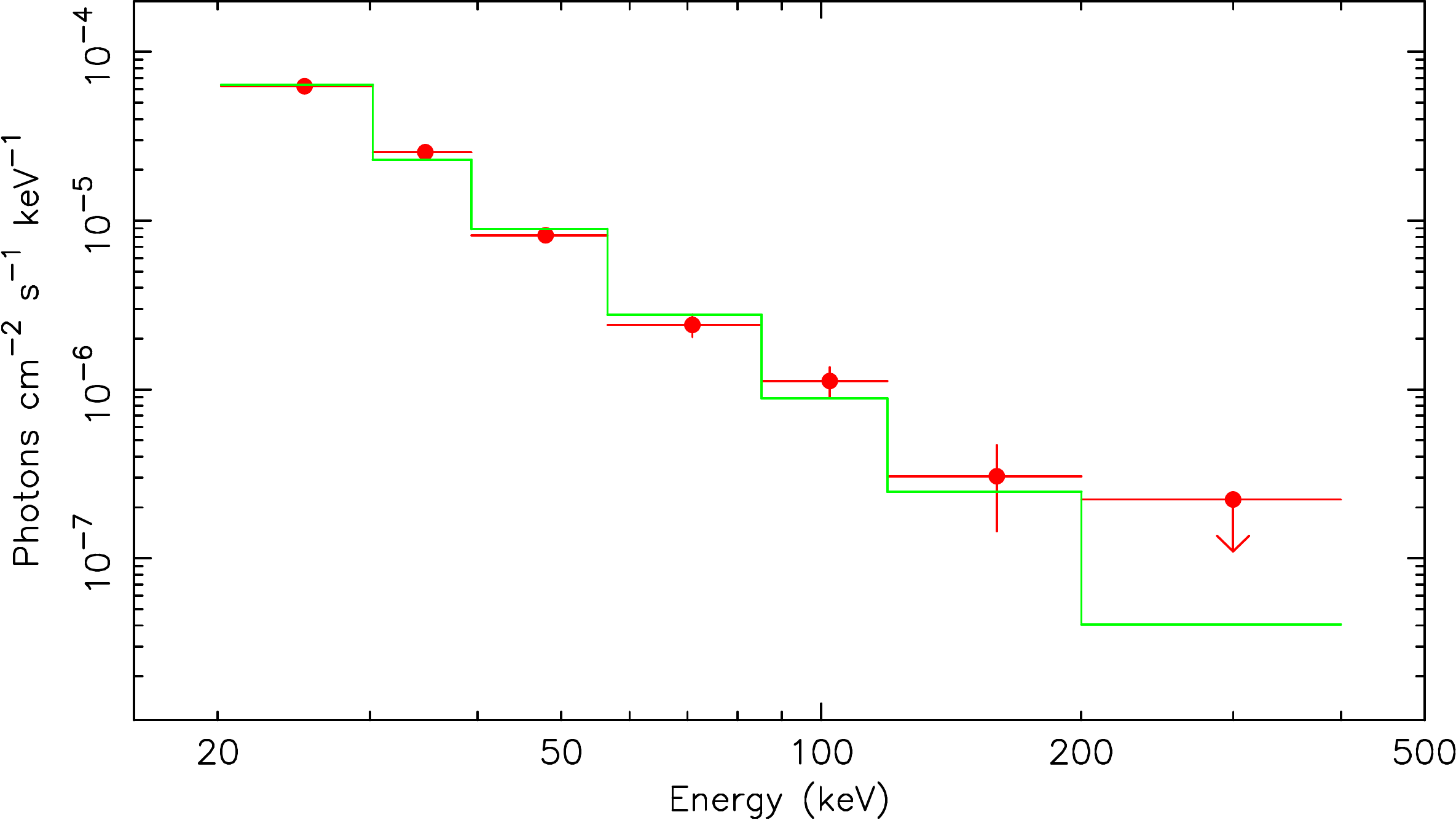}
\caption{Average 20--400 keV spectrum of the source IGR~J17456$-$2901 associated with a compact region around Sgr~A*, obtained by IBIS/ISGRI in 2003--2004. A powerlaw fit with $\Gamma\approx 3.0$ is shown in green. (From \citealt{2006ApJ...636..275B})}
\label{fig:belanger_sgraspec}
\end{figure}

Nonetheless, a persistent hard X-ray source, IGR~J17456--2901, has been detected and localized to within $1'$ of Sgr~A* by IBIS/ISGRI after 7\,Ms of observations, with a 20--400\,keV luminosity of about $5\times10^{35}$\,erg\,s$^{-1}$ (for a 8\,kpc distance to Sgr A*) and a spectral powerlaw slope $\Gamma\approx3$ (\citealt{2006ApJ...636..275B}, see Fig.~\ref{fig:belanger_sgraspec}). No significant variability was revealed on kilosecond to monthly timescales. \citet{2006ApJ...636..275B} concluded that this source is likely a compact (a few arcmin, or $\sim$5\,pc) yet extended region of diffuse emission surrounding Sgr~A*. Recently, {\it NuSTAR} partly resolved IGR~J17456--2901 into non-thermal X-ray filaments, molecular clouds and point sources \citep{2015ApJ...814...94M}. The authors further suggested that the remaining unresolved \lq central hard X-ray emission' is likely the integrated emission of numerous CVs (mostly intermediate polars), similarly to the large-scale Galactic ridge X-ray emission \citep{2007A&A...463..957K}. 

%----------------------------------------
%----------------------------------------

Sgr~A* itself is a variable source: its NIR and soft X-ray emission exhibits intensive flaring over a broad range of timescales with the amplitude ranging from tens to hundreds relative to the \lq quiescent' levels \citep[][]{2012ApJS..203...18W,2018ApJ...863...15W,2015ApJ...799..199N,2019arXiv190801781H}. Thanks to this, cross-correlation with the data of NIR/X-ray monitoring significantly improves opportunities of studying the hard X-ray emission of Sgr A* with \INTEGRAL\ despite the strong contamination by much brighter (known) sources \citep[e.g.][]{2014ApJ...786...46B}.   

Such a study has been conducted by \citet{2010AdSpR..45..507T} based on a multiwavelength campaign performed in April 2007. \INTEGRAL\ monitored the Galactic Centre region simultaneously with IR and soft X-ray telescopes for a total effective exposure $\sim 212$\,ks for IBIS/ISGRI and $\sim 46$\,ks for JEM-X at Sgr~A* position. No hard X-ray/soft $\gamma$-ray counterpart of a flare detected in IR and soft X-rays was  detected, but the presence of persistent emission was confirmed with high significance \citep{2010AdSpR..45..507T}. Measurement of the broadband spectral shape is important for testing the models of these flares, and \INTEGRAL\ provides a very useful possibility to constrain the spectral hardness and the presence of high-energy tails \citep{2006ApJ...644..198Y,2010AdSpR..45..507T}. Therefore, further analysis of \INTEGRAL\ data covering even more intensive flaring episodes \citep[e.g.][]{2019arXiv190801781H} will continue to be useful for the multiwavelength characterization of these events. 

%\cite{2010AdSpR..45..507T} --
%Soft gamma-ray constraints on a bright flare from the Galactic Center supermassive black hole. --
%--------------------------------------------------------------------------------------

Orders of magnitude more powerful flares must have been experienced by Sgr~A* on historical timescales, as suggested by the observed light-travel-time-delayed reflection of its X-ray emission off molecular clouds in the Central Molecular Zone. First indications of such \lq echoes' were obtained with the \textit{GRANAT} observatory, which revealed that the morphology of the hard X-ray emission from the direction of the Galactic Centre closely follows the projected distribution of the dense molecular gas in this region \citep{1993ApJ...407..606S}. The subsequent discovery by {\it ASCA} of an iron fluorescent line at 6.4\,keV associated with the brightest clouds \citep{1996PASJ...48..249K} and detailed studies of the morphology and spectra of this emission with {\it Chandra}, {\it XMM-Newton}, {\it Suzaku}, and {\it NuSTAR} \citep{2007ApJ...656L..69M,2009PASJ...61S.241I,2010ApJ...714..732P,2010ApJ...719..143T,2012A&A...545A..35C,2013A&A...558A..32C,2013PASJ...65...33R,2015ApJ...815..132Z,2017MNRAS.465...45C,2017MNRAS.468.2822K,2018A&A...610A..34C,2018A&A...612A.102T,2019MNRAS.484.1627K,2013ASSP...34..331P} have verified its X-ray reflection origin and opened the way for in-depth exploration of the past Sgr~A* activity and properties of the surrounding dense gas (see \citealt{2017MNRAS.471.3293C,2019BAAS...51c.325C} for a review of the current status and prospects for the future). 

\begin{figure}
\centering
\includegraphics[width=0.75\columnwidth]{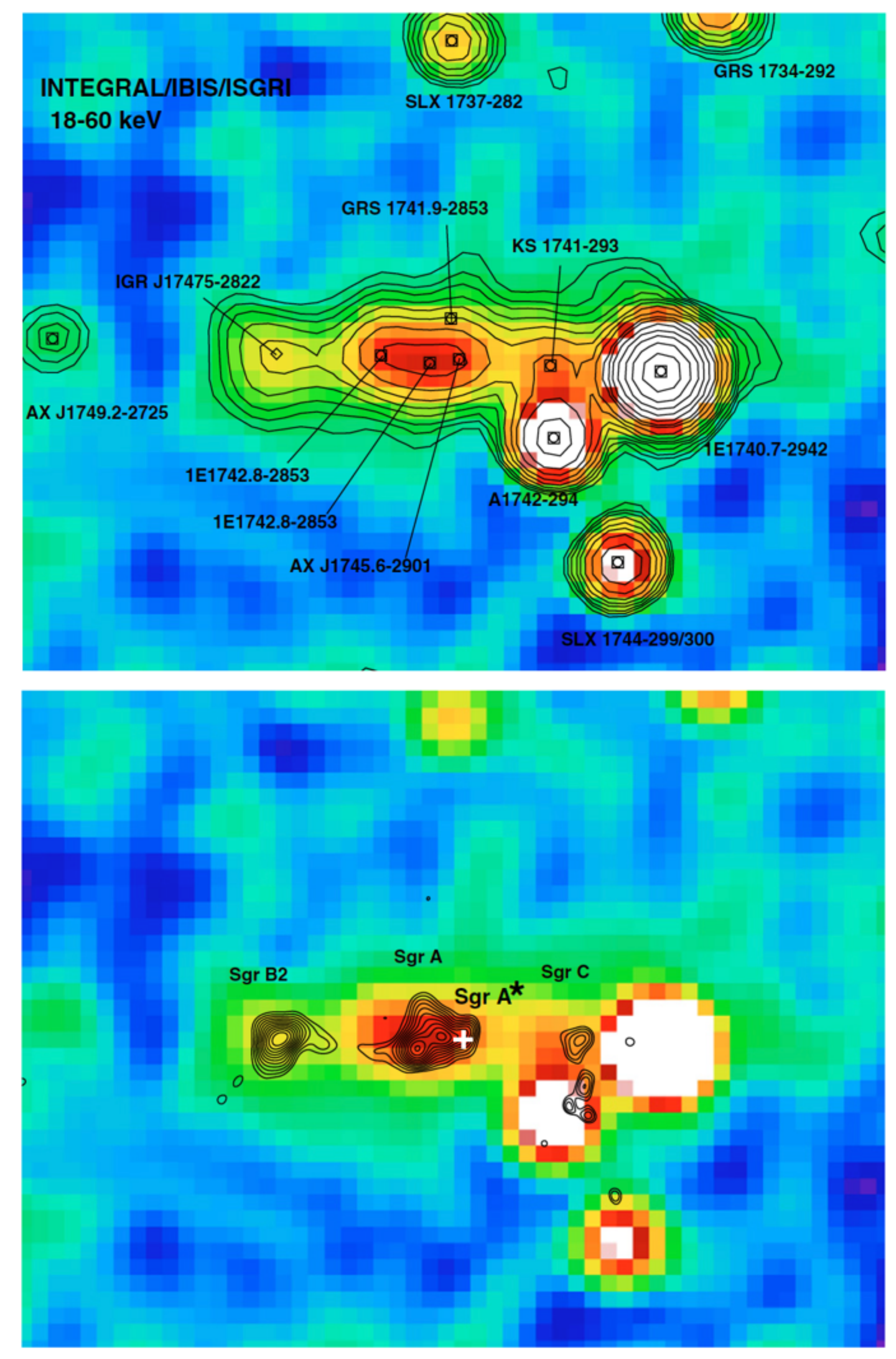}
\includegraphics[width=0.8\columnwidth,viewport=50 200 585 700]{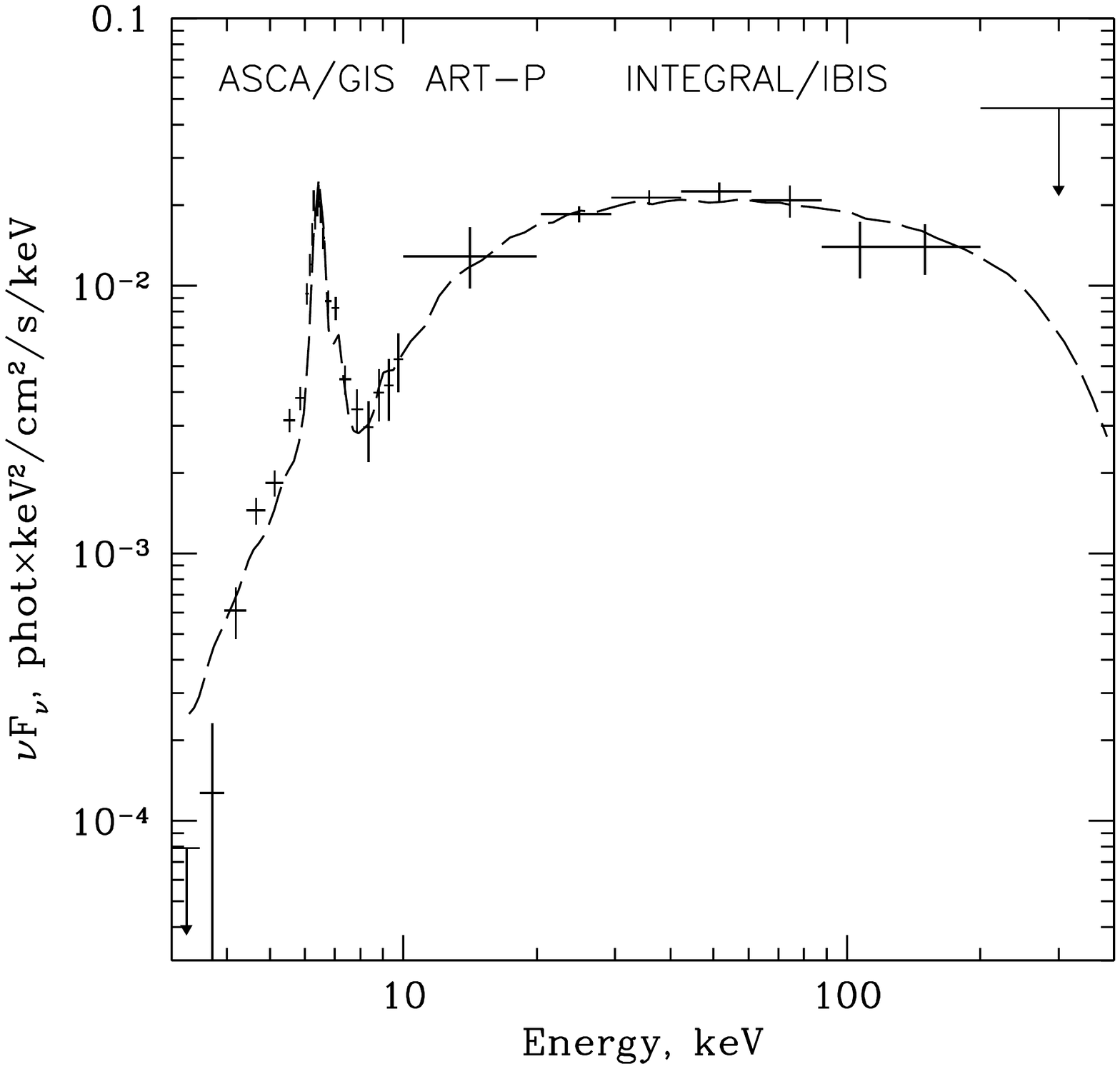}
\caption{Top and middle panels: 18--60\,keV image of a $3.5\times2.5\deg$ Galactic Centre region  obtained with \INTEGRAL\ in 2003--2004. In the top panel, contours of the signal-to-noise ratio are shown along with the names of the known X-ray sources. In the middle panel, contours of the surface brightness distribution in the 6.4\,keV fluorescent line of neutral iron measured by {\it ASCA} are overplotted, with the locations of the most prominent molecular complexes indicated. The location of Sgr~A* is marked with the white cross. Hard X-ray emission from the most massive molecular complex, Sgr~B2, is clearly seen  (identified as the IGR~J12475--2822 source). Bottom panel: broad band X-ray spectrum of IGR~J17475--2822 based on data of {\it ASCA}/GIS, {\it GRANAT}/ART-P and \INTEGRAL/IBIS. The best-fit reflection model is shown with the dashed line. (From \citealt{2004A&A...425L..49R})}
\label{fig:revnivtsev_sgrb2}
\end{figure}

The discovery of hard X-ray emission from the direction of the giant (a few~$10^6$\,$M_{\odot}$) molecular cloud Sgr~B2  by \INTEGRAL\ (\citealt{2004A&A...425L..49R}, see Fig.~\ref{fig:revnivtsev_sgrb2}) has provided further strong evidence that the illuminating flare was associated with the central SMBH rather than with some other transient source. Indeed, the estimated characteristics of the flare are very similar to the emission properties of low-luminosity AGN in terms of the spectral shape (a powerlaw with $\Gamma=1.8$) and luminosity ($L\approx1.5\times10^{39}$\,erg\,s$^{-1}$ in the 2--200\,keV energy band), as opposed to e.g. ultraluminous X-ray sources, which have similar luminosities but significantly softer spectra  \citep[see][]{2013A&A...558A..32C,2013ASSP...34..331P,2017MNRAS.468..165C}. Furthermore, since low-luminosity AGN are common in local galaxies, bright flares such as the one recorded in the past by \INTEGRAL\ may be quite usual for our central SMBH and occur again in the near future.

%----------------------------------------
%----------------------------------------

%\cite{2010ApJ...719..143T} --
%Fading Hard X-ray Emission from the Galactic Center Molecular Cloud Sgr B2.

\begin{figure}
\centering
\includegraphics[width=0.86\columnwidth, viewport=30 30 530 360]{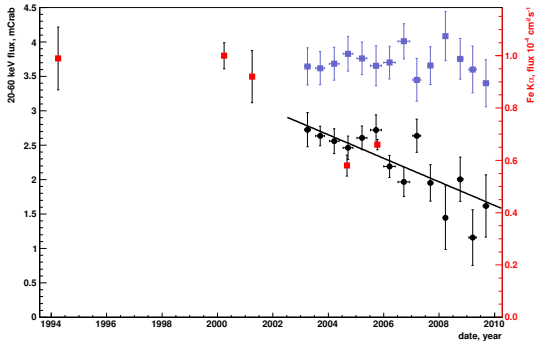}
\caption{Light curves of the 20--60\,keV emission (left axis, mCrab units) from Sgr~B2 (black circles) and a comparison source (Ophiuchus cluster, blue squares) based on \INTEGRAL\ observations from 2003 to 2009. Decay of the hard X-ray flux is clearly visible and broadly consistent with the decay of the flux in the 6.4\,keV iron fluorescent line as measured by {\it ASCA}, {\it Chandra}, {\it XMM-Newton}, and {\it Suzaku} (red squares, right axis) from 1996 to 2006 \citep{2009PASJ...61S.241I}. (From \citealt{2010ApJ...719..143T})}
\label{fig:terrier_sgrb2lc}
\end{figure}

Further \INTEGRAL\ observations together with contemporary {\it XMM-Newton} data revealed a fading of the hard X-ray emission from the Sgr~B2 cloud (\citealt{2010ApJ...719..143T}, see Fig.~\ref{fig:terrier_sgrb2lc}). Although models invoking cosmic rays to explain the soft X-ray emission with the iron fluorescent line along with the hard X-ray continuum are capable of reproducing the observed spectral shape, accounting for the detected variability is very problematic in these scenarios \citep[e.g.][]{2018ApJ...863...85C}, so the X-ray reflection origin of the observed emission gained even more support. 

The characteristic timescale of the measured flux decay, $\sim$8 years, is consistent with the light-crossing-time of the molecular cloud's core, indicating a relatively sharp switching off of the primary flare's emission. Using the estimates of the Sgr~B2 position relative to Sgr~A*, this allowed the time of the flare to be bound between 75 and 155 years ago, in good agreement with the latest estimates based on data from {\it Chandra} \citep{2017MNRAS.465...45C} and {\it XMM-Newton}  \citep{2018A&A...612A.102T}. The resulting remarkable picture of Sgr~A* experiencing outbursts of hard X-rays at least 5 orders of magnitude more luminous than the present \lq quiescent' level and lasting less than a few years provides a unique view of the \lq everyday' life of a low-luminosity AGN in our Galactic backyard.    

%----------------------------------------
%----------------------------------------

%Another advantage of the hard X-ray band is negligible impact of the X-ray absorption involved. Although the majority if the clouds have column densities not exceeding a few $\times10^{23}$ cm$^{-2}$, i.e. the Thompson optical depth $\tau_T\lesssim0.1$, X-ray reflection form the biggest molecular complexes and compact dense cores might be subject to severe photo-absorption in the soft X-ray band. The best example of this kind is the giant molecular complex Sgr~B2 with the mass of a few $\times10^6$ M$_{\odot}$ and extremely rich internal structure of the dense cores and filaments.   

The exceptional coverage of the Galactic Centre region by \INTEGRAL\ over the course of the mission makes it possible to put more attention to the details of the emission variability. Although the angular resolution of \INTEGRAL\ does not permit the exploration of the compact clumps of reflected emission (which are best suited for reconstruction of the primary source's light curve), the fading profile of the largest clouds (e.g. Sgr B2) becomes dominated by the contribution of the doubly scattered emission at the latest times \citep[e.g.][]{1998MNRAS.297.1279S,2011ApJ...740..103O,2016A&A...589A..88M}. Thanks to the low level of photoabsorption in the hard X-ray band, the doubly-scattered albedo is higher than in the standard 4--8\,keV band, and this is further enhanced by the down-scattering of higher energy photons due to Compton recoil. This long-lived signal is sensitive to the distribution of molecular gas in the extended envelopes of the giant clouds, which is of great interest as an intermediate state between the atomic and molecular phases of the interstellar medium \citep[e.g.][]{2020MNRAS.495.1414K}.

%%%%%%%%
\section{Conclusions}
%%%%%%%%

Seventeen years of broad-band X-ray observations with \INTEGRAL\ have provided a unique database for exploration of the Galactic LMXB population, which is still ongoing. The X-ray maps obtained with the IBIS and JEM-X instruments clearly show that LMXBs trace the distribution of old stars in the Milky Way, concentrating towards the Galactic Centre. Thanks to \INTEGRAL, we now have a large, highly complete sample of LMXBs ($\sim$170 identified objects and $\sim$40 further candidates among the currently unidentified objects) with hard X-ray luminosities from $\gtrsim 10^{37}$ down to $\sim 10^{35}$\,erg\,s$^{-1}$. This sample has been used to reliably measure the parameters (the break luminosity and the slopes of the faint and bright parts) of the LMXB luminosity function for the first time, which is crucial for constraining theories of binary stellar evolution and accretion onto relativistic compact objects. 

\INTEGRAL\ has allowed us to significantly increase the samples of LMXBs of specific types, such as Atoll sources, bursters, UCXBs, accretion-powered millisecond pulsars and SyXBs. This has important inter-connections with advances in other branches of astronomy. In particular, UCXBs are expected to be strong sources of low-frequency gravitational waves that can be detected with the future space experiment \emph{LISA}. In addition, thanks to the multi-year monitoring with \INTEGRAL, we have significantly expanded our knowledge of various energetic phenomena associated with accretion of matter onto NSs and BHs, such as X-ray spectral transitions, and physical processes occurring on the surface of NSs, such as thermonuclear X-ray bursts. 

Last but not least, the long-term monitoring of the Galactic Centre with \INTEGRAL\ has shed light on the activity of Sgr A* in the recent past, confirming previous indications that our SMBH experienced a major accretion episode just $\sim$100~years ago and adding further important details to the emergent picture: namely, that the flare ended fairly abruptly and that its broad-band X-ray spectral properties were similar to those of (persistent) low-luminosity AGN. Therefore, our SMBH is not dormant all the time and there is an exciting possibility to observe a new major outburst directly (rather than through X-ray reflection) in not-so-distant future, perhaps during the remaining lifetime of \INTEGRAL.   

As the \INTEGRAL\ observatory continues to operate in orbit, it is important to think about its exploitation for further LMXBs studies in the coming years. Although the accumulated exposure in the Galactic plane and in the Galactic Centre regions is already very high ($\sim$10 and $\sim$40\,Ms per position, respectively) and it will be difficult to significantly deepen the already existing \INTEGRAL\ maps, continued monitoring of the Milky Way by \INTEGRAL\ can certainly provide new valuable information, such as discoveries of new transients and further insights into the long-term behaviour of known LMXBs. 

Over the next several years, there may also be interesting synergies with the all-sky X-ray survey by the \emph{SRG} observatory (launched in July 2019), with its eROSITA and ART-XC grazing-incidence X-ray telescopes. This survey is expected to gradually reach an unprecedented sensitivity of a few $10^{-13}$\,erg\,s$^{-1}$\,cm$^{-2}$ in the 2--10\,keV energy band over the whole sky after 8 six-month scans \citep{2012arXiv1209.3114M}, while similar or somewhat deeper maps will be available for selected regions of the Galaxy already after a series of performance-verification observations at the beginning of the mission \citep{2019AstL...45...62M}. The discovery potential of the \emph{SRG} survey is huge. In particular, it may double the existing sample of Galactic LMXBs \citep{2014A&A...567A...7D}. Despite the substantially lower sensitivity, \INTEGRAL\ hard X-ray observations of the Galaxy may nicely complement the \emph{SRG} data obtained at lower energies, in particular in revealing and studying transient phenomena associated with LMXBs and constructing a maximally full sample thereof.

The Galactic plane and Galactic Centre are rich and ever-changing regions hosting persistent as well as transient sources, spanning over a wide range of X-ray intensities and physical mechanisms. Quantifying the spatial distribution, activity and properties of these sources is essential for population studies and for understanding the evolution of our own Galaxy. \INTEGRAL\ has helped us zoom into the high energy sky, unveiling an impressive wealth of Galactic phenomenology and physics and enabling international and multi-wavelength collaborations. There are all reasons to believe it will continue to do so in the coming years.

\section*{Acknowledgments}

Based on observations with \INTEGRAL, an ESA project with instruments and a science data center funded by ESA member states (especially the PI countries: Denmark, France, Germany, Italy, Spain, and Switzerland), the Czech Republic, and Poland and with the participation of Russia and the USA. The \INTEGRAL\ teams in the participating countries acknowledge the continuous support from their space agencies and funding organizations, in particular the Italian Space Agency ASI (via different agreements including the latest one, 2019-35HH) and the Russian Academy of Sciences. 
MDS acknowledges financial contribution from the agreement ASI/INAF n.2017-14-H.0 and from the INAF main-stream grant (PI T. Belloni). 

We dedicate this review to our late colleague and friend Mikhail Revnivtsev, who has provided a huge contribution to the success of the \INTEGRAL\ mission and was behind many of the results presented in the review.   

\section*{Distance measurements for the INTEGRAL LMXBs}
(D1) \cite{2017A&A...599A..88D}, 
(D2) \cite{2016ApJ...831...89S}, 
(D3) \cite{2018A&A...616A..12G},
(D4) \cite{2010A&A...514A..65K}, 
(D5) \cite{2008ApJS..179..360G}, 
(D6) \cite{2016A&A...586A.142A},
(D7) \cite{2015ApJ...812..149W}, 
(D8) \cite{2005A&A...441..675I}, 
(D9) \cite{2015ApJ...800L..12R},
(D10) \cite{2008A&A...485..183I}, 
(D11) \cite{2017RAA....17..108G}, 
(D12) \cite{2002A&A...392..885C},
(D13) \cite{1989ApJ...338.1024P},
(D14) \cite{2009ApJS..181..238C}, 
(D15) \cite{2015PASJ...67...30S}, 
(D16) \cite{2008A&A...482..113M},
(D17) \cite{2015ApJ...806..265H}, 
(D18) \cite{2004ApJ...616L.139W},
(D19) \cite{2011ApJ...730...75O},
(D20) \cite{1997ApJS..109..177C},
(D21) \cite{2007A&A...470..331M},
(D22) \cite{1999ApJ...512L.121B},
(D23) \cite{2018ATel11272....1C},
(D24) \cite{1997ApJS..109..177C},
(D25) \cite{1998ApJ...492..342C},
(D26) \cite{2018ApJ...859...88K},
(D27) \cite{2005A&A...444..821L},
(D28) \cite{1995Natur.375..464H},
(D29) \cite{2014ApJ...793...79L},
(D30) \cite{2007AJ....133.1287V},
(D31) \cite{2009ApJ...699...60L},
(D32) \cite{2010ATel.2814....1C},
(D33) \cite{2004ApJ...609..317H},
(D34) \cite{1989ESASP.296..185P},
(D35) \cite{2016A&A...596A..21I},
(D36) \cite{2017ApJ...836..111K},
(D37) \cite{2004MNRAS.354..355J},
(D38) \cite{2005A&A...440..287I},
(D39) \cite{2006AJ....132.2171L},
(D40) \cite{2007ESASP.622..373G},
(D41) \cite{2003MNRAS.342..909M},
(D42) \cite{2001MmSAI..72..757C},
(D43) \cite{2002A&A...392..885C},
(D44) \cite{2010MNRAS.401..223A},
(D45) \cite{1994A&A...290..803D},
(D46) \cite{2007A&A...469L..27C},
(D47) \cite{2003A&A...399..663K},
(D48) \cite{2011MNRAS.415.2373S},
(D49) \cite{2006ApJ...641..479H},
(D50) \cite{2018A&A...613A..22B},
(D51) \cite{2015A&A...579A..56B},
(D52) \cite{2010MNRAS.404.1591D},
(D53) \cite{2005A&A...434.1069M},
(D54) \cite{2007ApJ...657L..97K},
(D55) \cite{2008A&A...484...43F},
(D56) \cite{2012MNRAS.422.2661S},
(D57) \cite{2017AstL...43..656M},
(D58) \cite{1996IAUC.6337....1D},
(D59) \cite{2014MNRAS.441..640B},
(D60) \cite{2011AstL...37..597C},
(D61) \cite{2014ApJ...792..109D},
(D62) \cite{2015ApJ...799..123B},
(D63) \cite{2013MNRAS.432.1133M},
(D64) \cite{2017MNRAS.464..840P},
(D65) \cite{2012ApJ...745L...7S},
(D66) \cite{2004A&A...416..311W},
(D67) \cite{2011MNRAS.414L.104D},
(D68) \cite{2010A&A...510A..81C},
(D69) \cite{2007A&A...470.1043O},
(D70) \cite{2006ApJ...639L..31B},
(D71) \cite{2009MNRAS.393..126W},
(D72) \cite{2011ATel.3568....1L},
(D73) \cite{2018ApJ...866...53L},
(D74) \cite{2010MNRAS.409.1136A},
(D75) \cite{2013ApJ...770...10C},
(D76) \cite{1995ApJ...447L..33V},
(D77) \cite{2015A&A...578L..11M},
(D78) \cite{2005A&A...439..575I},
(D79) \cite{2019MNRAS.490.2300M},
(D80) \cite{2002A&A...392..885C},
(D81) \cite{2006ApJ...652..559G},
(D82) \cite{2015MNRAS.450.2915W},
(D83) \cite{2000ApJ...536..891N},
(D84) \cite{1999MNRAS.306..417B}, %GX13+1
(D85) \cite{2009A&A...501....1C},
(D87) \cite{1996AJ....112.1487H},
(D88) \cite{2016ATel.8769....1A},
(D89) \cite{1982ApJ...262..253M},
(D90) \cite{2016ApJ...818..135C},
(D91) \cite{2014MNRAS.439.1390R}, %MAXI1836
(D92) \cite{2000AJ....120.3102C},
(D93) \cite{2012ApJ...759....8D},
(D94) \cite{2012PASJ...64...72S},
(D95) \cite{2005MNRAS.360..825Z},
(D96) \cite{2006A&A...453..295M},
(D97) \cite{2009ApJ...706L.230M}

\section*{References}

\bibliography{lmxb}

\end{document}